\documentclass[]{aa} % for a referee version
\usepackage{graphicx}
\usepackage{txfonts}
\usepackage{numprint}
\usepackage[squaren,Gray]{SIunits}
\usepackage{color}

\newcommand{\ex}{\mathbf{e}_{\rm x}}
\newcommand{\ey}{\mathbf{e}_{\rm y}}
\newcommand{\ez}{\mathbf{e}_{\rm z}}

\newcommand{\rlight}{r_{\rm L}}

\newcommand{\me}{m_{\rm e}}

\newcommand{\LL}{Landau-Lifshitz }
\newcommand{\RR}{radiation reaction }

\DeclareRobustCommand{\rchi}{{\mathpalette\irchi\relax}}
\newcommand{\irchi}[2]{\raisebox{\depth}{$#1\chi$}} % inner command, used by \rchi

\begin{document} 

\title{Particle acceleration and radiation reaction in a strongly magnetized rotating dipole}

  % \subtitle{}

\author{J. P\'etri
	\inst{1}
	%\fnmsep\thanks{Just to show the usage of the elements in the author field}
}

\institute{Universit\'e de Strasbourg, CNRS, Observatoire astronomique de Strasbourg, UMR 7550, F-67000 Strasbourg, France.\\
\email{jerome.petri@astro.unistra.fr}         
}

\date{Received ; accepted }

% \abstract{}{}{}{}{} 
% 5 {} token are mandatory
 
  \abstract
   {Neutron stars are surrounded by ultra-relativistic particles efficiently accelerated by ultra strong electromagnetic fields. These particles copiously emit high energy photons through curvature, synchrotron and inverse Compton radiation. However so far, no numerical simulations were able to handle such extreme regimes of very high Lorentz factors and magnetic field strengths close or even above the quantum critical limit of \numprint{4.4e9}~T.}
   {It is the purpose of this paper to study particle acceleration and radiation reaction damping in a rotating magnetic dipole with realistic field strengths of \numprint{e5}~T to \numprint{e10}~T typical of millisecond and young pulsars as well as of magnetars.}
   {To this end, we implemented an exact analytical particle pusher including radiation reaction in the reduced Landau-Lifshitz approximation where the electromagnetic field is assumed constant in time and uniform in space during one time step integration. The position update is performed using a velocity Verlet method. We extensively tested our algorithm against time independent background electromagnetic fields like the electric drift in cross electric and magnetic fields and the magnetic drift and mirror motion in a dipole. Eventually, we apply it to realistic neutron star environments.}
   {We investigated particle acceleration and the impact of radiation reaction for electrons, protons and iron nuclei plunged around millisecond pulsars, young pulsars and magnetars, comparing it to situations without radiation reaction. We found that the maximum Lorentz factor depends on the particle species but only weakly on the neutron star type. Electrons reach energies up to $\gamma_e \approx \numprint{e8}-\numprint{e9}$ whereas protons energies up to $\gamma_p \approx \numprint{e5}-\numprint{e6}$ and iron up to $\gamma \approx \numprint{e4}-\numprint{e5}$. While protons and irons are not affected by radiation reaction, electrons are drastically decelerated, reducing their maximum Lorentz factor by 2 orders of magnitude. We also found that the radiation reaction limit trajectories fairly agree with the reduced Landau-Lifshitz approximation in almost all cases.}
   {}


   \maketitle
%
%-------------------------------------------------------------------

\section{Introduction}

Neutron stars are known to harbour ultra-strong magnetic fields close to or even above the quantum critical limit of $B_c\approx\numprint{4.4e9}$~T. The subclass of magnetars usually sustains field strengths well above this value of~$B_c$. These stars are therefore able to accelerate leptons and hadrons to extremely relativistic regimes of very high Lorentz factors $\gamma \approx \numprint{e9}$. In such an extreme environment, radiation reaction is expected to drastically perturb their trajectory compared to the pure Lorentz force motion. High energy and very high energy photons are produced and sometimes detected on Earth by Cerenkov telescopes.

Nevertheless, so far a quantitatively accurate study of this acceleration and radiation reaction mechanisms has failed due to the incapability of current numerical algorithms to handle such strong fields. The problem is circumvent by artificially decreasing the magnetic field strength and other relevant physical parameters like the Lorentz factor and meanwhile increasing the associated Larmor radius. Unfortunately, the highly non linearity of the problem renders any extrapolation to realistic fields risky. The only satisfactory results must come from faithfull simulations employing appropriate length and time scales met around neutron stars.

The combination of strong fields and large Lorentz factors leads naturally to strong radiation reaction damping of the charged particle motion. Those trajectories have been computed in the past for test particles like for instance by \cite{finkbeiner_effects_1989} in the pulsar vacuum field. \cite{finkbeiner_applicability_1990} discussed the validity of the Lorentz-Dirac equation and the Landau-Lifshitz approximation used in such computations. \cite{herold_generation_1985} integrated the equation of motion with radiation reaction in the ultra-relativistic regime and showed the difference between radiative damping and no damping for an aligned rotator. They also gave an estimate of the maximum Lorentz factor.

Exact analytical solutions of the Landau-Lifshitz equations have been found for monochromatic plane wave as reported by \cite{piazza_exact_2008} and \cite{hadad_effects_2010}. For constant and uniform electromagnetic fields, solutions are known since the work of \cite{heintzmann_exact_1973}. The latter are special solutions found by removing the temporal and spatial derivatives from the Landau-Lifshitz approximation. This simplified version is sometimes called the reduced Landau-Lifshitz equation (LLR). We will use this approximation to advance in time the position and velocity of charged particles.

Pusher based on exact analytical solutions have been implemented by several authors. For instance \cite{laue_acceleration_1986} evolved particles in an orthogonal magnetic dipole whereas \cite{ferrari_acceleration_1974} investigated particle motion in a dipole field, neglecting the displacement current. Recently \cite{petri_relativistic_2020} developed an algorithm to evolve particles in a strong electromagnetic field. \cite{tomczak_particle_2020} applied it to a magnetic dipole associated to strongly magnetized rotating neutron stars. 
\cite{gordon_superponderomotive_2017} and \cite{gordon_pushing_2017} showed how to implement a fully covariant particle pusher and gave some hints to include radiation reaction. Later \cite{gordon_special_2021} developed a special unitary pusher for extreme fields achieving computation costs comparable to the Boris algorithm \citep{boris_relativistic_1970}.

In the ultra-relativistic regime, radiation reaction almost exactly balance the electric field acceleration leading to a particle velocity only depending on the local electromagnetic field configuration. As shown by \cite{mestel_axisymmetric_1985}, the Lorentz factor can then be deduced from the trajectory curvature. \cite{kelner_synchro-curvature_2015} carefully studied the synchro-curvature radiation of ultra-relativistic particles evolving in a strongly curved electromagnetic field. The pitch angle plays a central role in controlling the synchrotron versus curvature regime.

Several different but not equivalent approaches have been designed to include radiation reaction in a particle pusher for ultra strong electromagnetic fields. \cite{vranic_classical_2016} offers a comprehensive study of the most widely used techniques to implement the radiation reaction force in standard Lorentz force pushers. However, numerical algorithms solving explicitly the \LL equation face some issues to satisfy conservation laws for long time runs. Nevertheless time-symmetric implicit methods seem to give better results \citep{elkina_improving_2014}. Interestingly, exact analytical solutions of the reduced \LL equation have been found several decades ago by \cite{heintzmann_exact_1973} for a constant electromagnetic field. These expressions are used by \cite{li_accurately_2021} for implementation in a PIC code following a projection onto an electric and a magnetic sub-space \citep{boghosian_covariant_1987}. \cite{petri_particle_2021} also applied this exact solution to the acceleration of particles in a low frequency strong amplitude electromagnetic plane wave as that launched by a strongly magnetized rotating neutron star. 

In this paper we study particle acceleration in a realistic neutron star environment, using the exact scaling between the neutron star spin and the cyclotron frequency. In section~\ref{sec:EOM} we recall the equation of motion as derived by \LL and its exact analytical solution, the appropriate normalization and the algorithm. Section~\ref{sec:Tests} presents extensive tests of our algorithm in static fields showing its second order in time convergence. Section~\ref{sec:Etoile} describes an astrophysical application to neutron star electrodynamics and the upper limit of particle acceleration efficiency. Section~\ref{sec:VRR} compares the radiation reaction limit regime to the exact motion. Eventually conclusions are drawn in section~\ref{sec:Conclusions}.

\section{Equation of motion}
\label{sec:EOM}

The self-force produced by an accelerated charge is usually described by the Lorentz-Abraham-Dirac equation (LAD) \citep{abraham_prinzipien_1902, abraham_zur_1904, lorentz_theory_1916, dirac_classical_1938}. Unfortunately this self-force leads to runaway solutions because the associated equation of motion is of third order in time. Several remedies have been found to remove this unacceptable solutions. See for instance \cite{rohrlich_classical_2007} for some discussions. One approach often quoted in the literature is the \LL formulation, a perturbative expansion of the LAD equation \citep{landau_physique_1989}. In the remainder of this paper, we adopt this point of view.

\subsection{Landau-Lifshitz approximation}

In order to get rid of the LAD flaw, \cite{landau_physique_1989} derived an approximation valid in most configurations met in astrophysical applications. This new equation of motion is free of runaway instabilities and is largely employed in the plasma community. 
Their formulation leads to the following equation of motion
\begin{subequations}
	\label{eq:LL}
	\begin{align}
	\frac{du^i}{d\tau} & = \frac{q}{m} \, F^{ik} \, u_k + \frac{q \, \tau_{\rm m}}{m} \, g^i \\
	g^i & = \partial_\ell F^{ik} \, u_k \, u^\ell + 
	\frac{q}{m} \, \left( F^{ik} \, F_{k\ell} \, u^\ell + ( F^{\ell m} \, u_m ) \, ( F_{\ell k} \, u^k ) \, \frac{u^i}{c^2} \right) 
	\end{align}
\end{subequations}
where $q$ and $m$ are the particle charge and rest mass, $u^i$ its 4-velocity, $\tau$ its proper time, $F^{ik}$ the electromagnetic or Faraday tensor, $c$ the speed of light and $\tau_{\rm m}$ the light crossing time across the particle classical radius~$r_{\rm m}$ (within a factor unity)
\begin{equation}\label{eq:tau_m}
\tau_{\rm m} = \frac{q^2}{6\,\pi\,\varepsilon_0\,m\,c^3}.
\end{equation}
It is advantageous to express it in term of the electron classical radius~$r_{\rm e}$ crossing time amounting to 
\begin{equation}\label{eq:tau_e}
\tau_{\rm e} = \frac{2}{3} \, \frac{r_{\rm e}}{c} = \numprint{6.26e-24}~\SIunits{\second} .
\end{equation}
The typical time scale for the radiation reaction is therefore
\begin{equation}\label{eq:tau_me}
\tau_{\rm m} = \frac{2}{3} \, \frac{r_{\rm m}}{c} = \left( \frac{q^2 / e^2}{m/\me}\right) \, \tau_{\rm e}.
\end{equation}
For instance for protons, this time is three orders of magnitude less than for leptons
\begin{equation}\label{eq:tau_p}
\tau_{\rm p} = \frac{m_e}{m_p} \, \tau_{\rm e} = \numprint{3.41e-27}~\SIunits{\second} .
\end{equation}

Interestingly, exact analytical solutions have been computed for eq.\eqref{eq:LL} in some special configurations of electromagnetic fields, time dependent or time independent. We succinctly recall the useful results required for the present work.

\subsection{Exact analytical solutions}

An exact solution for LLR is based on the eigensystem expansion of the electromagnetic tensor ${F^i}_k$. Earlier results were given by \cite{heintzmann_exact_1973}. Here we follow the notation of \cite{li_accurately_2021}. Starting from the Lorentz force written as
\begin{equation}\label{eq:Lorentz}
\frac{du}{d\tau} = G \, u
\end{equation}
where the electromagnetic tensor $F$ has been replaced by $G=q\,F/m$ to absorb the charge over mass ratio,
we decompose the 4-velocity~$u$ in a magnetic and an electric part denoted respectively by $u_B$ and $u_E$ such that $u=u_E+u_B$. The real eigenvalues of ${G^i}_k$ are $\pm \lambda_E$ whereas the imaginary eigenvalues are $\pm i\,\lambda_B$, $\lambda_E$ and $\lambda_B$ being real and positive numbers, with dimensions similar to pulsation thus in $1/s$. Then, each vector $u_E$ and $u_B$ remains in a eigen-subspace satisfying
\begin{subequations}
	\begin{align}
	G \, u_E & = \pm \, \lambda_E \, u_E \\
	G \, u_B & = \pm \, i\,\lambda_B \, u_B .
	\end{align}
\end{subequations}
The vector components $u_E$ and $u_B$ are obtained by defining the projection operators onto the sub-spaces $E$ and $B$ by \citep{boghosian_covariant_1987}
\begin{subequations}
	\begin{align}
	P & = \frac{\lambda_B^2 \, I + G^2}{\lambda_E^2 + \lambda_B^2} \\
	Q & = \frac{\lambda_E^2 \, I - G^2}{\lambda_E^2 + \lambda_B^2}
	\end{align}
\end{subequations}
where $I$ is the identity matrix.
These operators are well defined only if $\lambda_E^2 + \lambda_B^2 \neq 0$. If both electromagnetic invariants vanish, we retrieve a null-like field which requires a different treatment as given for instance by \cite{petri_particle_2021}. In the non null-like field we get
\begin{subequations}
	\begin{align}
	u_E & = P \, u \\
	u_B & = Q \, u .
	\end{align}
\end{subequations}
The equation of motion decouples into two parts given by
\begin{subequations}
	\begin{align}
	\frac{d^2 u_E}{d\tau^2} = + \lambda_E^2 \, u_E \\
	\frac{d^2 u_B}{d\tau^2} = - \lambda_B^2 \, u_B .
	\end{align}
\end{subequations}
The exact analytical solutions with initial conditions $u_E^0 = P \, u^0$ and $u_B^0 = Q \, u^0$ are
\begin{subequations}
	\begin{align}
	u_E(\tau) & = u_E^0 \, \cosh(\lambda_E \, \tau) + G \, u_E^0 \, \frac{\sinh(\lambda_E \, \tau)}{\lambda_E} \\
	u_B(\tau) & = u_B^0 \, \cos(\lambda_B \, \tau) + G \, u_B^0 \, \frac{\sin(\lambda_B \, \tau)}{\lambda_B} .
	\end{align}
\end{subequations}
Adding the radiation reaction in the LLR limit leads to the exact expression
\begin{subequations}
	\label{eq:solution_exacte_totale}
	\begin{align}
	\frac{u_E(\tau)}{c} & = \frac{u_E^0 \, \cosh(\lambda_E \, \tau) + G \, u_E^0 \, \sinh(\lambda_E \, \tau) / \lambda_E}{\sqrt{|u_E^0|^2 + |u_B^0|^2 \, e^{-2\,\alpha\,\tau}}} \\
	\frac{u_B(\tau)}{c} & = \frac{u_B^0 \, \cos(\lambda_B \, \tau) + G \, u_B^0 \, \sin(\lambda_B \, \tau) / \lambda_B}{\sqrt{|u_B^0|^2 + |u_E^0|^2 \, e^{ 2\,\alpha\,\tau}}}	
	\end{align}
\end{subequations}
with $\alpha = \tau_{\rm m} \, (\lambda_E^2 + \lambda_B^2)$. These expressions are similar to the original formulas found by \cite{heintzmann_exact_1973}. The radiation reaction effect becomes perceptible after a time $\tau \approx 1/\alpha$. The component $u_E$ is associated to the accelerating motion induced by the electric field whereas the $u_B$ component is related to the gyro-motion in the magnetic field.
When $\alpha$ vanishes, the radiation reaction effect disappears. The denominators in $u_E$ and $u_B$ reduce to unity and the solution to the Lorentz force 4-velocity components are recovered.

\subsection{Normalisation}

The relevant physical parameters determining the particle trajectory is decided through some normalisation procedure incriminating the following useful quantities in order to write the equation of motion without dimensions. These primary fundamental variables are
\begin{itemize}
	\item the speed of light~$c$.
	\item a typical frequency $\omega$ involved in the problem.
	\item the particle electric charge $q$.
	\item the particle rest mass $m$.
\end{itemize}
From these quantities we derive a typical time and length scale as well as electromagnetic field strengths such that
\begin{itemize}
	\item the length scale $L_0 = c/\omega$.
	\item the time scale $T_0 = 1/\omega$.
	\item the magnetic field strength~$B_0 = m\,\omega/q$.
	\item the electric field strength~$E_0 = c\,B_0$.% = m\,c\,\omega/q$.
\end{itemize}
Normalized quantities will be overlaid with a tilde symbol.

The two important parameters defining the family of solutions are the field strength parameters $a_{B}$ and $a_{E}$ and the radiation reaction efficiency $\omega\,\tau_{\rm m}$ according to the following definitions
\begin{subequations}
	\label{eq:Parametres}
	\begin{align}
	a_{B} & = \frac{B}{B_0} = \frac{\omega_{\rm B}}{\omega} \\
	a_{E} & = \frac{E}{E_0} = \frac{\omega_{\rm E}}{\omega} \\
	b & = \omega\,\tau_{\rm m}.
%	\omega_{\rm B} & = \frac{q\,B}{m} = \frac{q\,E}{m\,c} .
	\end{align}
\end{subequations}
Introducing the weighted and normalized electromagnetic field tensor by~$\tilde{F}^{ik} = q \, F^{ik}/m\,\omega$ and a normalized time $\tilde{\tau} = \omega \, \tau$, the Landau-Lifshitz equation~\eqref{eq:LL} is rewritten without dimensions as
\begin{subequations}
\label{eq:LLnormee}
	\begin{align}
\frac{d\tilde{u}^i}{d\tilde{\tau}} & = \tilde{F}^{ik} \, \tilde{u}_k + b \, \tilde{g}^i \\
\tilde{g}^i & = \tilde{\partial}_\ell \tilde{F}^{ik} \, \tilde{u}_k \, \tilde{u}^\ell + \left( \tilde{F}^{ik} \, \tilde{F}_{k\ell} \, \tilde{u}^\ell + ( \tilde{F}^{\ell m} \, \tilde{u}_m ) \, ( \tilde{F}_{\ell k} \, \tilde{u}^k ) \, \tilde{u}^i \right) .
	\end{align}
\end{subequations}
The normalised and reduced Landau-Lifshits equation reads
\begin{equation}\label{eq:LLRnormee}
\frac{d\tilde{u}^i}{d\tilde{\tau}} = \tilde{F}^{ik} \, \tilde{u}_k + 
b \, \left( \tilde{F}^{ik} \, \tilde{F}_{k\ell} \, \tilde{u}^\ell + ( \tilde{F}^{\ell m} \, \tilde{u}_m ) \, ( \tilde{F}_{\ell k} \, \tilde{u}^k ) \, \tilde{u}^i \right) .
\end{equation}
The particle 4-velocity depends only on the strength parameters~$a_B$ and $a_E$ and on the radiation reaction strength parameter $b$. Therefore it is unnecessary to compute trajectories for different particles possessing the same numbers $a_B, a_E, b$. The only differences reflect in the physical time and space scales involved.

As a rule of thumb, we admit that radiation reaction is negligible whenever the time scale of damping, given by $1/\alpha$ becomes larger than the characteristic time scale of our system, that is $1/\omega$. Expressed in quantities without dimension, we get $\tau_{\rm m} \, \omega \, (a_E^2 + a_B^2) = b \, (a_E^2 + a_B^2) \ll 1$. Therefore the relevant parameter to quantify radiation reaction is not $b$ but the combination of $b$ and the strength parameters $a_B$ and $a_E$. Specific examples will be given in the test section \ref{sec:Tests}.

\subsection{Algorithm}

For the remainder of this paper, we use a Cartesian coordinate system labelled by $(x,y,z)$ and the corresponding Cartesian orthonormal basis $(\ex,\ey,\ez)$.

The velocity vector is integrated analytically following the previous discussion. Unfortunately, for the position vector, there exists no simple analytical expression, although some formulas can be found involving hypergeometric $_2F_1$ functions with complex arguments, see section~\ref{sec:Tests} for an example in a constant magnetic field. The update in particle position is therefore performed by the velocity-Verlet algorithm namely
\begin{subequations}
	\begin{align}
	\vec{u}^{n+1/2} & = \vec{L}(\Delta\tau/2, \vec{u}^n, \vec{E}(\vec{x}^n), \vec{B}(\vec{x}^n)) \\
	\vec{x}^{n+1} & = \vec{x}^n + \vec{u}^{n+1/2} \, \Delta\tau \\
	\vec{u}^{n+1} & = \vec{L}(\Delta\tau/2, \vec{u}^{n+1/2}, \vec{E}(\vec{x}^{n+1}), \vec{B}(\vec{x}^{n+1})) .
	\end{align}
\end{subequations}
The subscript $n$ refers to the proper time $\tau^n = n \,\Delta\tau$ and the same for half integer subscript $\tau^{n+1/2} = (n+1/2) \, \Delta\tau$. 
We found this method more robust than the full analytical update in velocity and position. Indeed for particles trapped in a dipole magnetic field, undergoing bouncing motion with banana orbits typical of magnetic confinement devices for thermonuclear fusion reactors or in Earth magnetosphere known as Van Allen belt, the stability and convergence properties of the velocity-Verlet algorithm is superior.

Before using our code to compute particle acceleration and radiation in the ultra-strong electromagnetic field of a dipole rotating in vacuum, we test it against exact analytical solutions in simple geometric configurations but with very high Lorentz factors and/or very high fields. Results will also be compared to the radiation reaction limit regime which is much less time consuming from a computational point of view but also less accurate in some configurations, section~\ref{sec:VRR}.

\subsection{Radiation reaction limit}

In ultra-strong electromagnetic fields as such present around neutron stars, radiation reaction plays an important role. In the asymptotic limit of ultra-relativistic motions, assuming that the radiation damping exactly balances the electric field acceleration, there exists a simple analytical expression for the particle velocity depending only on the local values of the fields \citep{mestel_axisymmetric_1985}. This velocity is decomposed into an electric drift motion, interpreted as the velocity required to switch to a frame where the electric and magnetic field are aligned, and a motion along this common direction in this new frame. Denoting the velocity vector for positive charges as $\mathbf{v}_+$ and that for negative charges as $\mathbf{v}_-$, we find
\begin{equation}
\label{eq:VRR}
\mathbf{v}_\pm = \frac{\mathbf{E} \wedge \mathbf{B} \pm ( E_0 \, \mathbf{E} / c + c \, B_0 \, \mathbf{B})}{E_0^2/c^2+B^2} .
\end{equation}
It corresponds to particles moving exactly at the speed of light. $E_0$ and $B_0$ are the strength of the electric and magnetic field in the frame where they are aligned. They are obtained from the electromagnetic invariants $\mathcal{I}_1 = \mathbf E^2 - c^2 \, \mathbf B^2 = E_0^2 - c^2 \, B_0^2$ and $\mathcal{I}_2 = c \, \mathbf E \cdot \mathbf B = c\,E_0 \, B_0$. Imposing $E_0\geq0$ we find
\begin{subequations}
	\label{eq:E0B0}
	\begin{align}
	E_0^2 & = \frac{1}{2} \, (\mathcal{I}_1 + \sqrt{\mathcal{I}_1^2 + 4 \, \mathcal{I}_2^2 }) \\
	c\,B_0 & = \textrm{sign} (\mathcal{I}_2) \, \sqrt{E_0^2 - \mathcal{I}_1} .
	\end{align}
\end{subequations}
We will compared the simulation results obtained from this simple prescription with the exact integration of the equation of motion according to LLR.

Applying this radiation reaction limit to neutron star magnetospheres, the velocity in eq.~\eqref{eq:VRR} can be slightly simplified because of the presence of a plasma, the parallel electric field component (with respect to the magnetic field direction) being efficiently screened. In such a configuration, $|\mathcal{I}_2| \ll |\mathcal{I}_1|$ and $\mathcal{I}_1<0$. The velocity then reduces to
\begin{equation}\label{eq:VRRapprox}
\mathbf{v}_\pm \approx \frac{\mathbf{E} \wedge \mathbf{B}}{B^2} \pm \textrm{sign} (\mathbf E \cdot \mathbf B) \,\frac{\sqrt{c^2\,B^2 - E^2}}{B^2} \, \mathbf{B} .
\end{equation}
The first term corresponds to the electric drift speed whereas the second term is associated to the motion along the magnetic field lines, the particle gyro-motion being absent in this picture. We note that the velocity component along the magnetic field reverses sign when crossing a point where $\mathbf E \cdot \mathbf B$ changes sign. These regions are able to trapped particles depending on their charge and on the $(\mathbf E \cdot \mathbf B)$ configuration in the neighbourhood of this surface \citep{finkbeiner_effects_1989}.

Several limiting cases are also useful to discuss. First, if the electric field vanishes, $E_0=0$, the radiated power vanishes too and the particle moves along the field lines with $\mathbf{v}_\pm = \pm c \, \mathbf{B}/B$. Second if the electric field is orthogonal to the magnetic field, $\mathbf E \cdot \mathbf B=0$ and $E<c\,B$, the particle motion is decomposed into an electric drift and a motion along $\mathbf{B}$ such that
\begin{equation}\label{eq:VRREinfB}
\mathbf{v}_\pm = \frac{\mathbf{E} \wedge \mathbf{B}}{B^2} \pm \frac{\sqrt{c^2\,B^2 - E^2}}{B^2} \, \mathbf{B} .
\end{equation}
This expression holds well within the light-cylinder of a force-free magnetosphere.

\section{Tests}
\label{sec:Tests}

We checked our algorithm against simple electromagnetic field configurations containing either only an electric field or a magnetic field or a cross electromagnetic field.
Although the exact solutions are simple expressions, from a numerical point of view it is of paramount importance to ensure the code to be able to handle very high strength parameters and Lorentz factors as those met in neutron star magnetospheres, that is about $a_B \approx 10 ^{20}$ and $\gamma \approx 10^{10}$. Our main purpose in this section is to check that the results are not affected by round off errors.

%\rev{1. Section 3 is dedicated to testing the proposed numerical recipe to solve the 	particle trajectories under radiation reaction. While the results look good in 	general, the standard procedure to quantify the appropriateness of a certain 	numerical recipe is to perform so-called "convergence tests", where one varies 	the size of numerical time-steps and studies the deviation of the solution from 	the analytic expectation as a function of the discrete time-step. For the 	algorithm described in this paper, although the momentum part is given by an 	exact solution of the LLR equation, the position part is updated with a discrete 	time-step, and thus will be subject to discretization error. It would be useful to quantify this error in the evaluation of the numerical algorithm by performing a convergence test to establish an expectation of how quickly this algorithm converges to the analytic solution. Such a test can also be useful to highlight that this method requires resolving the gyrofrequency, and quantify how much it breaks down when this requirement is not met.	}
%\subsection{Constant and uniform fields}

\subsection{Constant electric field}

In a constant electric field, a charged particle is permanently accelerated in the direction of the electric field while it loses energy. Specializing the general solution \eqref{eq:solution_exacte_totale} to a pure electric field aligned with the $z$ axis such that $\vec{E} = E \, \ez$ we get
\begin{subequations}
	\label{eq:quadri_vitesse_electrique_constant}
	\begin{align}
	\label{eq:quadri_vitesse_electrique_constant_gamma}
	\frac{u^t}{c} & = \gamma(\tau) =  \frac{\gamma_0 \, c \, \cosh(\omega_{\rm E} \, \tau) + u_z^0 \,  \sinh(\omega_{\rm E} \, \tau) }{\sqrt{\gamma_0^2 \,c^2 - u_\parallel^2 - u_\perp^2 \, e^{-2\,\alpha\,\tau}} } \\
	\frac{u^x}{c} & = \frac{u_x^0}{\sqrt{(\gamma_0^2 \,c^2 - u_\parallel^2)\, e^{2\,\alpha\,\tau} - u_\perp^2 } } \\
	\frac{u^y}{c} & = \frac{u_y^0}{\sqrt{(\gamma_0^2 \,c^2 - u_\parallel^2)\, e^{2\,\alpha\,\tau} - u_\perp^2 } } \\
	\frac{u^z}{c} & = \frac{u_z^0 \, \cosh(\omega_{\rm E} \, \tau) + \gamma_0 \, c \, \sinh(\omega_{\rm E} \, \tau)}{\sqrt{\gamma_0^2 \,c^2 - u_\parallel^2 - u_\perp^2 \, e^{-2\,\alpha\,\tau}} } 
	\end{align}
\end{subequations}
with $u_\parallel=u_z^0$ the initial 4-velocity component along $\vec{E}$, $u_\perp$ the initial 4-velocity component perpendicular to $\vec{E}$ and $\alpha=\tau_{\rm m}\,\omega_{\rm E}^2$. %Apart from the change in the gyro-frequency, the magnetic field strength impacts only the time scale for the decay via the exponential terms of arguments $2\,\alpha\,\tau$.

For a particle starting at rest, $u_\parallel = u_\perp = 0$ and $\gamma_0=1$, the 4-velocity simplifies drastically into
\begin{equation}
\label{eq:quadri_vitesse_electrique_constant_repos}
u^i = c \, \left(\cosh(\omega_{\rm E} \, \tau), 0, 0, \sinh(\omega_{\rm E} \, \tau) \right).
\end{equation}
This 4-velocity does not depend on the radiation reaction intensity. It accelerates as if it only experiences the Lorentz force. This peculiar situation is well known and discussed in length by \cite{fulton_classical_1960} for a charge and its related classical radiation in an uniformly accelerating field.

The 4-position is given by introducing the two complex functions with help on the hypergeometric functions $_2F_1$ \citep{olver_nist_2010} such that
\begin{subequations}%\label{key}
	\begin{align}
	J_1(\tau) & = \frac{e^{(\alpha+\omega_{\rm E})\,\tau}}{\alpha + \omega_{\rm E}} \, _2F_1 \left(\frac{1}{2}, 1+\frac{1}{2\,b}; \frac{3}{2}  + \frac{1}{2\,b}; \frac{\gamma^2 \, e^{2\, \alpha \, \tau } }{\gamma^2-1} \right) \\
	J_2(\tau) & = \frac{e^{(\alpha-\omega_{\rm E})\,\tau}}{\alpha - \omega_{\rm E}} \, _2F_1 \left(\frac{1}{2}, 1-\frac{1}{2\,b}; \frac{3}{2}  - \frac{1}{2\,b}; \frac{\gamma^2 \, e^{2\, \alpha \, \tau } }{\gamma^2-1} \right) .
	\end{align}\end{subequations}	
Then the time and position are given by
\begin{subequations}
	\label{eq:quadriposition_electric}
	\begin{align}
	t & = - \frac{\gamma  \sqrt{\gamma ^2 \left(e^{2 \alpha  \tau }-1\right)+1}}{2 \left(\gamma
		^2-1\right)} \, (J_2(\tau) + J_1(\tau))+ C_0 \\
	x/c & = 0 \\
	y/c & = \frac{1}{\alpha} \, \arctan \left(\sqrt{\frac{\gamma^2 \, \left(e^{2\, \alpha \, \tau }-1\right)+1}{\gamma
				^2-1}}\right)+ C_2\\
	z/c & = \frac{\gamma  \sqrt{\gamma ^2 \left(e^{2 \alpha  \tau }-1\right)+1}}{2 \left(\gamma
		^2-1\right)} \, (J_2(\tau) - J_1(\tau)) + C_3
	\end{align}
\end{subequations}
with $C_0,C_2,C_3$ complex constants of integration to satisfy the initial conditions.

Returning to eq.~\eqref{eq:quadri_vitesse_electrique_constant}, the typical electric acceleration time scales is $\tau_{\rm acc} \sim 1/\omega_{\rm E}$. On the other side, the radiation damping time scale is $\tau_{\rm rad} \sim 1/\alpha$. The ratio between both time scales is therefore $\tau_{\rm acc}/\tau_{\rm rad} \sim \tau_{\rm m}\,\omega_{\rm E} = b$. As expected, for small damping parameters $b\ll1$, the acceleration time is much shorter than the radiative damping and the particle accelerated as if it would not radiate, until the time $\tau_{\rm rad} \sim \tau_{\rm acc}/b \gg \tau_{\rm acc}$. Note that this rough estimate needs to be corrected by taking into account the initial Lorentz factor as discussed below.

\begin{figure}
	\centering
	\includegraphics[width=\linewidth]{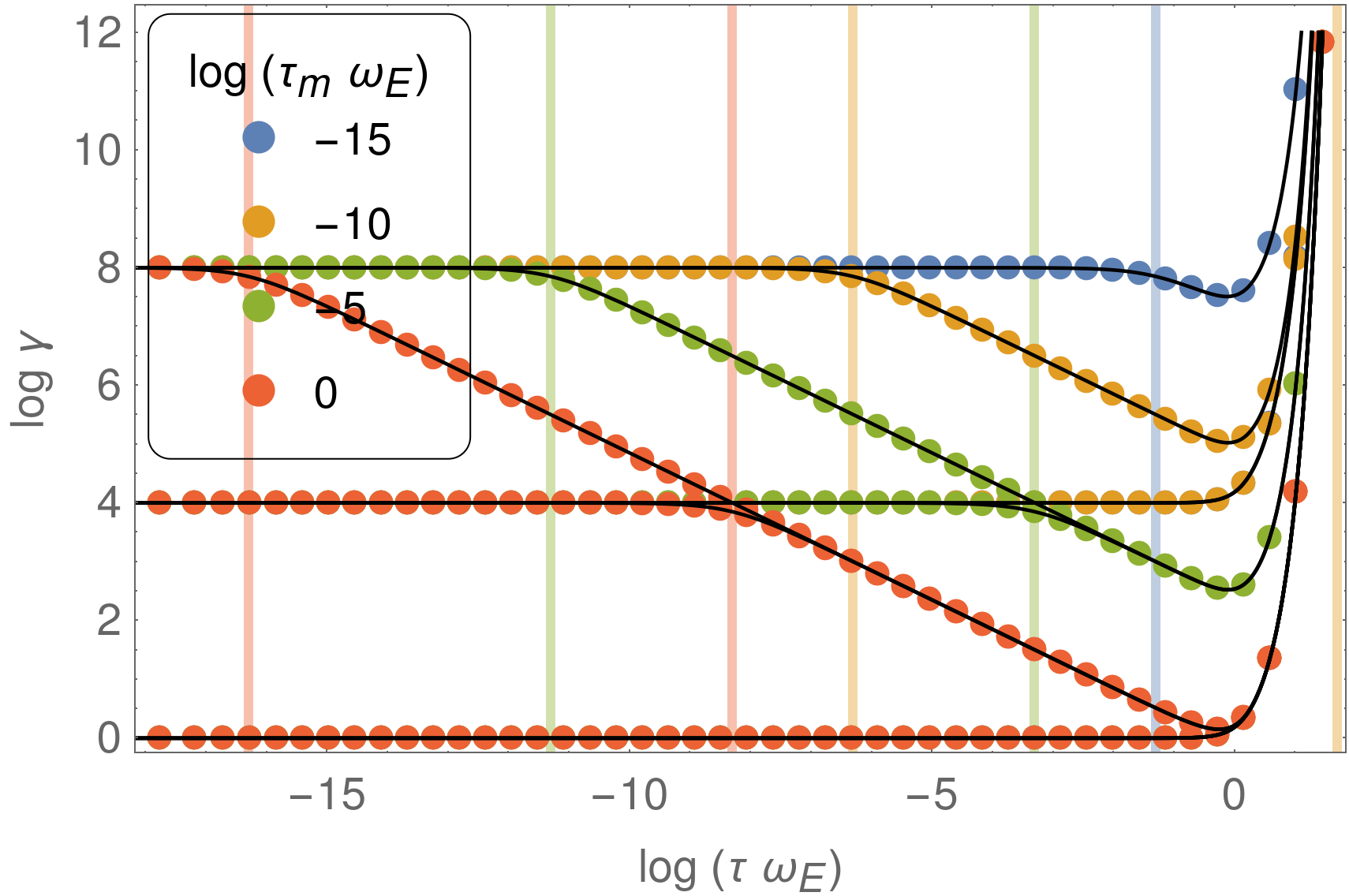}
	\caption{Increase of the Lorentz factor due to radiation reaction for different initial Lorentz factor~$\log\gamma_0=\{0,4,8\}$ and different damping factor~$\log b = \{0,-5,-10,-15\}$. The vertical lines show the time when the damping sets in before the electric acceleration phase starts. Dotted colour points show the simulation results and the black solid lines correspond to the exact analytical solutions.}
	\label{fig:test_electrique_constant_gamma}
\end{figure}

For performing the simulations, we use the characteristic frequency $\omega_{\rm E}$ as normalisation leading to a normalized proper time $\tilde{\tau} = \omega_{\rm E} \, \tau$. Therefore the only relevant parameter apart from the initial conditions is $b = \tau_{\rm m}\,\omega_{\rm E}$ and $\alpha\, \tau = b \, \tilde{\tau}$. Particles starting at rest or possessing an initial velocity directed along the electric field do not suffer from the radiative force. Consequently, as a typical example, particles are starting with an initial velocity perpendicular to the electric field, meaning $u_\parallel=0$ and $u_\perp \neq 0$. We chose different initial Lorentz factors in the set $\log \gamma_0 = \{0,4,8\}$. The damping factor is given by $\log b=\{0,-5,-10,-15\}$. As output of the simulations, we plot the Lorentz factor increasing according to eq.~\eqref{eq:quadri_vitesse_electrique_constant_gamma} and shown in Fig.~\ref{fig:test_electrique_constant_gamma}. For a particle starting at rest, whatever the damping parameter~$b$, the solution is always equal to \eqref{eq:quadri_vitesse_electrique_constant_repos} with an acceleration arising around the time $\omega_{\rm E}\,\tau \sim 1$. This configuration is very particular and is not impacted by radiation reaction. More interesting are the particles starting with a substantial kick velocity and high Lorentz factors $\gamma_0 \gg 1$. The time derivative of the Lorentz factor is always negative for $\gamma_0>1$ because
\begin{equation}%\label{key}
\frac{d\gamma(\tau)}{d\tau} = - \alpha \, \gamma_0 \, (\gamma_0^2-1) < 0
\end{equation}
meaning that the particle first decelerates due to the radiative friction. At large times, when $\alpha\,\tau \gg 1$ and $\omega_{\rm E} \, \tau \gg 1$, the Lorentz factor behaves as $\gamma(\tau) \approx \cosh(\omega_{\rm E} \, \tau)$, loosing its information about the initial state. It resembles to the motion of particle starting at rest, independently of $\gamma_0$. This is because the perpendicular motion is strongly damped, $\lim\limits_{\tau \to +\infty} u_\perp \to 0$ and only the parallel velocity $u_\parallel$ survives at large times with $\lim\limits_{\tau \to +\infty} u_\parallel \to c \, \sinh(\omega_{\rm E}\,\tau)$. In between, the normalized time remains small and the Lorentz factor can be approximated by
\begin{equation}%\label{key}
\gamma(\tau) \approx \frac{\gamma_0 \, \cosh (\omega_{\rm E} \, \tau)}{\sqrt{1+2\,\gamma_0^2 \, \alpha \, \tau}} .
\end{equation}
Therefore, before the acceleration phase starts, there is a deceleration step arising at time $\omega_{\rm E}\,\tau \sim 1/2\,\gamma_0^2\,b$. These values agree with the curves in figure~\ref{fig:test_electrique_constant_gamma}. If $\gamma_0^2\,b \lesssim 1$, the radiation reaction force has no time to set in and the motion tends to a purely accelerated regime given by eq.\eqref{eq:quadri_vitesse_electrique_constant_repos}. This is for instance the case with $\gamma_0 = 10^4$ and $b=10^{-10}$, orange dots, or $\gamma_0 = 10^8$ and $b=10^{-15}$ which is just on the edge of this condition, showing a weak deceleration right before the electric boost, blue dots. The perpendicular momentum decrease is not necessarily significant before the acceleration, it is controlled by $b$ and $\gamma_0$ because at time $\omega_{\rm E}\,\tau \approx 1$ it braked to a momentum
\begin{equation}\label{eq:uperp}
 u_\perp(\tau) \approx \frac{u_\perp^0}{\sqrt{1+2\,\gamma_0^2\,b}} .
\end{equation}
Thus again, radiation reaction impacts the motion if $\gamma_0^2\,b \gtrsim 1$. Consequently, it is the combination $\gamma_0^2\,b$ that controls the damping efficiency, not $b$ alone found from the simple arguments above.

\begin{figure}
	\centering
	\includegraphics[width=\linewidth]{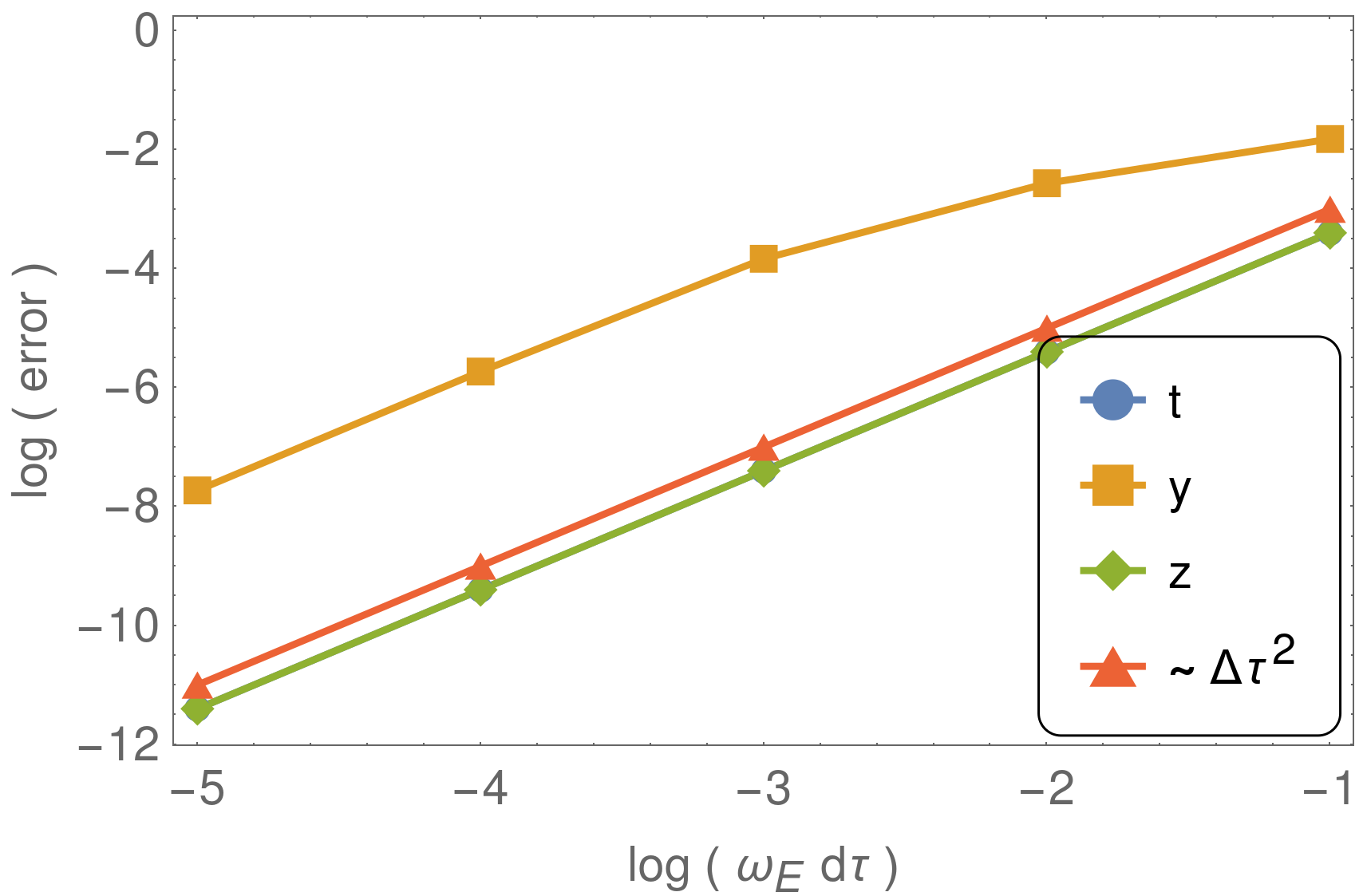}
	\caption{Relative error of the position $y,z$ and time $t$ as shown in the legend. The error decreases with second order in $\Delta \tau$ as given by the green line $\Delta \tau^2$ for $\log \gamma_0=4$ and $\log b=-5$. $t$ and $z$ errors overlap and are undistinguishable.}
	\label{fig:erreuraccelerationa0g4}
\end{figure}

Because the 4-position of the particle is computed numerically and not analytically according to eq.~\eqref{eq:quadriposition_electric}, it is important to estimate the convergence rate of our scheme. To this end Fig.~\ref{fig:erreuraccelerationa0g4} shows the error in the $y$ and $z$ position and time~$t$ with decreasing proper time step $\Delta \tau$ for $\log \gamma_0=4$ and $\log b=-5$ in blue, orange, green and red respectively. The second order expectations are depicted by the green line. We conclude that the decrease in the relative error follows a second order in time scheme as expected from the velocity-Verlet algorithm exposed in section~\ref{sec:EOM}.

\subsection{Constant magnetic field}

A charged particle orbiting in a constant magnetic field loses energy and decays until it rests. The rate of decay is controlled by the magnetic field strength only. The exact solution for the 4-velocity in a magnetic field directed along the $z$ axis with $\vec{B} = B \, \ez$ is given by
\begin{subequations}
	\label{eq:quadri_vitesse_orbite_larmor}
	\begin{align}
	\label{eq:quadri_vitesse_orbite_larmor_gamma}
	\frac{u^t}{c} & = \gamma(\tau) = \frac{\gamma_0 \, c}{\sqrt{\gamma_0^2 \,c^2 - u_\parallel^2 - u_\perp^2 \, e^{-2\,\alpha\,\tau}} } \\
	\frac{u^x}{c} & = \frac{u_x^0 \, \cos (\omega_B \, \tau) + u_y^0 \, \sin (\omega_B \, \tau)}{\sqrt{(\gamma_0^2 \,c^2 - u_\parallel^2)\, e^{2\,\alpha\,\tau} - u_\perp^2 } } \\
	\frac{u^y}{c} & = \frac{u_y^0 \, \cos (\omega_B \, \tau) - u_x^0 \, \sin (\omega_B \, \tau)}{\sqrt{(\gamma_0^2 \,c^2 - u_\parallel^2)\, e^{2\,\alpha\,\tau} - u_\perp^2 } } \\
	\frac{u^z}{c} & = \frac{u_z^0}{\sqrt{\gamma_0^2 \,c^2 - u_\parallel^2 - u_\perp^2 \, e^{-2\,\alpha\,\tau}} } 
	\end{align}
\end{subequations}
with $u_\parallel=u_z^0$ the initial 4-velocity component along $\vec{B}$, $u_\perp$ the initial 4-velocity component perpendicular to $\vec{B}$ and $\alpha=\tau_{\rm m}\,\omega_{\rm B}^2$. Apart from the change in the gyro-frequency, the magnetic field strength impacts only the time scale for the decay via the exponential terms of arguments $2\,\alpha\,\tau$.

For performing simulations, we use the characteristic frequency $\omega_{\rm B}$ as normalisation with $\tilde{\tau} = \omega_{\rm B} \, \tau$. Therefore the only relevant parameter apart from the initial conditions is $b=\tau_{\rm m}\,\omega_{\rm B}$ and $\alpha\, \tau = b \, \tilde{\tau}$. The length scale is therefore given in units of the non-relativistic Larmor radius
\begin{equation}\label{eq:Rayon_Larmor}
r_{\rm B} = \frac{c}{\omega_{\rm B}} .
\end{equation}
%For a low self-force intensity, the energy goes into synchrotron radiation braking the particle with a Lorentz factor evolving in time according to
%\begin{equation}
%\gamma(t) = \frac{e^{2\,K\,t} - C}{e^{2\,K\,t} + C}
%\end{equation}
%with the integration constant
%\begin{equation}
%C = \frac{1-\gamma_0}{1+\gamma_0}
%\end{equation}
%and
%\begin{equation}
%K = 2 \, \frac{\sigma_T \, U_B \, \sin^2\alpha}{m\,c}
%\end{equation}
%with $\alpha$ the pitch angle.

Integrating the 4-velocity vector, an exact analytical expression for the 4-position is computed with help on the hypergeometric functions $_2F_1$. Introducing the complex functions
\begin{subequations}%\label{key}
\begin{align}
H_1(\tau) & = e^{+i\,\omega_{\rm B}\,\tau}\, _2F_1 \left(\frac{1}{2}, +\frac{i}{2\,b}; 1+\frac{i}{2\,b}; \frac{\gamma^2 \, e^{-2\, \alpha \, \tau } }{\gamma^2-1} \right) \\
H_2(\tau) & = e^{-i\,\omega_{\rm B}\,\tau}\, _2F_1 \left(\frac{1}{2}, -\frac{i}{2\,b}; 1-\frac{i}{2\,b}; \frac{\gamma^2 \, e^{-2\, \alpha \, \tau } }{\gamma^2-1} \right) .
\end{align}\end{subequations}	
the solution reads
	\begin{subequations}
		\label{eq:quadri_position_orbite_larmor}
		\begin{align}
		\label{eq:quadri_position_orbite_larmor_gamma}
		t & = \frac{1}{\alpha} \, \tanh^{-1} \, \left(\frac{\gamma \, e^{\alpha \,  \tau }}{\sqrt{\gamma ^2 \, \left(e^{2	\, \alpha \, \tau} - 1\right) + 1 } } \right) + C_0 \\
	 x / r_{\rm B} & = \frac{H_1(\tau) + H_2(\tau)}{2\,i} + C_1  \\
	% & = \frac{e^{i\,\omega_{\rm B}\,\tau}\, _2F_1(\frac{1}{2},\frac{i}{2\,b}, 1+\frac{i}{2\,b}, \frac{\gamma^2 \, e^{-2\, \alpha \, \tau } }{\gamma^2-1}) + e^{-i\,\omega_{\rm B}\,\tau}\, _2F_1(\frac{1}{2}, -\frac{i}{2\,b}, 1-\frac{i}{2\,b}, \frac{\gamma^2 \, e^{-2\, \alpha \, \tau } }{\gamma^2-1})}{2\,i} + C_1 \\
	 y /  r_{\rm B}  & = \frac{H_1(\tau) - H_2(\tau)}{2} + C_2 \\ %& = \Im[P(\tau)] \\
	 %& = \frac{e^{i\,\omega_{\rm B}\,\tau}\, _2F_1(\frac{1}{2},\frac{i}{2\,b}, 1+\frac{i}{2\,b}, \frac{\gamma^2 \, e^{-2\, \alpha \, \tau } }{\gamma^2-1}) - e^{-i\,\omega_{\rm B}\,\tau}\, _2F_1(\frac{1}{2}, -\frac{i}{2\,b}, 1-\frac{i}{2\,b}, \frac{\gamma^2 \, e^{-2\, \alpha \, \tau } }{\gamma^2-1})}{2} + C_2 \\
	z /  r_{\rm B} & = 0
		\end{align}
	\end{subequations}
where the $C_i$ with $i\in[0..2]$ are complex constants of integration to satisfy the initial conditions.

\begin{figure}
	\centering
	\includegraphics[width=\linewidth]{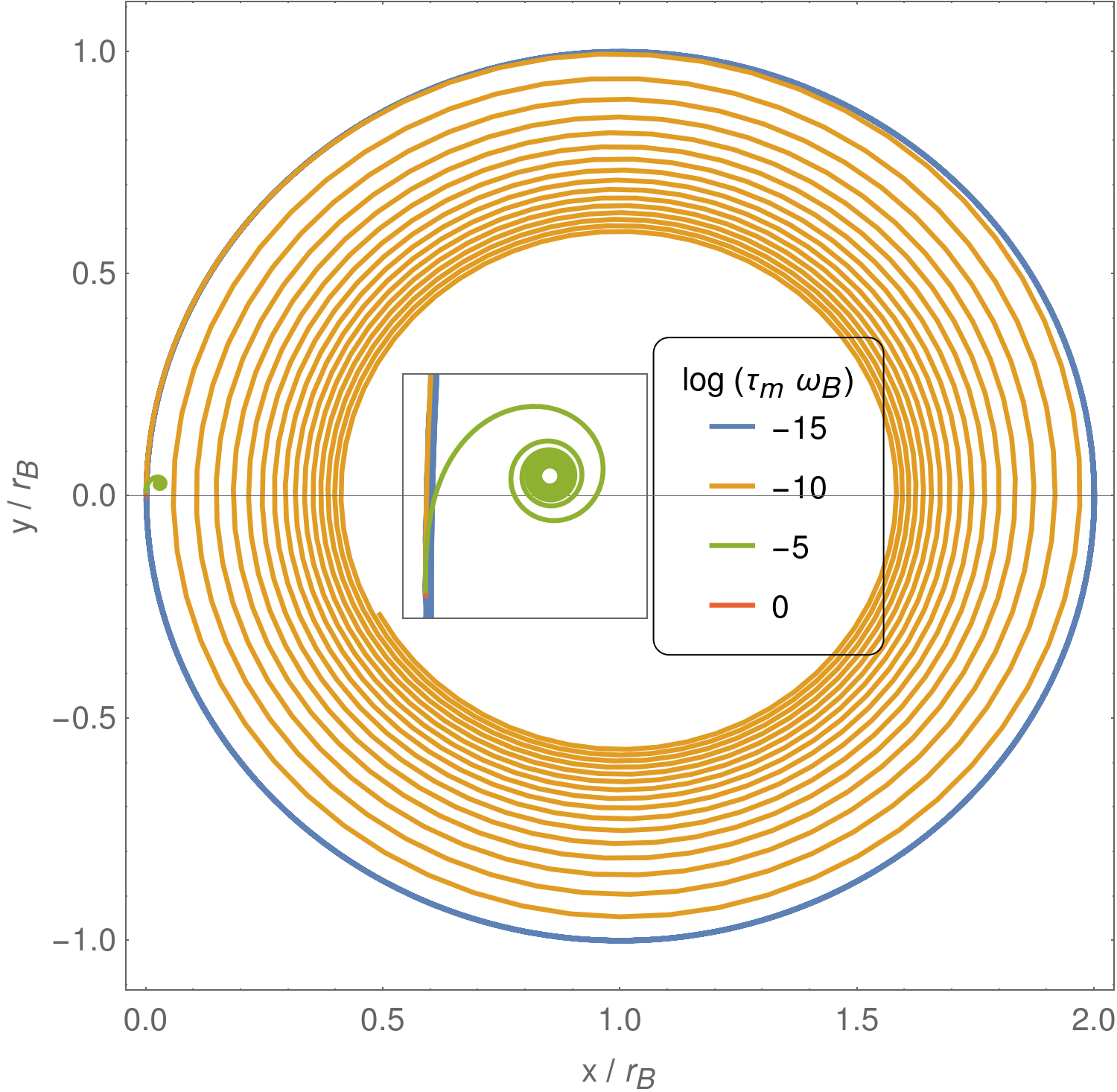}
	\caption{Particle orbit in an uniform and constant magnetic field and subject to radiation reaction. The initial Lorentz factor is $\log \gamma_0 = 4$. The inset shows the strong damped motion in green and even stronger damping in red where the spiralling is not seen.}
	\label{fig:orbite_larmor}
\end{figure}

\begin{figure}
	\centering
	\includegraphics[width=\linewidth]{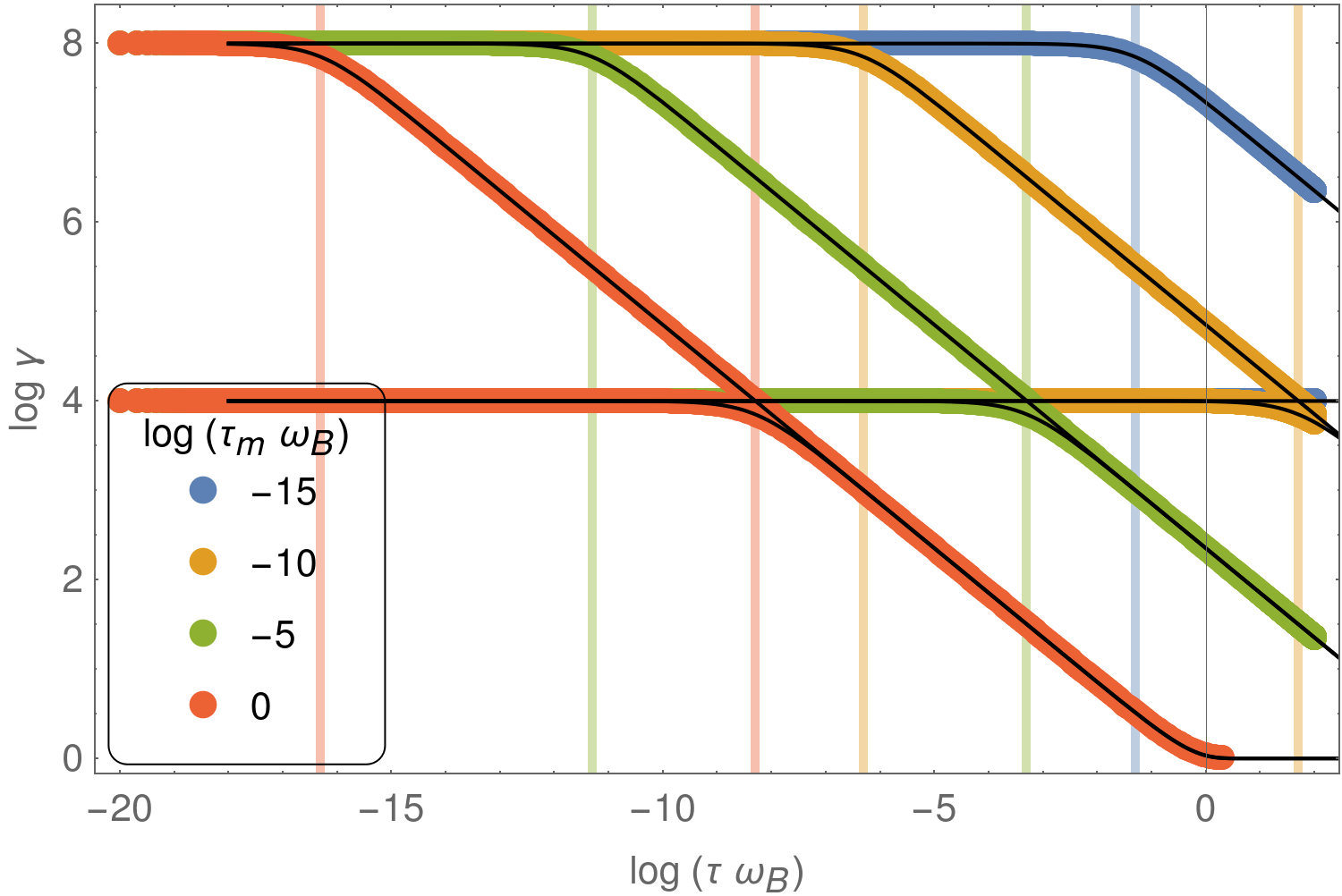}
	\caption{Decrease of the Lorentz factor due to radiation reaction in an uniform and constant magnetic field associated to the orbits shown in fig.~\ref{fig:orbite_larmor}. Dotted colour points show the simulation results and the black solid lines correspond to the exact analytical solutions.}
	\label{fig:lorentz_magnetique}
\end{figure}

The particle trajectory follow a spiral as shown in Fig.~\ref{fig:orbite_larmor}. The particle comes to rest after a typical time $\omega_{\rm B} \, \tau_\infty \gg 1/b$.
The corresponding Lorentz factor decreases according to eq.~\eqref{eq:quadri_vitesse_orbite_larmor_gamma} and is shown in Fig.~\ref{fig:lorentz_magnetique}. The time when damping sets in is given approximately by $2\,\alpha\,\gamma_0^2\,\tau \approx1$. These times are shown as coloured vertical lines in the Fig.~\ref{fig:lorentz_magnetique}. If the particle moves along the field line, it experiences no damping and keeps a uniform motion.

\begin{figure}
	\centering
	\includegraphics[width=\linewidth]{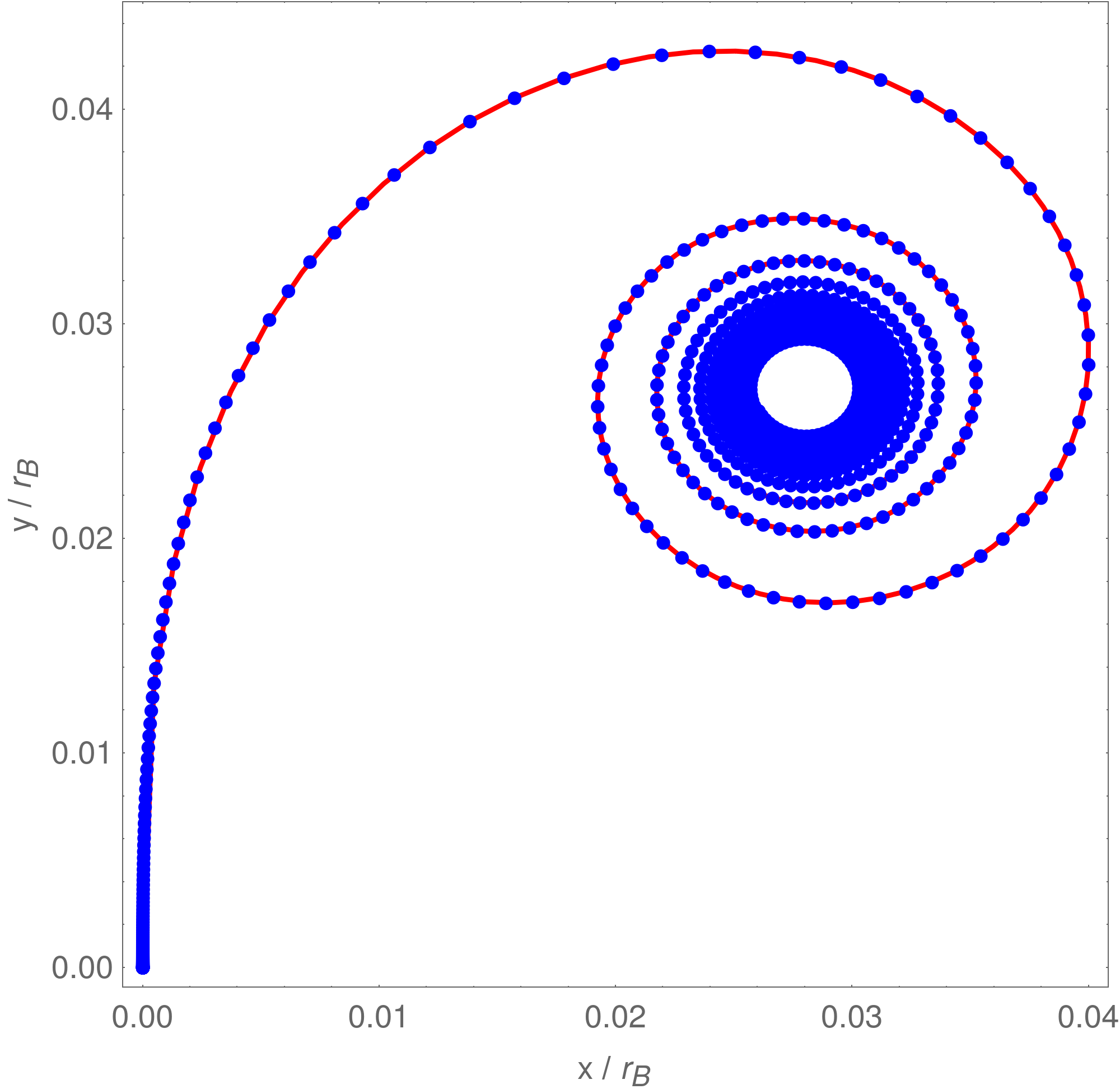}
	\caption{Comparison between the analytical solution, eq.~\eqref{eq:quadri_position_orbite_larmor}, in red solid line and the numerical simulation in blue dots for $\log \gamma_0=4$ and $\log b=-5$.}
	\label{fig:testmagnetiqueconstanttrajectoirecomparaison}
\end{figure}

\begin{figure}
	\centering
	\includegraphics[width=\linewidth]{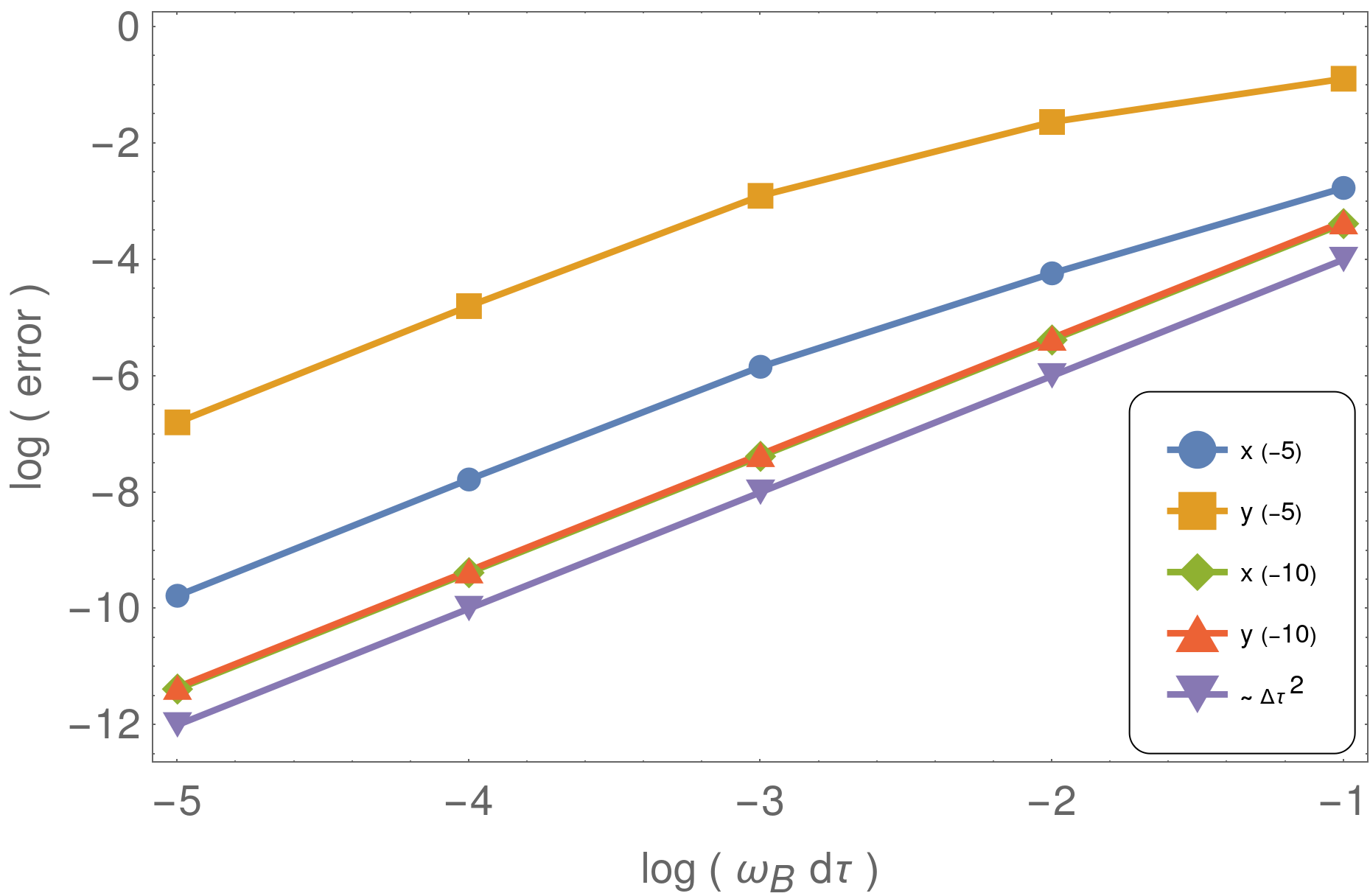}
	\caption{Relative error of the position $x$ and $y$ as shown in the legend. The error decreases with second order in $\Delta \tau$ as given by the violet line $\Delta \tau^2$ for $\log \gamma_0=4$ and $\log b=\{-5,-10\}$.}
	\label{fig:erreurspiralea0g4}
\end{figure}

A comparison between the analytical trajectory in red solid line and the numerical integration in blue dots is shown in Fig.~\ref{fig:testmagnetiqueconstanttrajectoirecomparaison}. A more quantitative agreement is proven in Fig.~\ref{fig:erreurspiralea0g4} where the relative error decreases with respect to the proper time step~$\Delta\tau$. Here also the method is second order in time as expected.

\subsection{Cross electric and magnetic field}

The cross electric and magnetic field configuration is a stringent test for an ultra relativistic particle pusher. If the electric field strength is less than the magnetic field strength $E<c\,B$, then an appropriate Lorentz transform brings the problem to a frame where the electric field vanishes. We therefore return to the situation of the last section with a constant and uniform magnetic field. For sufficient long time, the only remaining motion is the electric drift at speed $\vec{v}_{\rm E} = \vec{E} \wedge \vec{B} /B^2$. Therefore the velocity is $\beta_{\rm E} = v_{\rm E}/c = E/cB$ and the corresponding final Lorentz factor $\gamma_\infty = (1-\beta_{\rm E}^2)^{-1/2}$.

\begin{figure}
	\centering
	\includegraphics[width=\linewidth]{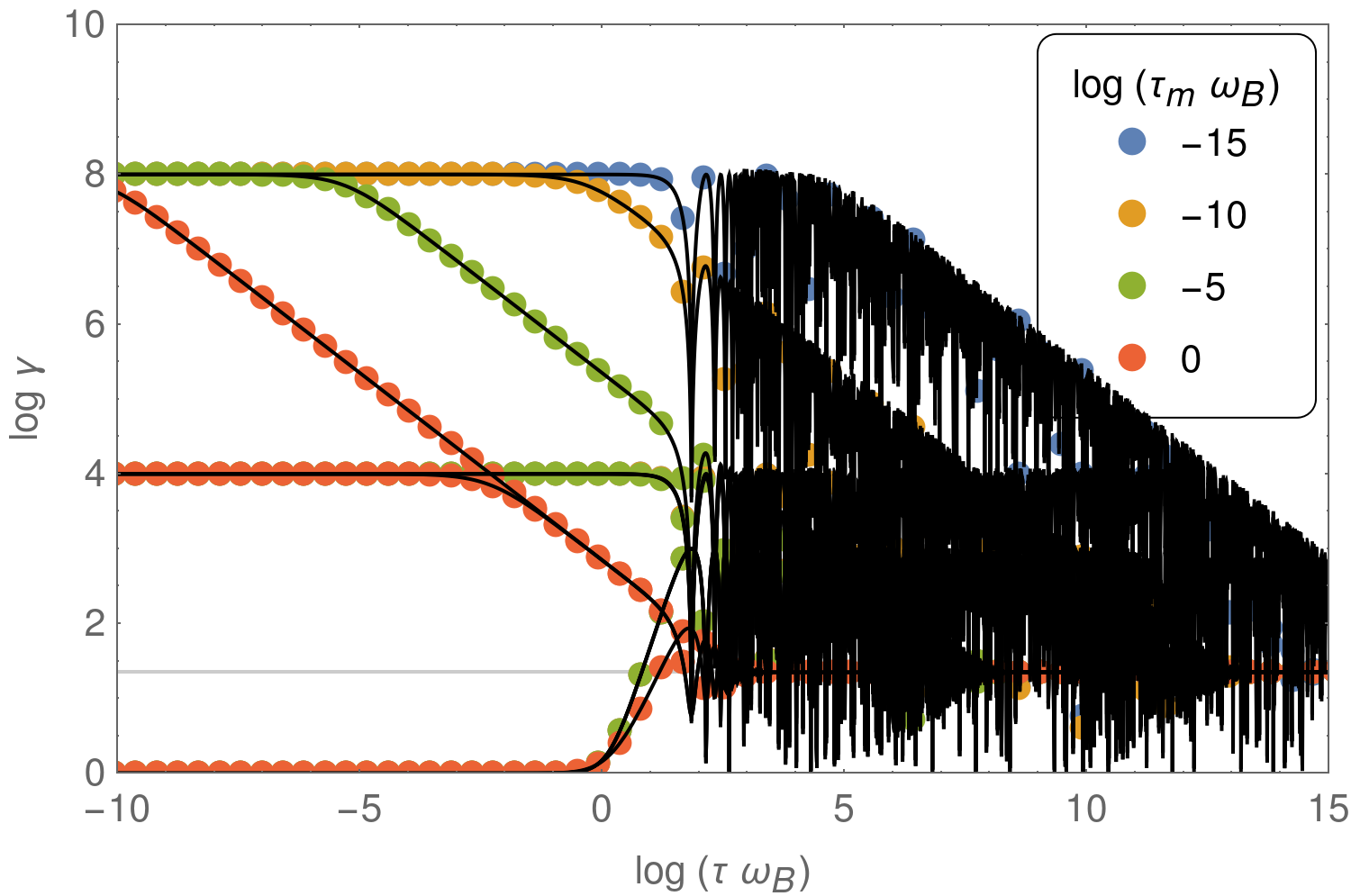}
	\caption{Decrease of the Lorentz factor due to radiation reaction in a cross electric and magnetic field. Dotted colour points show the simulation results and the black solid lines correspond to the exact analytical solutions.}
	\label{fig:lorentz_derive}
\end{figure}

We performed simulations with $E/cB=0.999$ and initial Lorentz factors $\log \gamma_0=\{0, 4,8\}$. The final Lorentz factor is $\gamma_\infty \approx 22.3$.
Fig.~\ref{fig:lorentz_derive} shows the Lorentz factor with colour dots compared to the analytical expression shown in black solid lines. The agreement is excellent and demonstrates the high efficiency of our algorithm to capture ultra-relativistic motion with high precision.

\begin{figure}
	\centering
	\includegraphics[width=\linewidth]{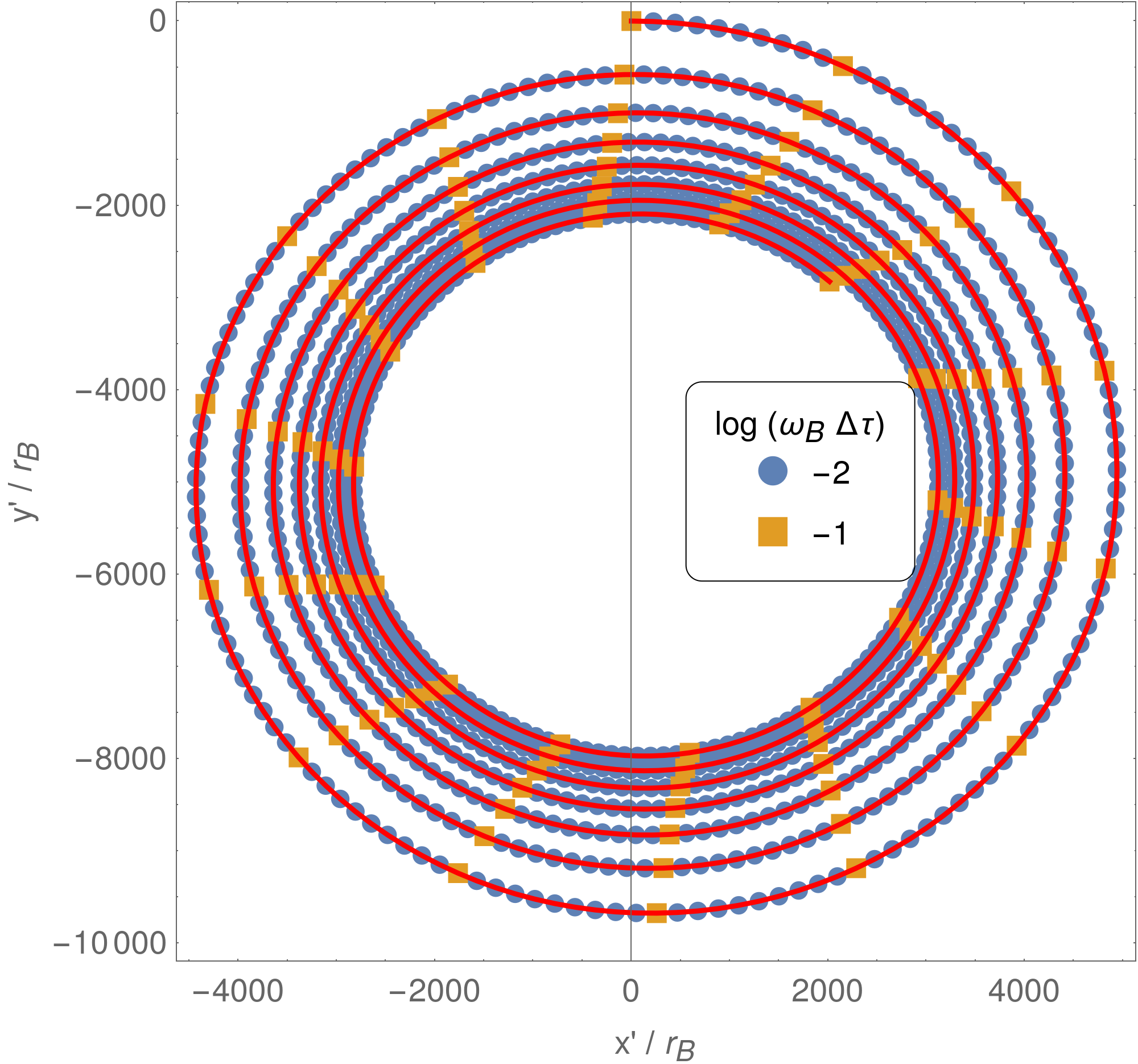}
	\caption{Orbit in the electric drift frame $(x',y')$ for $\log \gamma_0=4$ and $\log b = -5$ for different proper time steps $\log (\omega_{\rm B} \, \Delta\tau) = \{-1,-2\}$.}
	\label{fig:testderiveconstanttrajectoirecomparaison}
\end{figure}

\begin{figure}
	\centering
	\includegraphics[width=\linewidth]{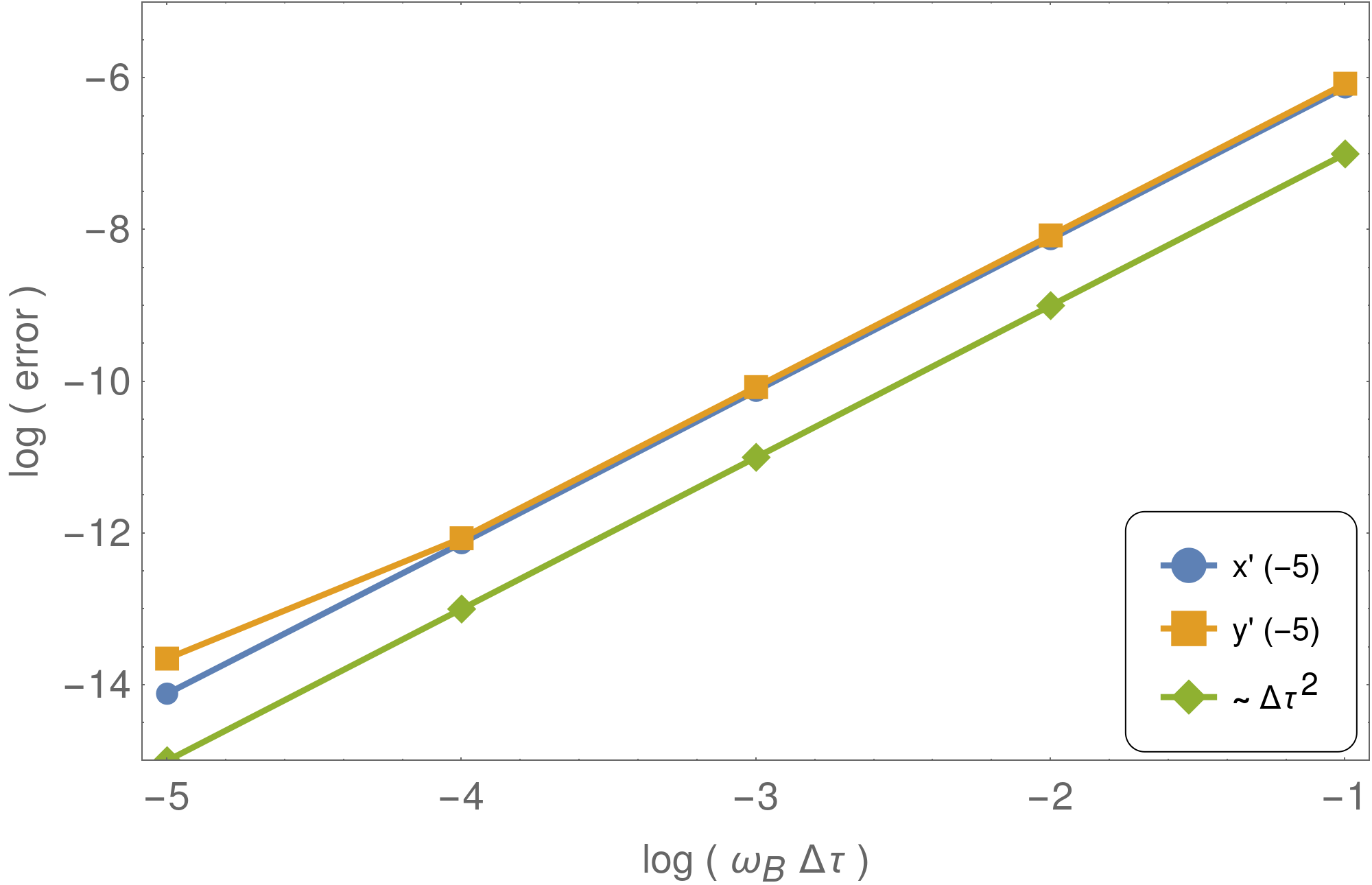}
	\caption{Relative error of the particle position associated to the electric drift motion shown in Fig.~\ref{fig:testderiveconstanttrajectoirecomparaison}.}
	\label{fig:erreurderivea0g4}
\end{figure}

The quantitative agreement is checked by transforming the trajectory to the electric drift frame denoted by the coordinates~$(x',y')$. In this frame the orbital radius is decreasing as shown in Fig.~\ref{fig:testderiveconstanttrajectoirecomparaison} for $\log \gamma_0=4$ and $\log b = -5$, using different proper time steps such as $\log (\omega_{\rm B} \, \Delta\tau) = \{-1,-2\}$ respectively in orange and blue dots. The analytical solution found from appropriate parameters in eq.~\eqref{eq:quadri_position_orbite_larmor} is overlapped in red solid lines.
Fig.~\ref{fig:erreurderivea0g4} shows the relative error in the $x'$ and $y'$ position depending on the proper time step~$\Delta\tau$. The scheme converges to second order in proper time step.

\subsection{Parallel electric and magnetic field}

Another interesting configuration not reducible to any of the previous one is a parallel electric and magnetic field. In this case the second electromagnetic invariant does not vanish $\mathcal{I}_2 \neq 0$. Consequently, there exist no frame where either the electric or magnetic field vanishes. The electric and magnetic velocity components $u_E$ and $u_B$ decouple into an acceleration along the common direction and a gyration around the same direction. Assuming this direction to be $\ez$, the 4-velocity reads
\begin{subequations}
	\begin{align}
	\frac{u^t}{c} & = \gamma(\tau) =  \frac{\gamma_0 \, c \, \cosh(\omega_{\rm E} \, \tau) + u_z^0 \,  \sinh(\omega_{\rm E} \, \tau) }{\sqrt{\gamma_0^2 \,c^2 - u_\parallel^2 - u_\perp^2 \, e^{-2\,\alpha\,\tau}} } \\
	\frac{u^x}{c} & = \frac{u_x^0 \, \cos (\omega_B \, \tau) + u_y^0 \, \sin (\omega_B \, \tau)}{\sqrt{(\gamma_0^2 \,c^2 - u_\parallel^2)\, e^{2\,\alpha\,\tau} - u_\perp^2 } } \\
	\frac{u^y}{c} & = \frac{u_y^0 \, \cos (\omega_B \, \tau) - u_x^0 \, \sin (\omega_B \, \tau)}{\sqrt{(\gamma_0^2 \,c^2 - u_\parallel^2)\, e^{2\,\alpha\,\tau} - u_\perp^2 } } \\
	\frac{u^z}{c} & = \frac{u_z^0 \, \cosh(\omega_{\rm E} \, \tau) + \gamma_0 \, c \, \sinh(\omega_{\rm E} \, \tau)}{\sqrt{\gamma_0^2 \,c^2 - u_\parallel^2 - u_\perp^2 \, e^{-2\,\alpha\,\tau}} } .
	\end{align}
\end{subequations}
We recognize the special cases of a pure electric field for $u^t,u^z$ and a pure magnetic field for $u^x,u^z$, the only difference being the value of $\alpha = \tau_{\rm m} \, ( \lambda_{\rm E}^2 + \lambda_{\rm B}^2)$, including a non vanishing electric and magnetic contribution.

After a transition time, the gyro-motion has been significantly damped and the electric acceleration has directed the velocity along its direction. The initial conditions are washed out and the particle moves like in a constant electric field with an almost constant acceleration leading to a hyperbolic motion, well know in special relativity kinematics. Let us estimate the duration of this transient stage. Either the particle is drastically accelerated before the gyration is damped or vice versa the orbit shrinks significantly before the electric field accelerated sensibly the particle. The situation depends on ordering of the eigenvalues $\lambda_{\rm E}$ and $\lambda_{\rm B}$.

\begin{figure}
	\centering
	\includegraphics[width=1\linewidth]{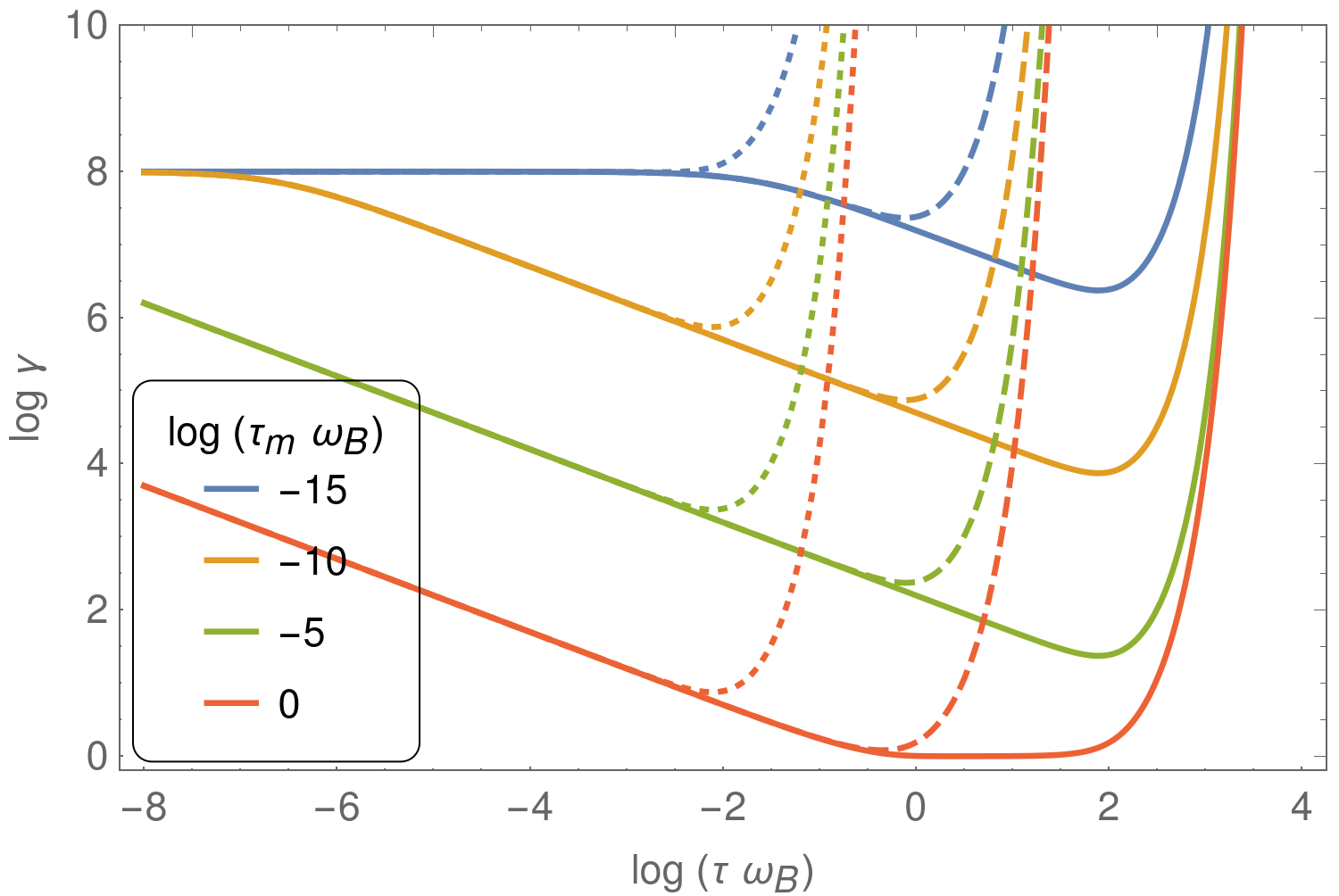}
	\caption{Analytic evolution of the Lorentz factor with initial condition~$\log\gamma_0=8$, different damping factor~$\log b = \{0,-5,-10,-15\}$ and different electric field strength $E_0$ relative to $B_0$ such that $\log (E_0/B_0) = \{-2,0,2\}$ in solid lines, dashed lines and dotted lines respectively.}
	\label{fig:testconstantparallelegammaanalytique}
\end{figure}

\begin{figure}
	\centering
	\includegraphics[width=\linewidth]{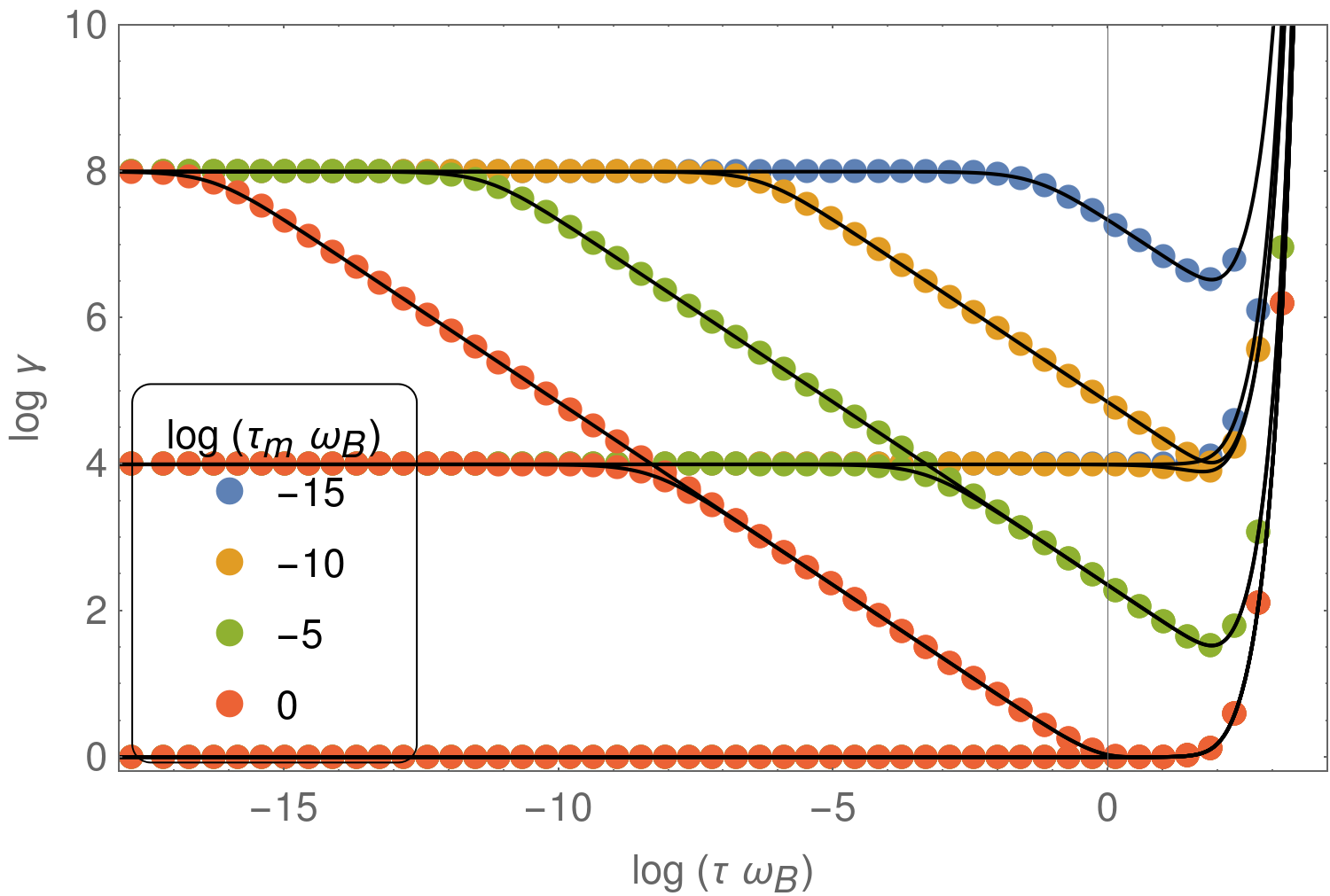}
	\caption{Evolution of the Lorentz factor for a parallel electromagnetic field configuration with initial Lorentz factors $\log\gamma_0=\{0,4,8\}$, an electric field strength $\log(E_0/B_0)=-2$ and $\log b = \{0,-5,-10,-15\}$. The black solid lines correspond to the exact analytical solutions.}
	\label{fig:testconstantparallelegamma}
\end{figure}

Fig.~\ref{fig:testconstantparallelegammaanalytique} shows the analytic evolution of the Lorentz for a particle starting with only a perpendicular velocity component such that $\log\gamma_0=8$. The damping factor is set to $\log b = \{0,-5,-10,-15\}$ and the electric field strength $E_0$ is varied relative to $B_0$ such that $\log (E_0/B_0) = \{-2,0,2\}$ and depicted in solid lines, dashed lines and dotted lines respectively. For a weak electric field $E_0 \ll B_0$ the particle trajectory follows a spiral motion similar to the previous case of a pure magnetic field until it almost rests. At later times the electric field starts to accelerate it quickly to ultra-relativistic speeds on a timescale $1/\omega_{\rm E} \gg 1/\omega_{\rm B}$. The particle performs many orbits before being deflected along the parallel direction ($\ez$). Increasing $E_0$ will decrease this time scale and the particle follow the common $\vec{E}$ and $\vec{B}$ direction before performing many gyrations. In the opposite limit of a strong electric field $E_0 \gg B_0$ electric acceleration quickly sets in. Fig.~\ref{fig:testconstantparallelegamma} shows some results of numerical simulations with a weak electric field $\log(E_0/B_0)=-2$, pertinent for almost force-free neutron star magnetospheres, and initial Lorentz factors $\log\gamma_0=\{0,4,8\}$ and $\log b = \{0,-5,-10,-15\}$.

Whenever there exist a magnetic field aligned electric field component, at late times particles are always accelerated along the common direction. The timescale required is estimated by reckoning the proper time at which the parallel 4-velocity component becomes comparable to the perpendicular 4-velocity component for an initial velocity perpendicular to the magnetic field line. This represents the worst case, useful to be compared to the radiation reaction limit regime.

\subsection{Almost cross field}

\begin{figure}
	\centering
	\includegraphics[width=0.9\linewidth]{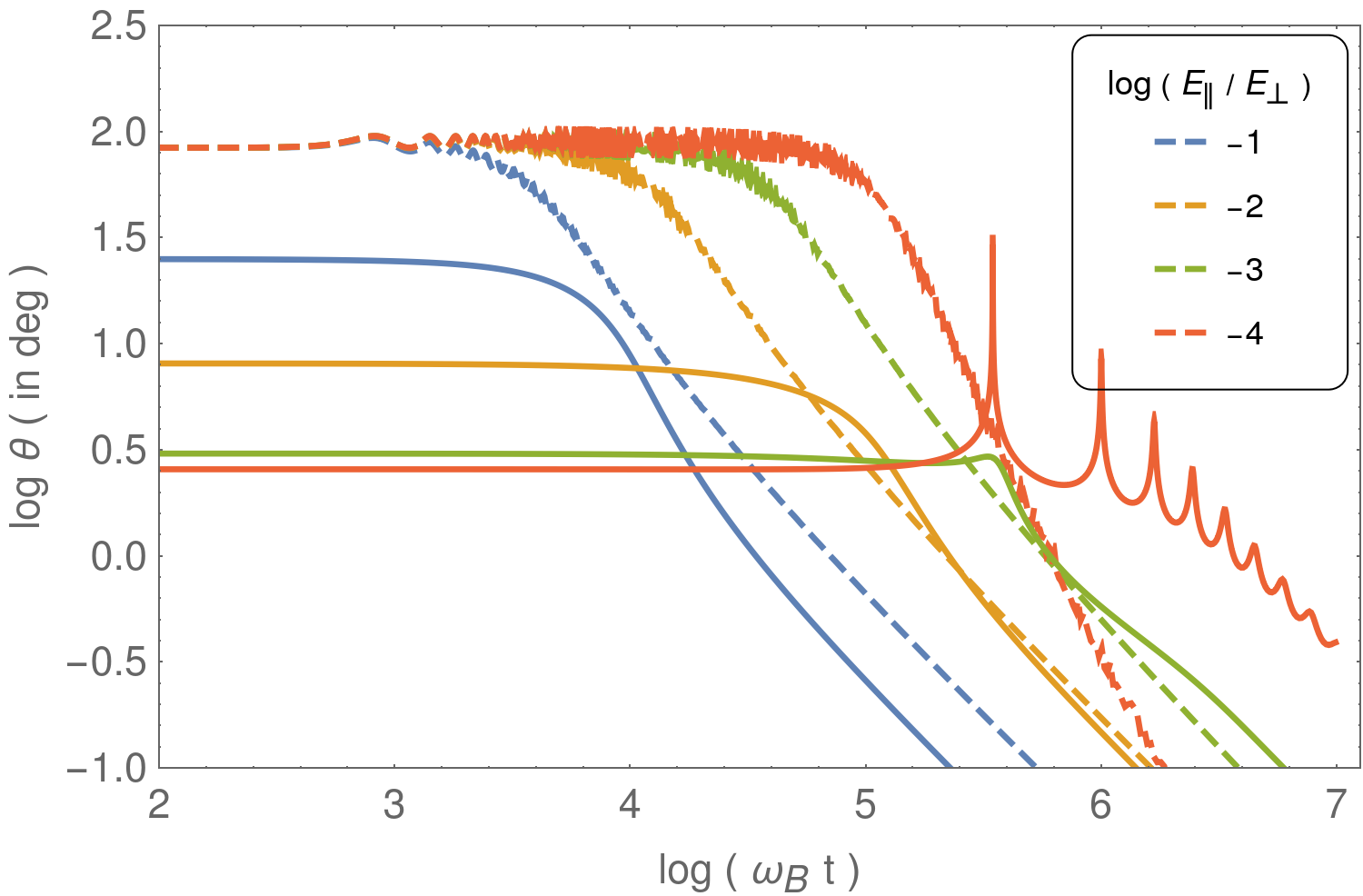}
	\caption{Alignment of the velocity vector with the radiation reaction limit direction given by $\theta=0\degree$ for different strengths of the parallel electric field component $E_\parallel$ compared to the perpendicular component $E_\perp$ for $E_\perp/cB=0.1$ in dashed lines and $E_\perp/cB=0.999$ in solid lines. The damping parameter is $\log b=-5$ and the initial Lorentz factor $\log \gamma_0 = 4$.}
	\label{fig:testderiveconstantparalleledt-4g4b-5}
\end{figure}

In a plasma filled magnetosphere, the electric field is efficiently screened meaning that the parallel component of the electric field is negligible with respect to its perpendicular component, $E_\parallel \ll E_\perp$. As an application towards this configuration, we computed the motion of a particle in an almost cross electric and magnetic field with $\log (E_\parallel \ll E_\perp) = \{-1,-2,-3,-4\}$ and different ratio $E_\perp / c\,B = \{0.1, 0.999\}$. The particle starts with an initial velocity in a plane perpendicular to $\mathbf{B}$ and Lorentz factor $\log \gamma_0 = 4$ in a field with $\log b=-5$. Fig.~\ref{fig:testderiveconstantparalleledt-4g4b-5} shows the evolution to alignment of the velocity vector with the radiation reaction limit direction for a weak parallel electric field component as reported in the legend. Note that the angle $\theta$ should not be interpreted as the angle between the velocity vector and the magnetic field direction because the velocity in eq.\eqref{eq:VRR} is not along $\mathbf{B}$. We observe that the time required for alignment is insensitive to the ratio $E_\perp / c\,B$ but depends strongly on the ratio $E_\parallel / E_\perp$. As expected a weak parallel component tends to align slower the trajectory compared to a strong parallel component. If this alignment occurs on a length scale smaller than the magnetic field curvature radius, the radiation reaction limit could be used without significant loss of accuracy.

\subsection{Dipole magnetic field}

As a step towards realistic configurations, we also investigate particle motion in a static magnetic dipole. Unfortunately, no simple exact analytical expressions are available for checking the algorithm therefore no quantitative accurate converge test can be performed. First we study the magnetic drift in the equatorial plane of a dipole field. Next we look at trapped particles due to the mirror effect.

\subsubsection{Magnetic drift}

The magnetic drift motion in the equatorial plane of a dipole field is an interesting example to test our algorithm. The characteristic frequency is again $\omega_{\rm B}$ and the particle initial Lorentz factor is $\gamma_0=10^4$. The damping parameter is $\log b = \{0,-5,-10,-15\}$.

\begin{figure}
	\centering
	\includegraphics[width=\linewidth]{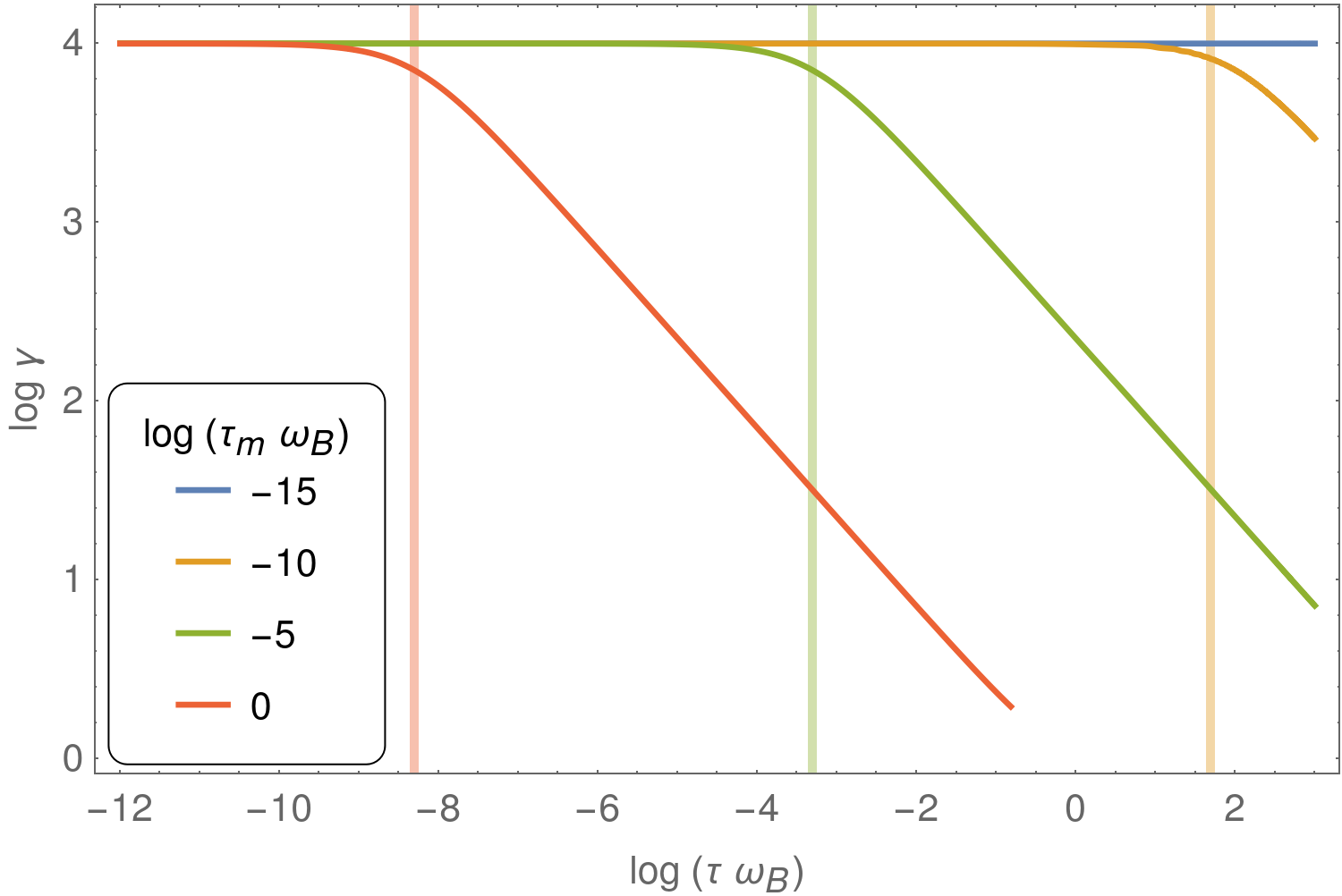}
	\caption{Decrease of the Lorentz factor due to radiation reaction when drifting in a dipole magnetic field with initial Lorentz factor $\log \gamma_0=4$ and different damping constants.}
	\label{fig:test_derive_magnetique_gamma}
\end{figure}

\begin{figure}
	\centering
	\includegraphics[width=\linewidth]{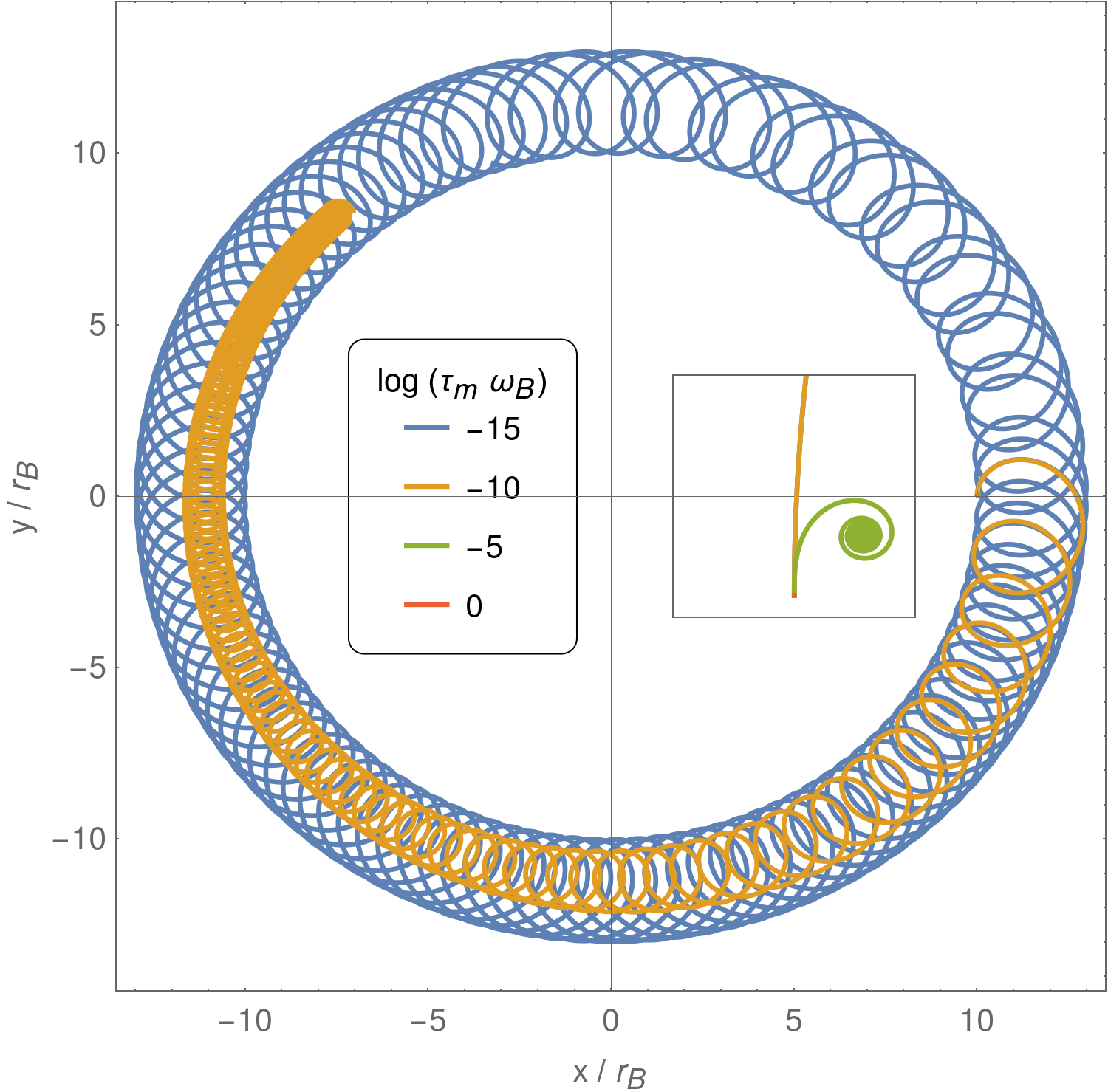}
	\caption{Particle trajectory in the equatorial plane of a dipole magnetic field and associated to Fig.~\ref{fig:test_derive_magnetique_gamma}. The inset shows the strong damped motion in green and even stronger damping in red where the spiralling is not seen.}
	\label{fig:test_derive_magnetique_trajectoire}
\end{figure}

Fig.~\ref{fig:test_derive_magnetique_gamma} shows the time evolution of the Lorentz factor that tends asymptotically to unity meaning the particle will rest. The particle returns to rest after a typical time controlled by $b$ and shown as coloured vertical lines.
Fig.~\ref{fig:test_derive_magnetique_trajectoire} highlights the corresponding particle trajectory in the equatorial plane. For the weakest damping, the motion remains circular for the guiding centre. For the strongest damping, in green and red, the particle suffers from drastic radiative friction and tends to rest on a very short time scale compared to the drifting motion and orbital motion.

\subsubsection{Magnetic mirror}

Due to the mirror effect, particles remain trapped in the dipole magnetic field of a star like around Earth in the van Allen belt. However when some dissipation occurs as for instance through radiation reaction, for high damping parameters particles quickly crash onto the surface of the magnetic object.

\begin{figure}
	\centering
	\includegraphics[width=\linewidth]{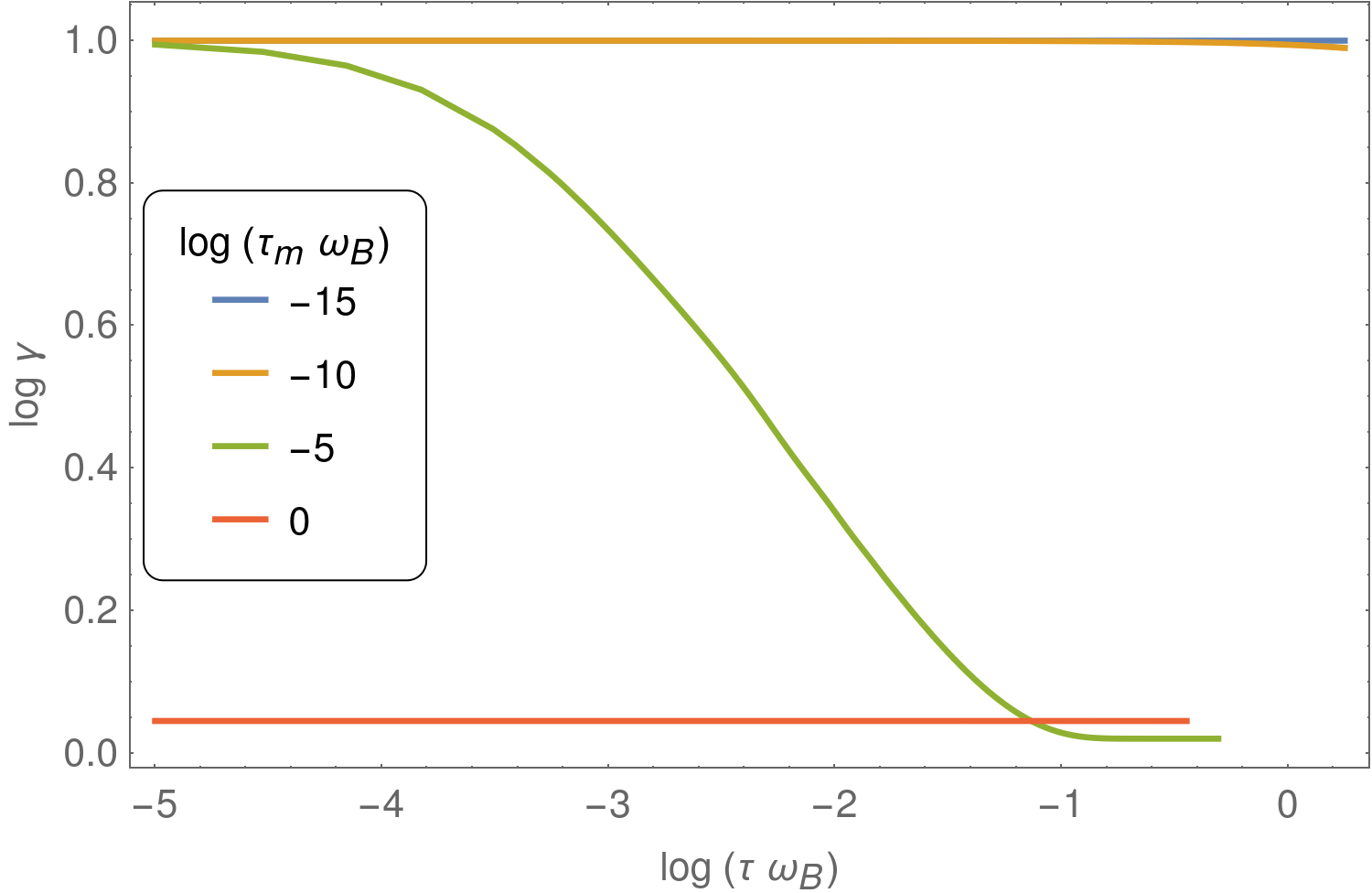}
	\caption{Decrease of the Lorentz factor of particles subject to the mirror effect.}
	\label{fig:test_miroir_magnetique_gamma}
\end{figure}

\begin{figure}
	\centering
	\includegraphics[width=\linewidth]{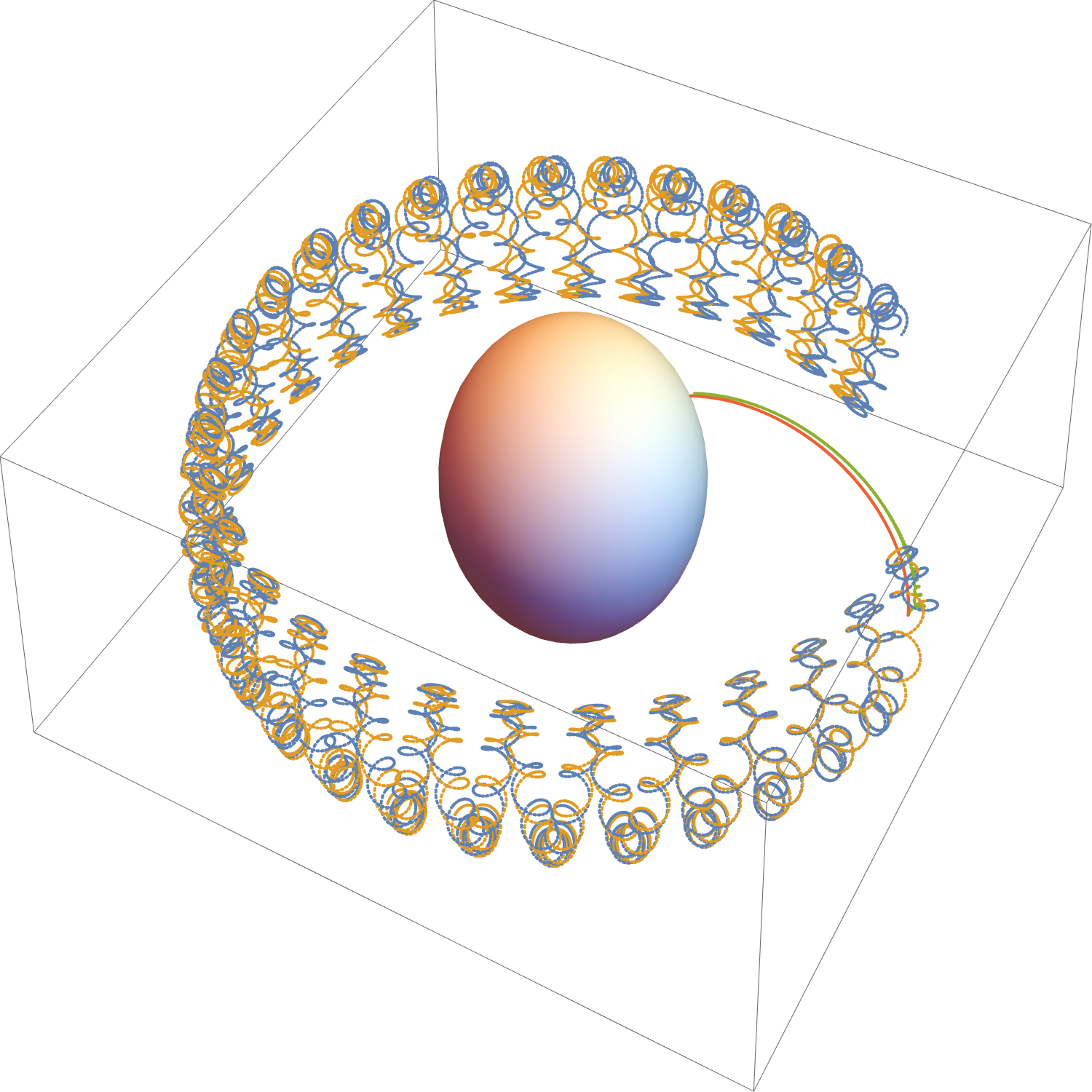}
	\caption{Particle trajectories in the dipole magnetic field and associated to Fig.~\ref{fig:test_miroir_magnetique_gamma}. For small damping parameter, orange and blue lines, the particle is trapped for a long time in the dipole, whereas for larger damping it quickly crashes onto the stellar surface.}
	\label{fig:test_miroir_magnetique_trajectoire}
\end{figure}

Fig.~\ref{fig:test_miroir_magnetique_gamma} shows the evolution of the Lorentz factor for particles moving in the magnetic dipole field. For weak damping parameters $\log (\tau_{\rm m}\,\omega_{\rm B}) \lesssim -10$, radiation reaction remains negligible and the particle motion is almost adiabatic with the three characteristic periodic motions: gyration around the magnetic field, bouncing between north and south magnetic pole and precession in the azimuthal direction, see blue and orange lines in Fig.~\ref{fig:test_miroir_magnetique_trajectoire}. For $\log (\tau_{\rm m}\,\omega_{\rm B}) = -5$, the cyclotron motion is rapidly damped and the particle falls onto the star, green trajectory. For $\log (\tau_{\rm m}\,\omega_{\rm B})=0$, the damping is even faster and the particle crashes onto the stellar surface, following a trajectory similar to the previous case, see red solid line in Fig.~\ref{fig:test_miroir_magnetique_trajectoire}.

\section{Application to neutron stars}
\label{sec:Etoile}

After checking and testing our new algorithm, we are ready to apply it to realistic extreme cases of rotating magnetized neutron stars. The neutron star radius is fixed to $R_*=12$~km.
The accurate configuration of the electromagnetic field is taken from a rotating magnetic dipole in vacuum and given by \cite{deutsch_electromagnetic_1955}.

\subsection{Relevant parameters without dimension}

As a typical frequency we choose the stellar angular frequency~$\omega = \Omega_*$ and consider three populations of neutron stars: young pulsars with period $P_*=1$~s and surface magnetic field strengths $B_*=10^8$~T, millisecond pulsars with period $P_*=5$~ms and $B_*=10^5$~T and magnetars with period $P_*=10$~s and $B_*=10^{10}$~T. These quantities and their associated normalised strength and damping parameters $a_{\rm B}$, $a_{\rm E}$ and $b$ for electrons, protons and irons are summarized in table~\ref{tab:Parametres}.
\begin{table*}
	\centering
	\begin{tabular}{lccccc}
		\hline
		Neutron star & $P_*$ (s) & $\log B_*$ (T) & $\log a_{\rm B}$ & $\log a_{\rm E}$ & $-\log b$ \\
		\hline
		millisecond & 0.005 & \numprint{5}  & 13.1 / \ 9.9 / \ 9.5 & 11.8 / \ 8.6 / \ 8.2 & 20.1 / 23.4 / 22.5 \\
		young       & 1     & \numprint{8}  & 18.4 / 15.2 / 14.8 & 14.8 / 11.6 / 11.2 & 22.4 / 25.7 / 24.8 \\
		magnetar    & 10    & \numprint{10} & 21.4 / 18.2 / 17.8 & 16.8 / 13.6 / 13.2 & 23.4 / 26.7 / 25.8 \\
		\hline
	\end{tabular}
	\caption{\label{tab:Parametres}Typical period and surface magnetic field strength of millisecond pulsars, young pulsars and magnetars. The relevant parameters without dimension are given by the strength parameters for the magnetic field~$a_B$ and for the electric field $a_E$ and the damping parameter $b$ for electrons / protons / iron nuclei.}
\end{table*}
The normalization frequency is arbitrary but from a microscopic point of view, the most relevant frequencies are related to the electromagnetic tensor eigenfrequencies. Therefore the low value of $b$ should not be misinterpreted as a weak feedback of radiation reaction. It is an artefact of the chosen typical frequency associated to the stellar rotation and which is many orders of magnitude smaller than the cyclotron frequency.

We distinguish three kind of particles: a first group crashing onto the stellar surface, a second group of trapped particles and a third group of escaping particles, all accelerated to high energies. Particles are considered trapped when they still have not crashed onto the surface or not yet escaped the light cylinder.
Particles are placed regularly within the light-cylinder, starting at rest or with an initial velocity vector oriented along the magnetic field line, directed toward the star or towards infinity, with a Lorentz factor equal to $\gamma_0=10^3$ or starting at rest. The neutron star obliquity is set to $\rchi = \{0\degr, 30\degr, 60\degr, 90\degr, 120\degr, 150\degr, 180\degr\}$. We found that the final results are not very sensitive to the initial Lorentz factor because charges are immediately accelerated in the direction of the electric field and therefore loose memory about their initial state. Our simulation results are thus summarized for particles starting at rest only.
We simulated a total number of 48 particles for each neutron star type and each obliquity, spread around three radii $r_0$, right at the surface $R_*$, approximately half-way between the surface and the light-cylinder (a geometric average) and at the light cylinder, thus $r_0=\{R_*, \sqrt{R_* \, \rlight}, \rlight\}$. For comparison we performed simulations with and without radiation reaction.

\subsection{Orders of magnitude}

Before presenting the accurate numerical simulations of particle trajectories and their radiation reaction in the vicinity of neutron stars, we remind the orders of magnitude of the maximum Lorentz factors expected when charges are accelerated in the electric potential produced by a rotating magnetized perfectly conducting star. The most optimistic view adopts the full potential drop between the pole and the equator as an estimate of the accelerating field thus
\begin{equation}\label{eq:gamma_max_max}
\gamma_{\rm max}^{\rm full} \approx \frac{q \, \Omega_*\,B_*\,R_*^2}{m\,c^2} = \frac{R_*}{\rlight} \, \frac{R_*}{r_{\rm B}}
\end{equation}
where $r_{\rm B} = c/\omega_{\rm B}$ is the non relativistic Larmor radius. If the accelerating potential is only available across the polar caps as expected from nearly force-free magnetosphere models, the maximum energy corresponds to
\begin{equation}\label{eq:gamma_max_calotte}
\gamma_{\rm max}^{\rm pc} \approx \frac{q \, \Omega_*^2\,B_*\,R_*^3}{m\,c^3} = \left( \frac{R_*}{\rlight} \right)^2 \, \frac{R_*}{r_{\rm B}} \approx \frac{R_*}{\rlight} \, \gamma_{\rm max}^{\rm full} 
\end{equation}
which is a factor $R_*/\rlight$ smaller than for the former case. Table~\ref{tab:Gamma_max} summarizes the maximum Lorentz factors for electrons, protons and irons around millisecond pulsars, young pulsars and magnetars. The values reported in this table for $\gamma_{\rm max}^{\rm full}$ are at best upper limits for the vacuum case. Only an accurate numerical integration of the equation of motion gives robust results as we now show.
\begin{table}
	\centering
\begin{tabular}{cccc}
	\hline
	Neutron star & \multicolumn{3}{c}{ $\log \gamma_{\rm max}^{\rm full}\ /\ \log \gamma_{\rm max}^{\rm pc}$} \\
	\hline 
	& electron & proton & iron \\
	\hline
	millisecond & 10.5 / 9.3 & 7.3 / 6.0 & 7.0 / 5.7 \\
	young       & 11.2 / 7.7 & 8.0 / 4.4 & 7.7 / 4.1 \\
	magnetar    & 12.2 / 7.7 & 9.0 / 4.4 & 8.7 / 4.1 \\
	\hline
\end{tabular}
	\caption{Maximum Lorentz factor orders of magnitude from conservative arguments about neutron star magnetospheres. Values for full potential drops are given on the left of the "/" symbol and for polar cap potential drops on the right in logarithmic scale. \label{tab:Gamma_max}}
\end{table}

\subsection{Escaping particles}

\begin{figure*}
	\centering
	\begin{tabular}{cc}
		\includegraphics[width=0.5\linewidth]{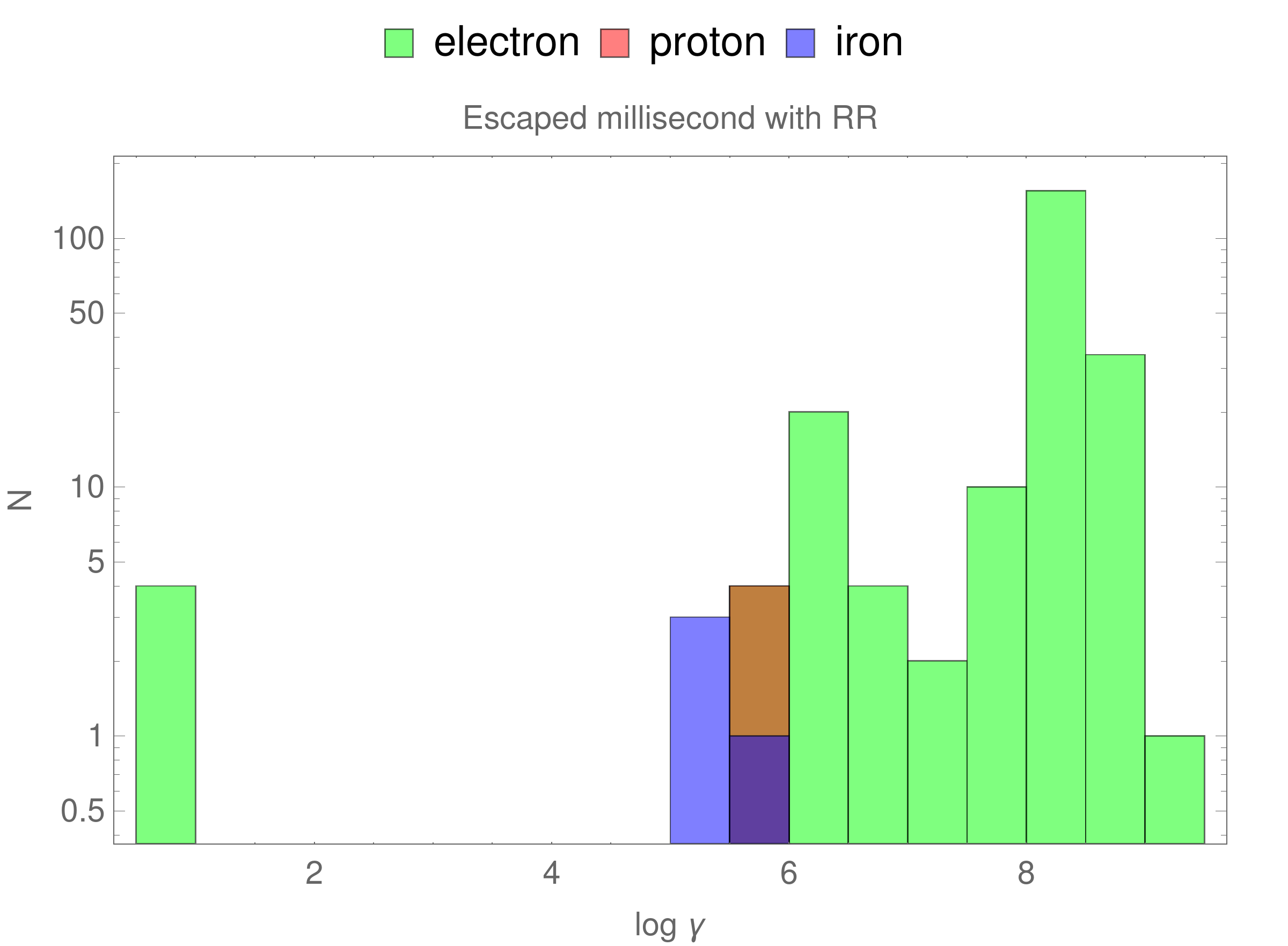} &	
		\includegraphics[width=0.5\linewidth]{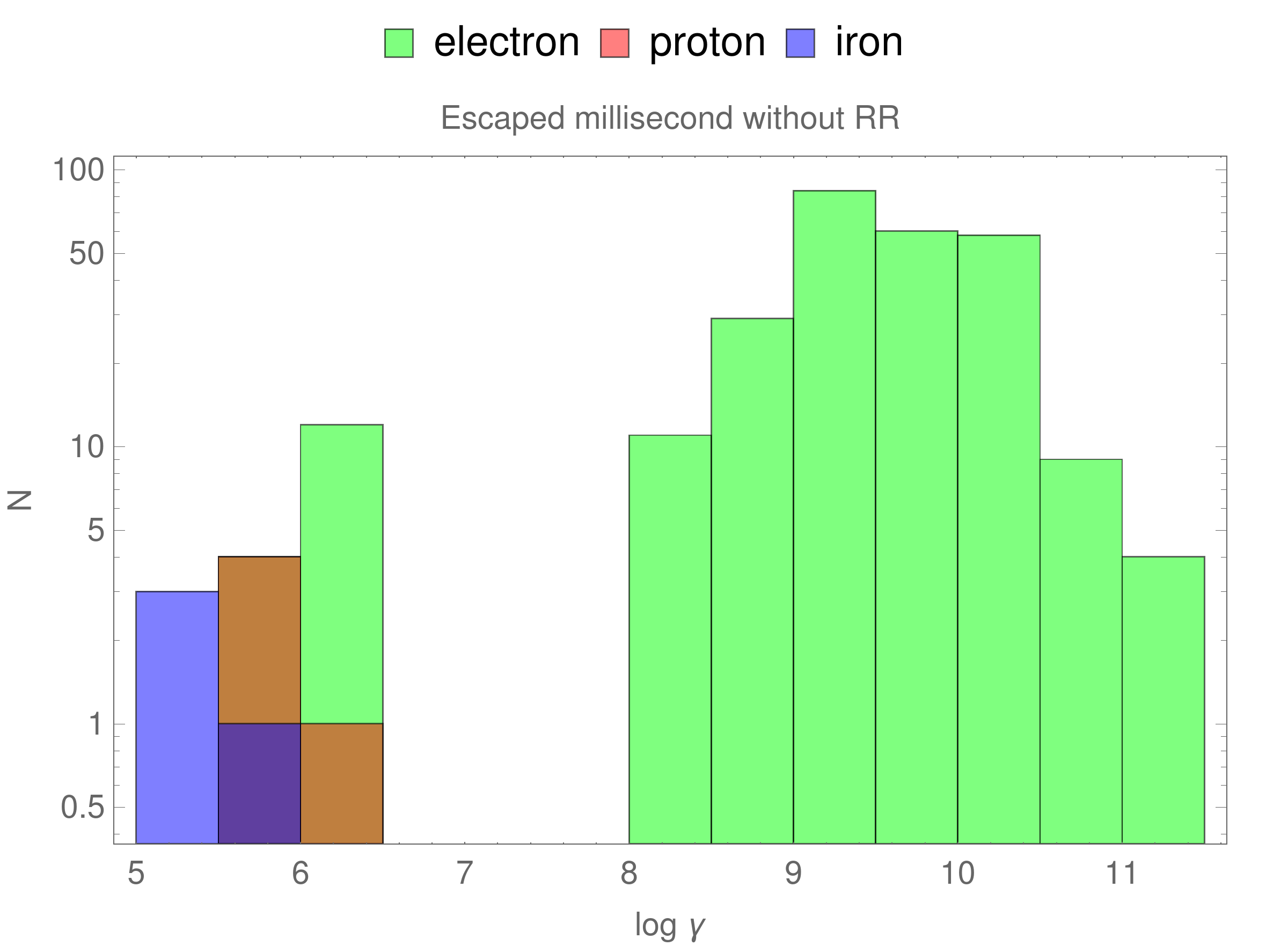} \\
		\includegraphics[width=0.5\linewidth]{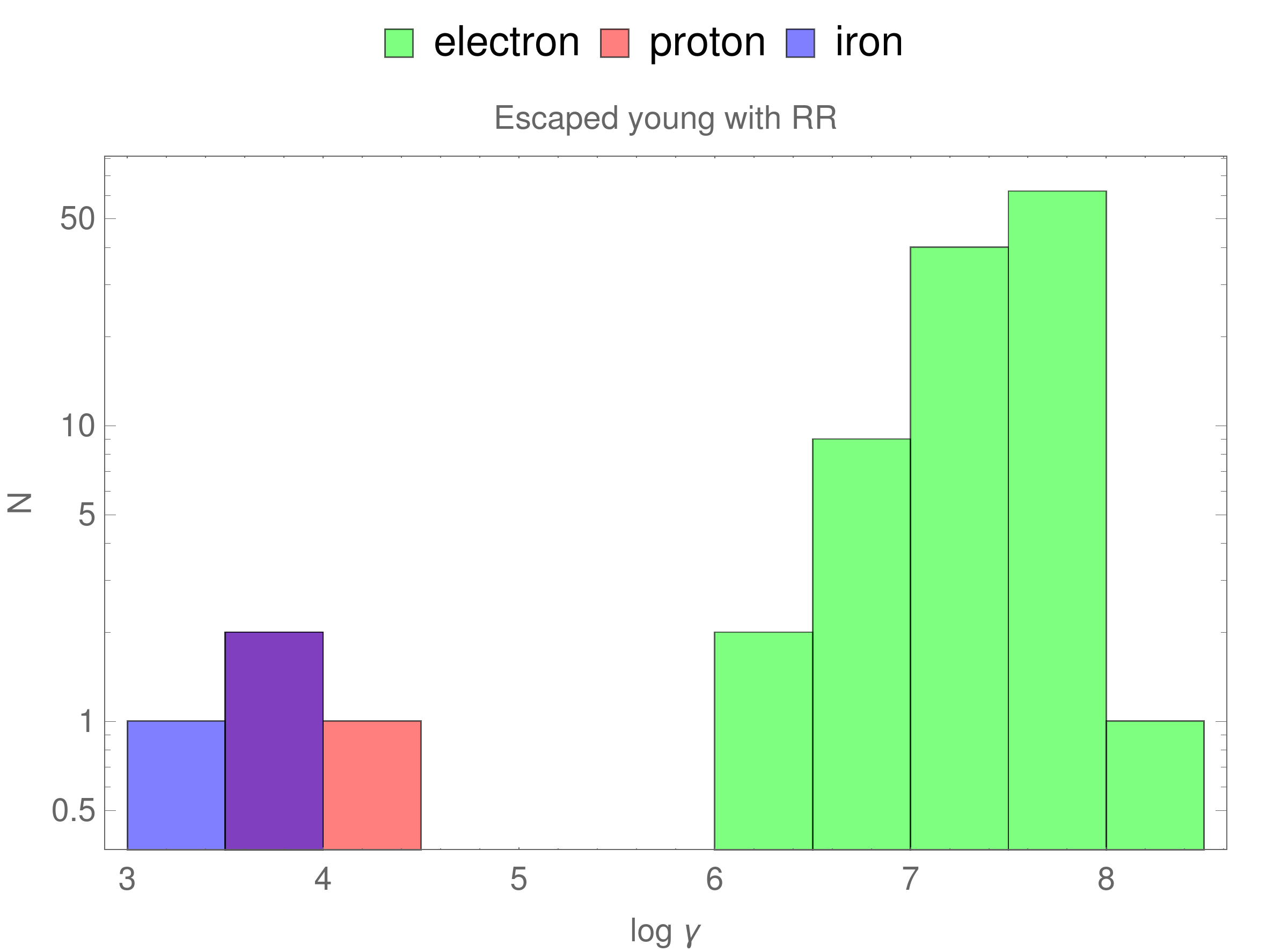} &	
		\includegraphics[width=0.5\linewidth]{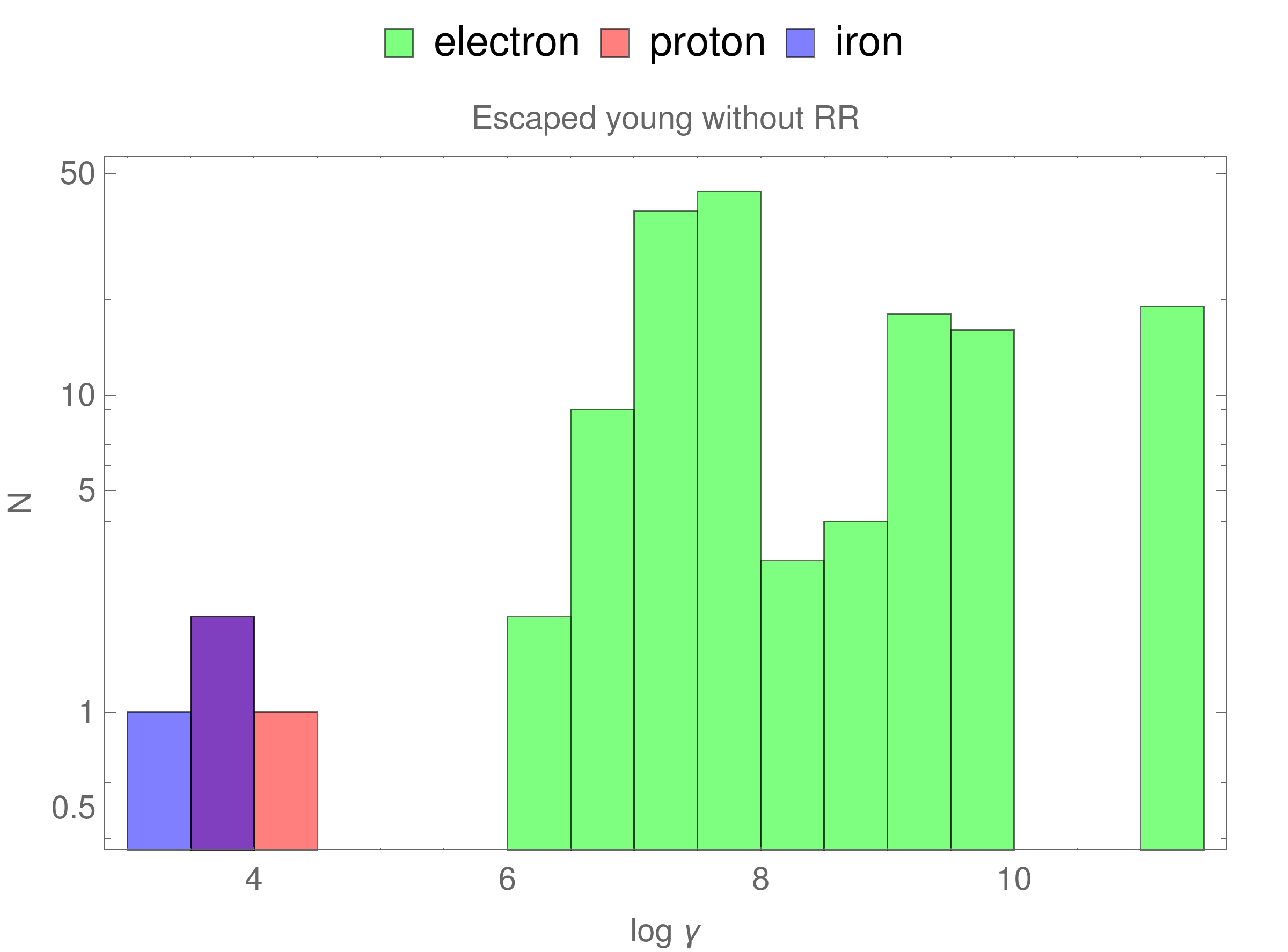} \\
		\includegraphics[width=0.5\linewidth]{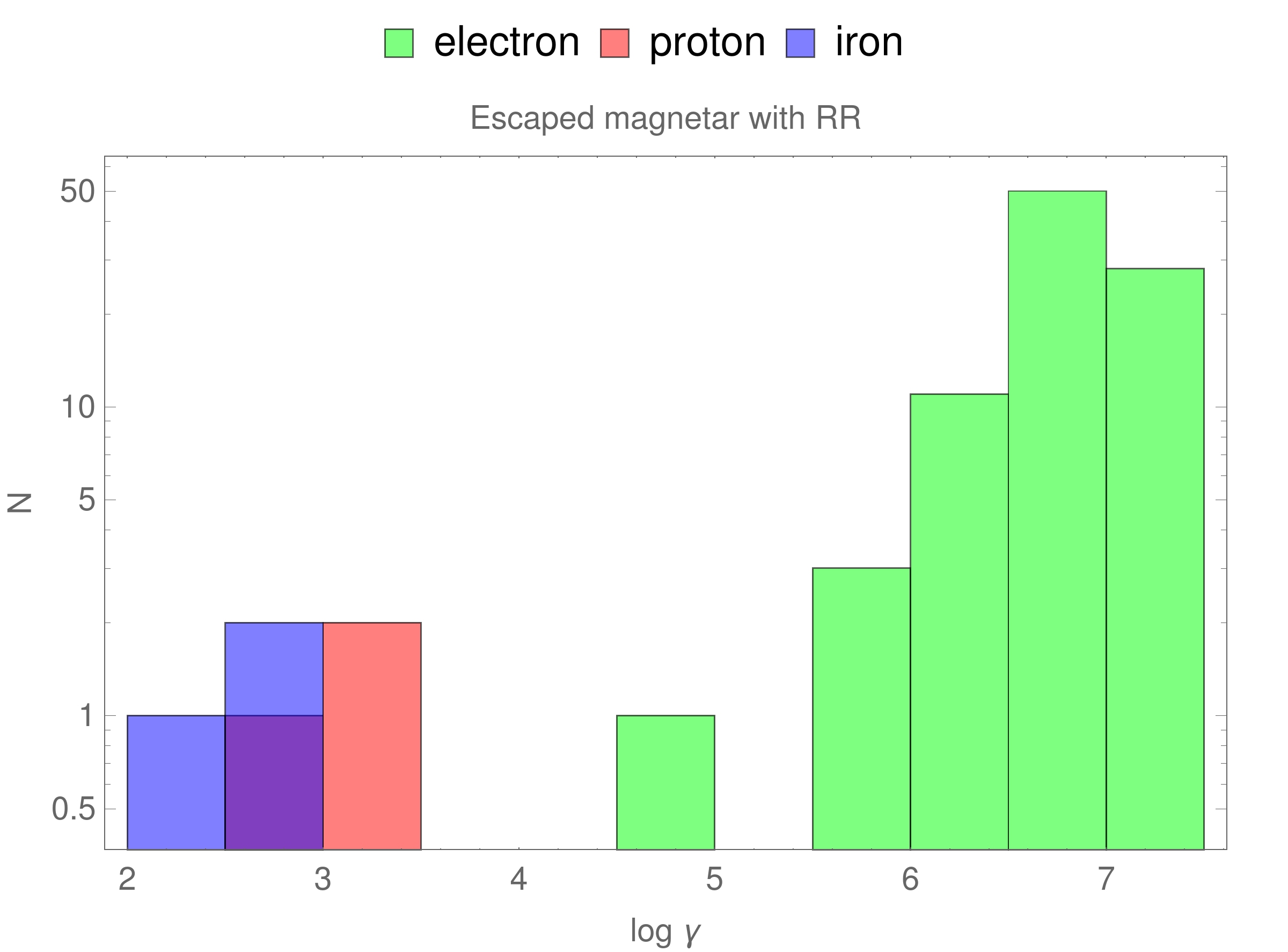} &
		\includegraphics[width=0.5\linewidth]{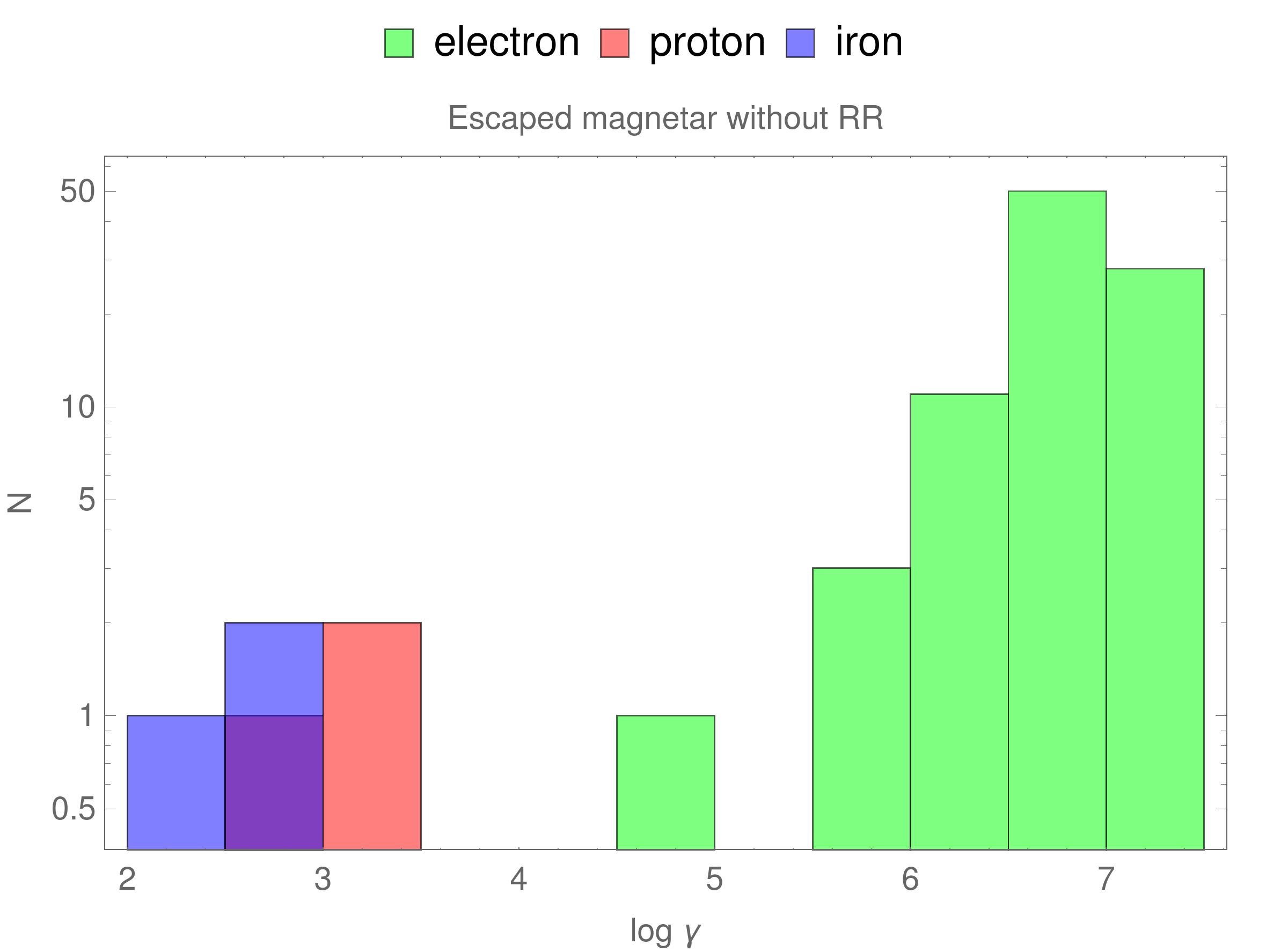}
	\end{tabular}
	\caption{Histogram of escaped particles for millisecond pulsars on the top row, for young pulsar on the middle row and for magnetars on the bottom row. Electron Lorentz factors are shown in green, proton in red and iron nuclei in blue. The left column includes radiation reaction (RR) whereas the right panel does not.}
	\label{fig:escaped_gamma_final_g1_cfl-2}
\end{figure*}

Particles reaching distances larger than $10\,\rlight$ are reputed to be leaving the neutron star magnetosphere. The run halts when the particle reaches larger distances. Fig.~\ref{fig:escaped_gamma_final_g1_cfl-2} shows the histogram of Lorentz factor for electrons in green, protons in red and iron nuclei in blue, irrespective of the magnetic field inclination angle~$\rchi$. The left column corresponds to a motion with radiation reaction (RR) whereas the right column to motion without radiation reaction. %First we notice that no particles escaped from a young pulsar or from a magnetar during the simulation time span. Second, 
First, electrons are the most effectively accelerated particles reaching final Lorentz factors up to $\gamma_{\rm f} \sim 10^9$ in the LLR approximation for millisecond pulsars. This is however two orders of magnitude less than without radiation reaction where $\gamma_{\rm f} \sim 10^{11}$. Second, as expected protons and iron nuclei acquire much less energy, only about $\gamma_{\rm f} \sim 10^6$ for millisecond pulsars, wherever LLR is used or not. For young pulsars, electrons also reach $\gamma_{\rm f} \sim 10^9$ in the LLR regime instead of $\gamma_{\rm f} \sim 10^{11}$ for the pure Lorentz force. Protons and iron nuclei are much less subject to radiation reaction, showing no impact on the maximum Lorentz factor remaining at $\gamma_{\rm f} \sim 10^{4}-10^{4.5}$. For magnetars, radiation reaction remains negligible irrespective of the nature of each species. Electrons reach energies up to $\gamma_{\rm f} \sim 10^{7.5}$ whereas protons and iron nuclei  $\gamma_{\rm f} \sim 10^{3}-10^{3.5}$. Therefore, \RR does not significantly perturb the trajectories of particles with lower charge over mass ratio $q/m$. Contrary to electrons, protons and irons do not suffer from radiation friction.
%The most efficient acceleration paths are associated to particles escaping the neutron star environment.

\subsection{Crashed particles}

\begin{figure*}
	\centering
	\begin{tabular}{cc}
	\includegraphics[width=0.5\linewidth]{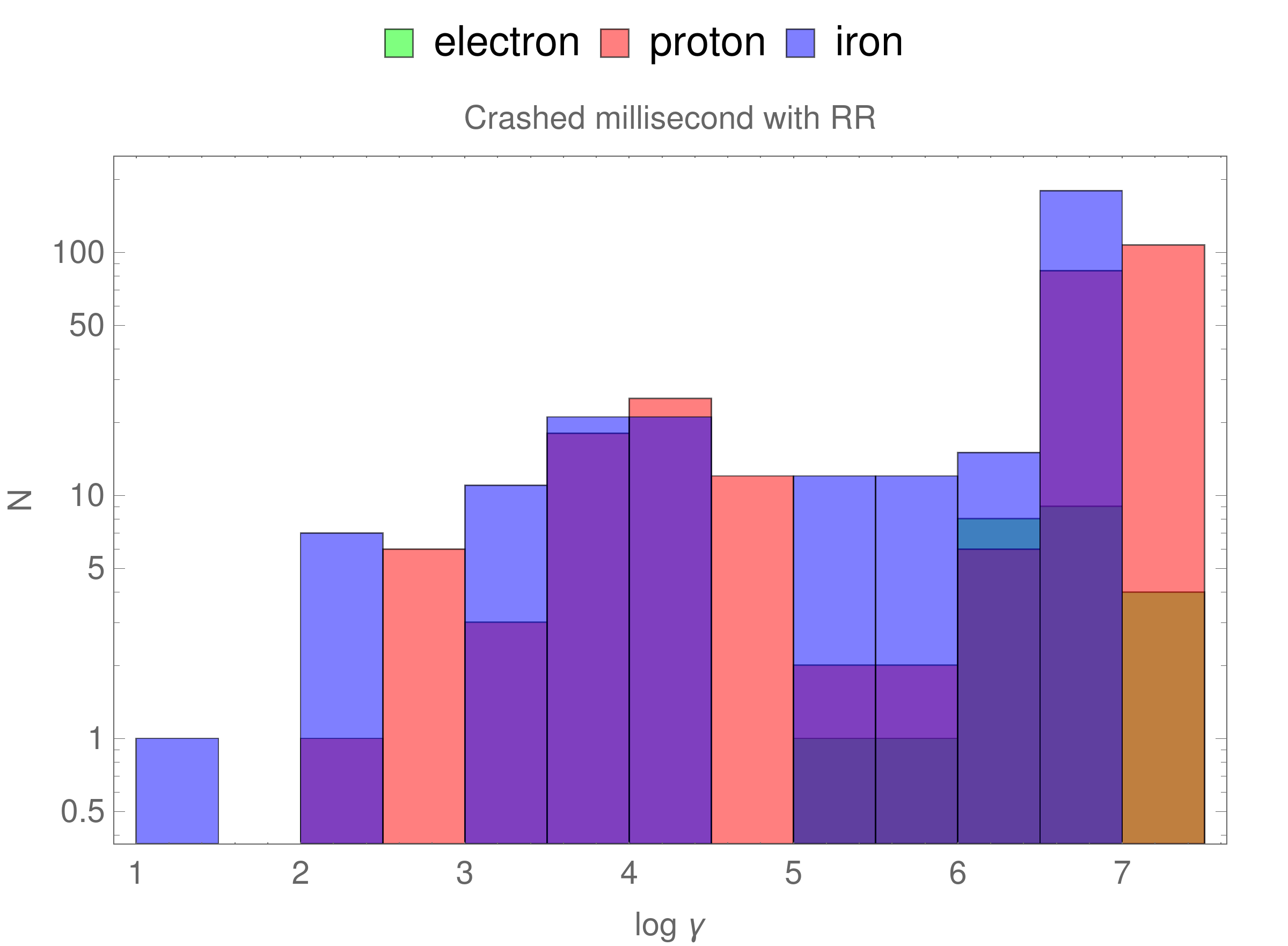} &
	\includegraphics[width=0.5\linewidth]{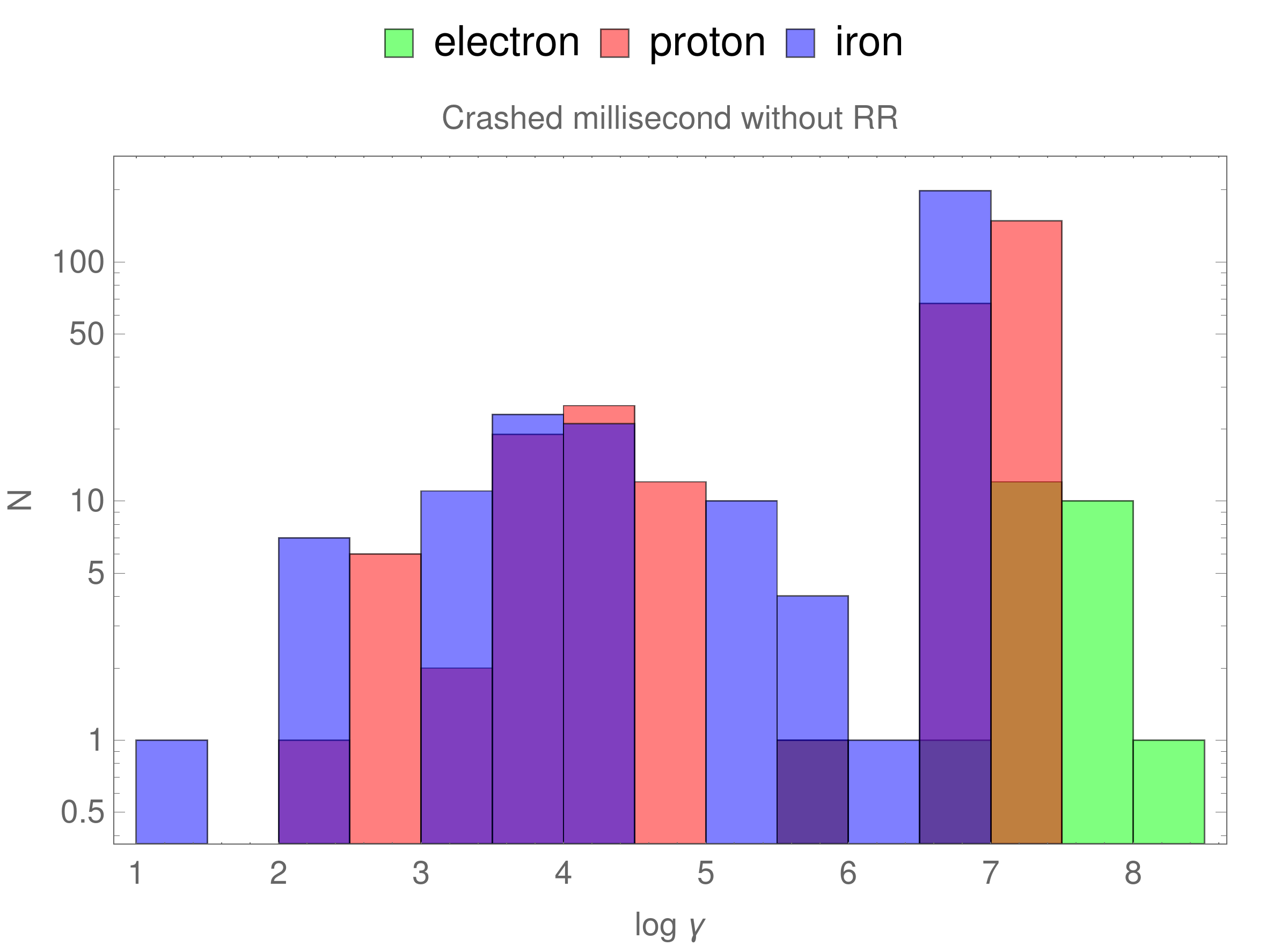} \\
	\includegraphics[width=0.5\linewidth]{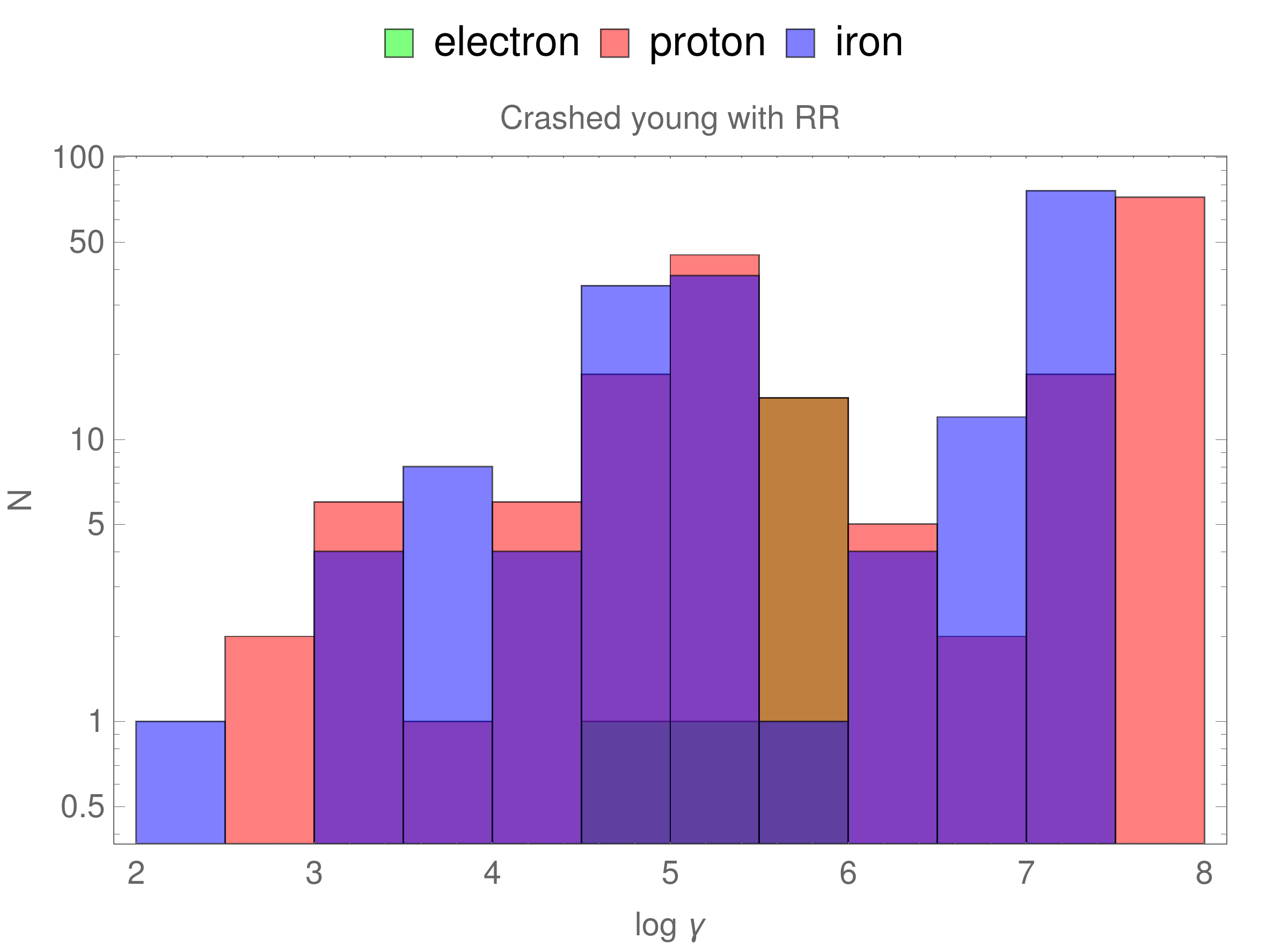} &
	\includegraphics[width=0.5\linewidth]{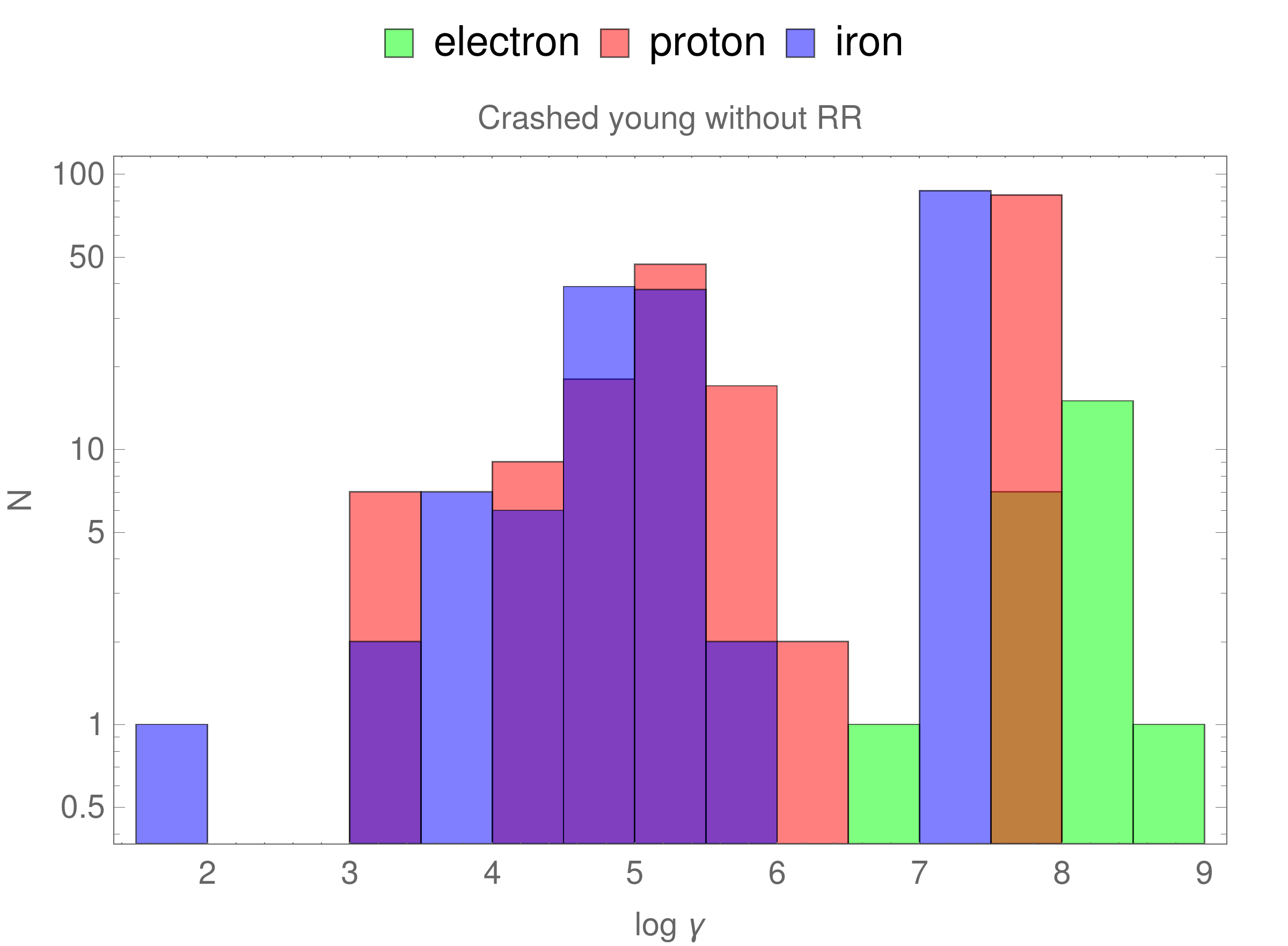} \\	
	\includegraphics[width=0.5\linewidth]{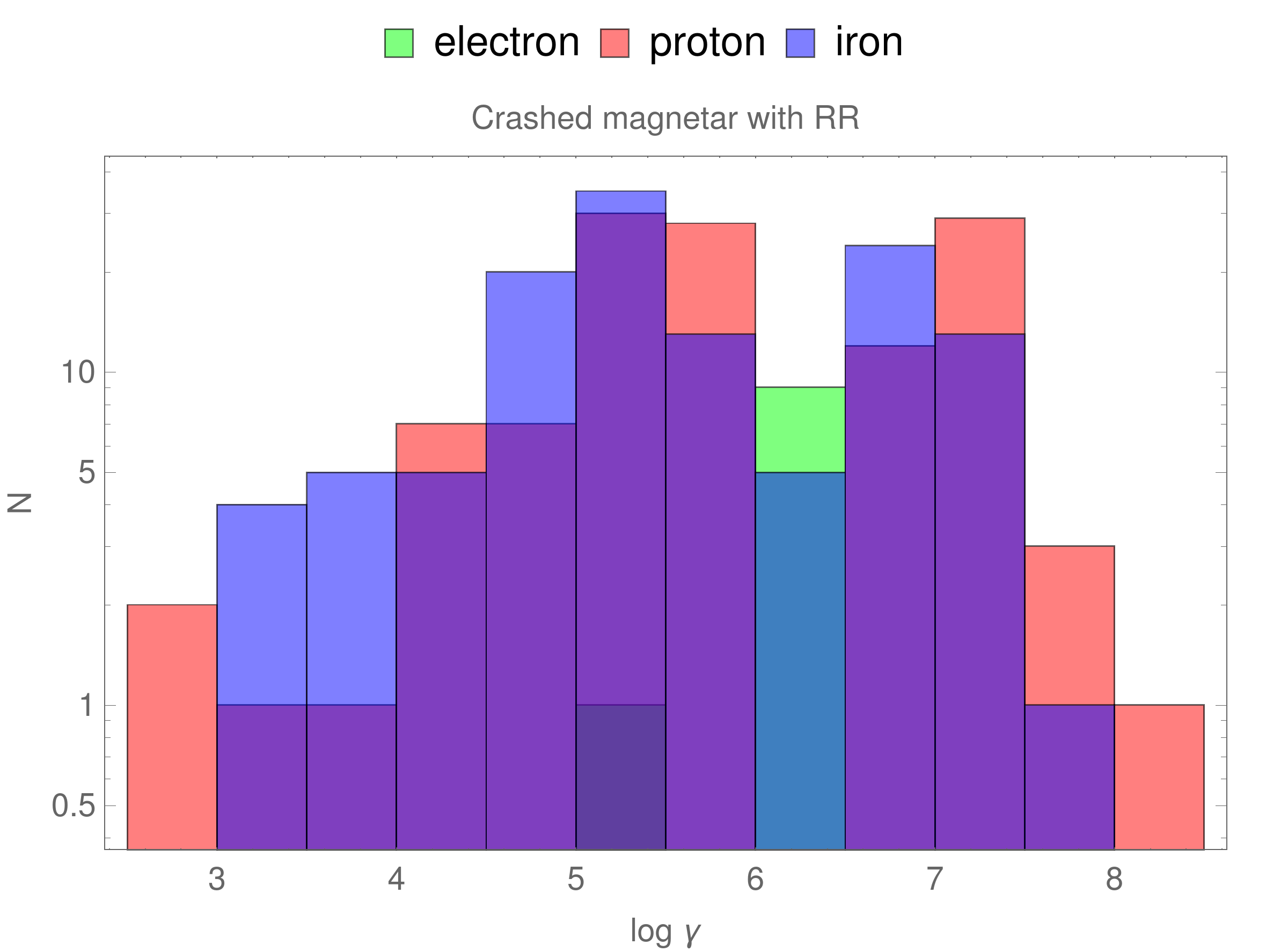} & 
	\includegraphics[width=0.5\linewidth]{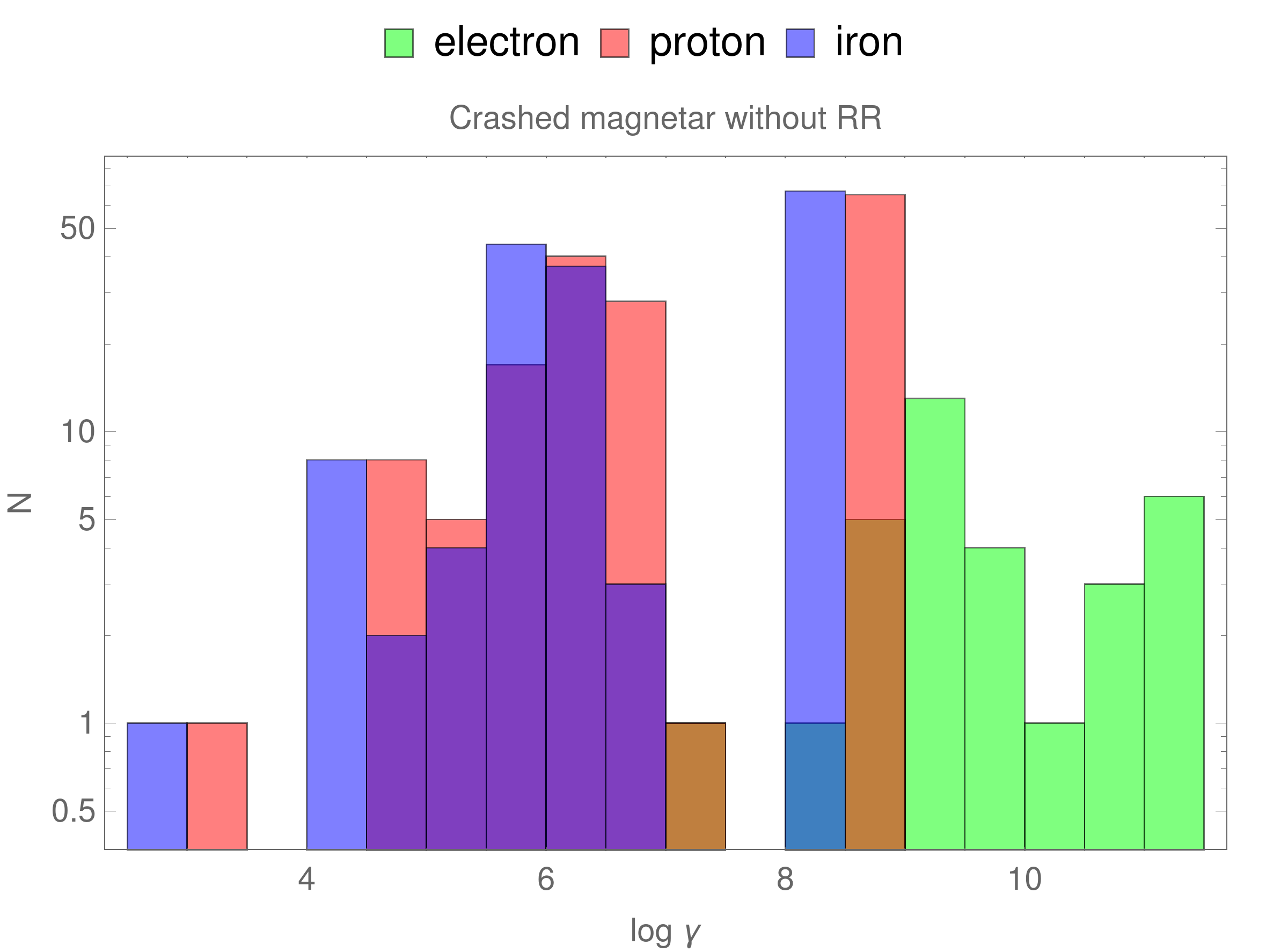}	
	\end{tabular}
	\caption{Same as Fig.~\ref{fig:escaped_gamma_final_g1_cfl-2} but for crashed particles.}
	\label{fig:crashed_gamma_final_g1_cfl-2}
\end{figure*}

Closer to the star, most species quickly crash onto the surface in a time much shorter than the neutron star spin period. Particles crashing onto the neutron star surface are easily recognized by the fact that their final position lies inside the star. %Although being accelerated in an electromagnetic field of increasing strength when moving towards the star, they reach moderate to high Lorentz factors $\gamma \sim 10^7-10^8$, less than those of trapped of escaping particles, 
Compared to escaping particles, the situation is now reversed, magnetars offering the highest energetic particles heating the surface and millisecond pulsars the lowest energetic particles, see left column of  Fig.~\ref{fig:crashed_gamma_final_g1_cfl-2}. This is accounted for by the lower surface magnetic field of millisecond pulsars, being three to five orders of magnitude lower than young pulsars or magnetars respectively. Neglecting \RR, electrons are able to reach Lorentz factors up to $\gamma_{\rm f} \sim 10^{11}$ for magnetars but only $\gamma_{\rm f} \sim 10^{8.5}$ for millisecond pulsars. The \RR impact is strongest for magnetars. However, protons and irons are not perturbed by \RR except sensibly for magnetars. Nevertheless, we observe that with \RR protons remain the most energetic particles with final Lorentz factors about $\gamma_{\rm f} \sim 10^{7.5}-10^{8.5}$ irrespective of the neutron star nature, millisecond, young or magnetar. For electrons the situation is drastically different. They radiate copiously, decreasing they Lorentz factor by three orders of magnitude comparing to the no \RR case in the magnetar environment. The decrease is less pronounced for young or millisecond pulsars but still perceptible. %Moreover, millisecond pulsars produce the highest energy crashing electrons with $\gamma_f \sim 10^7$ whereas young pulsars and magnetars produces ten times less energetic crashing electrons with  $\gamma_f \sim 10^6$. We attribute this decrease to a more efficient RR in these stronger magnetic fields.

\subsection{Trapped particles}

\begin{figure*}
	\centering
	\begin{tabular}{cc}
	\includegraphics[width=0.5\linewidth]{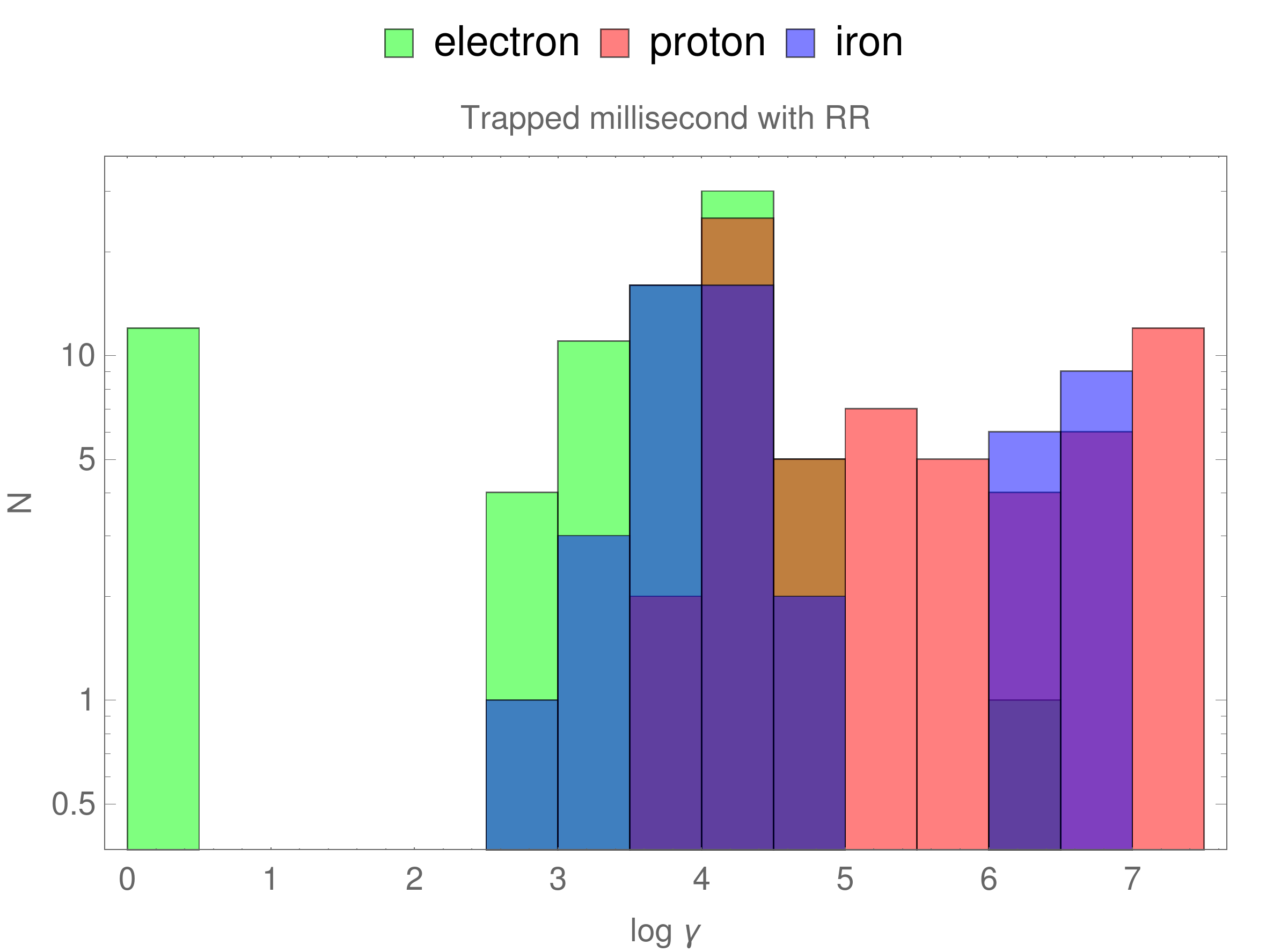} &
	\includegraphics[width=0.5\linewidth]{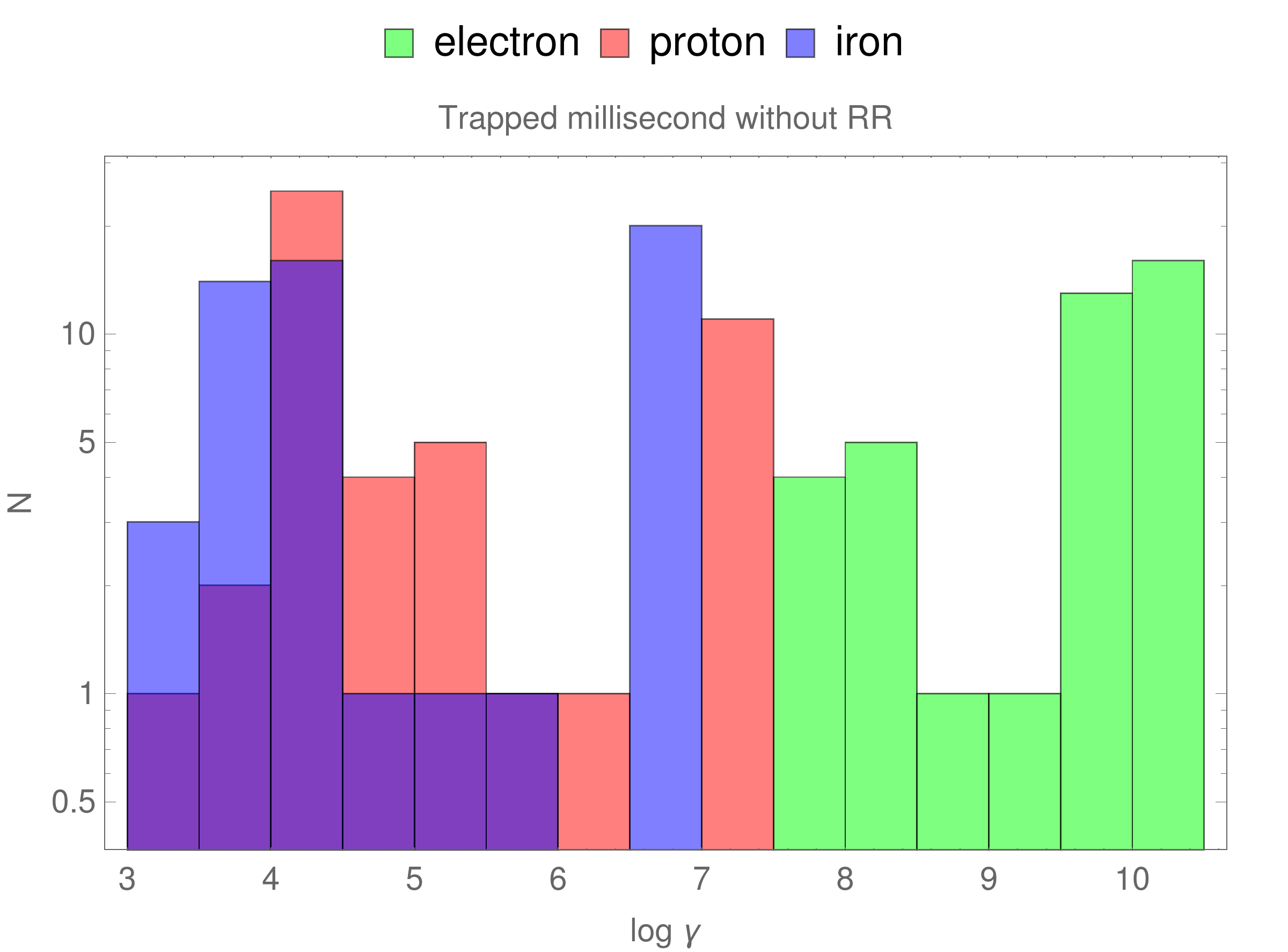} \\
	\includegraphics[width=0.5\linewidth]{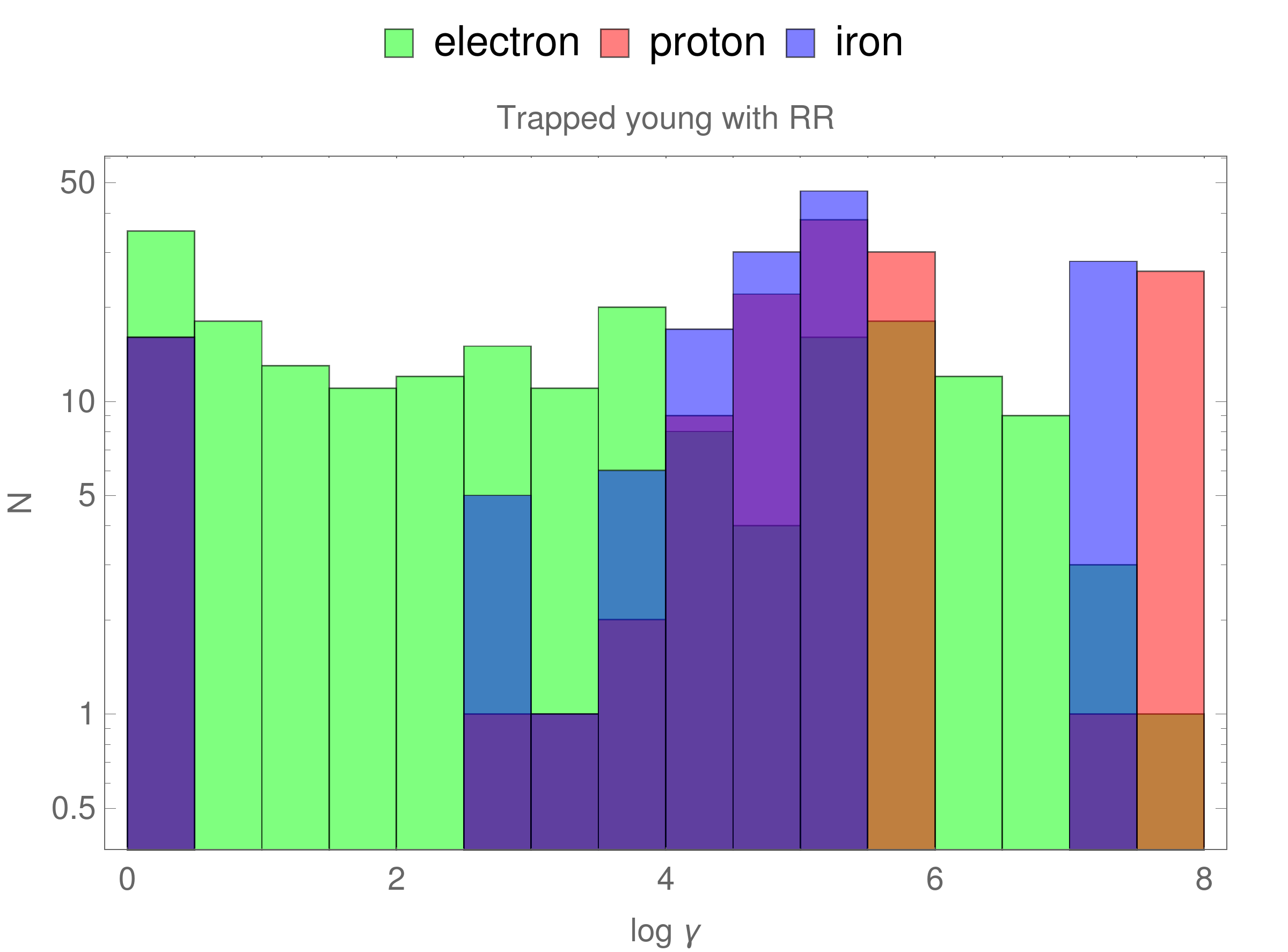} &
	\includegraphics[width=0.5\linewidth]{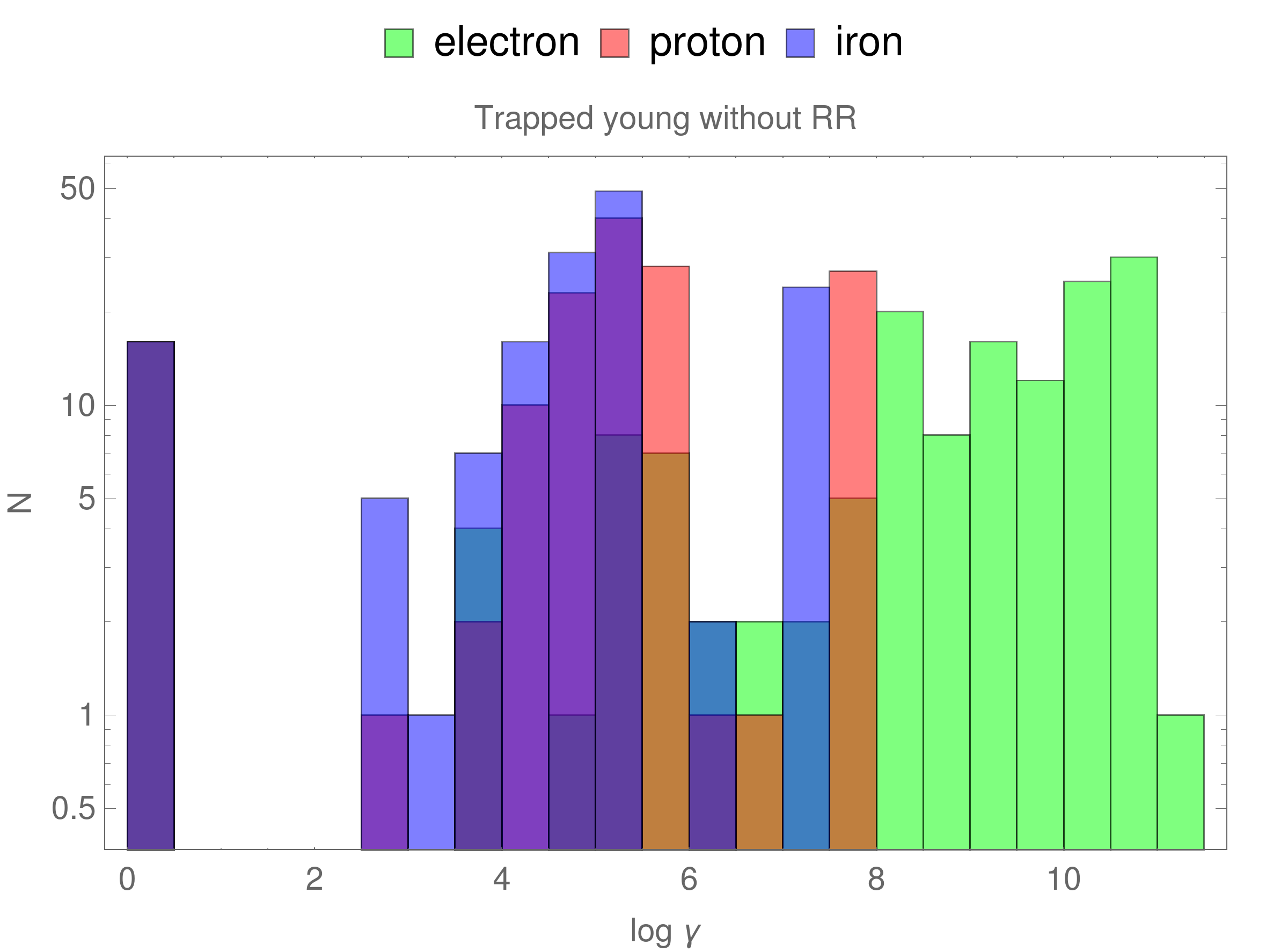} \\
	\includegraphics[width=0.5\linewidth]{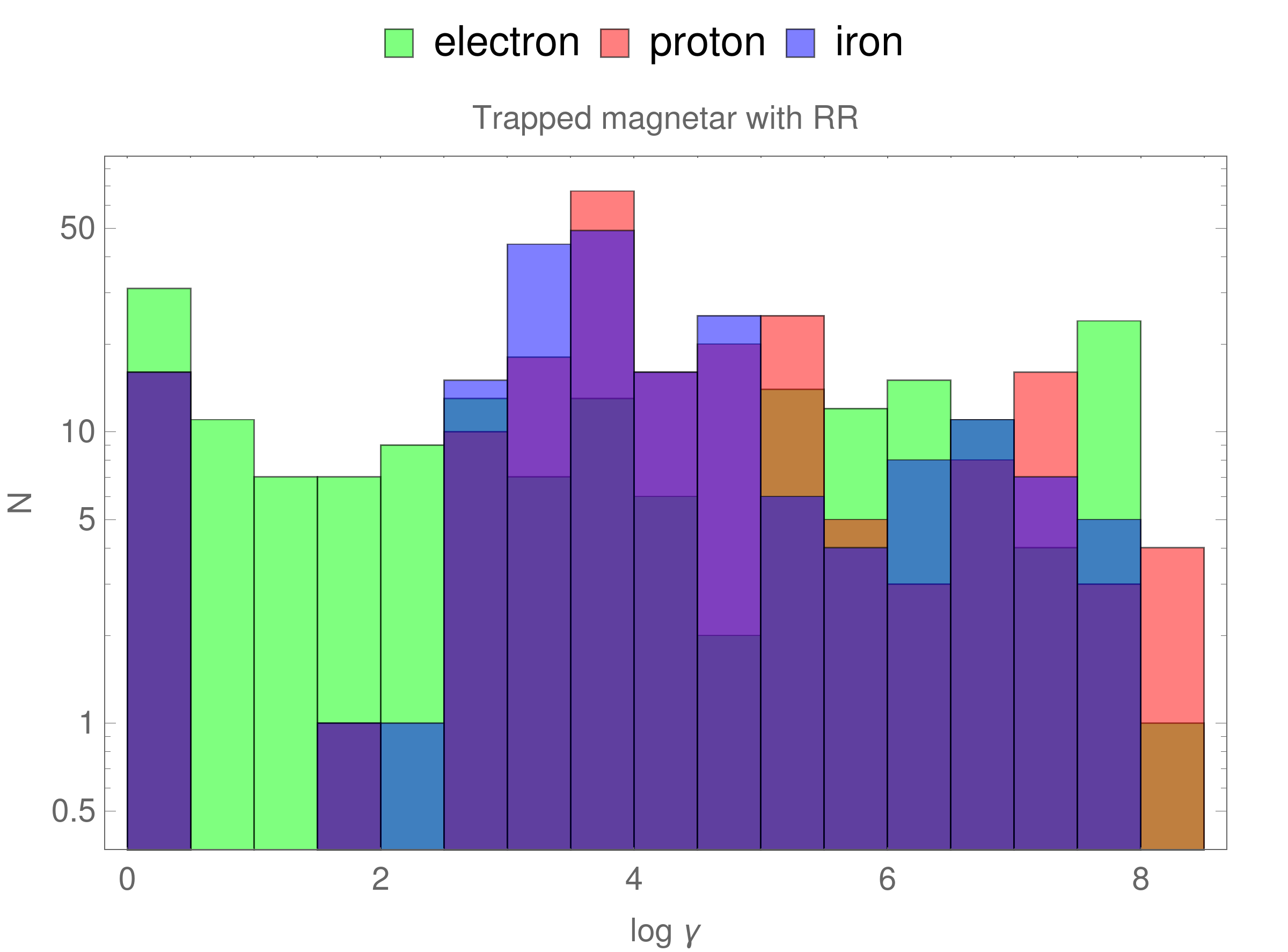} &	
	\includegraphics[width=0.5\linewidth]{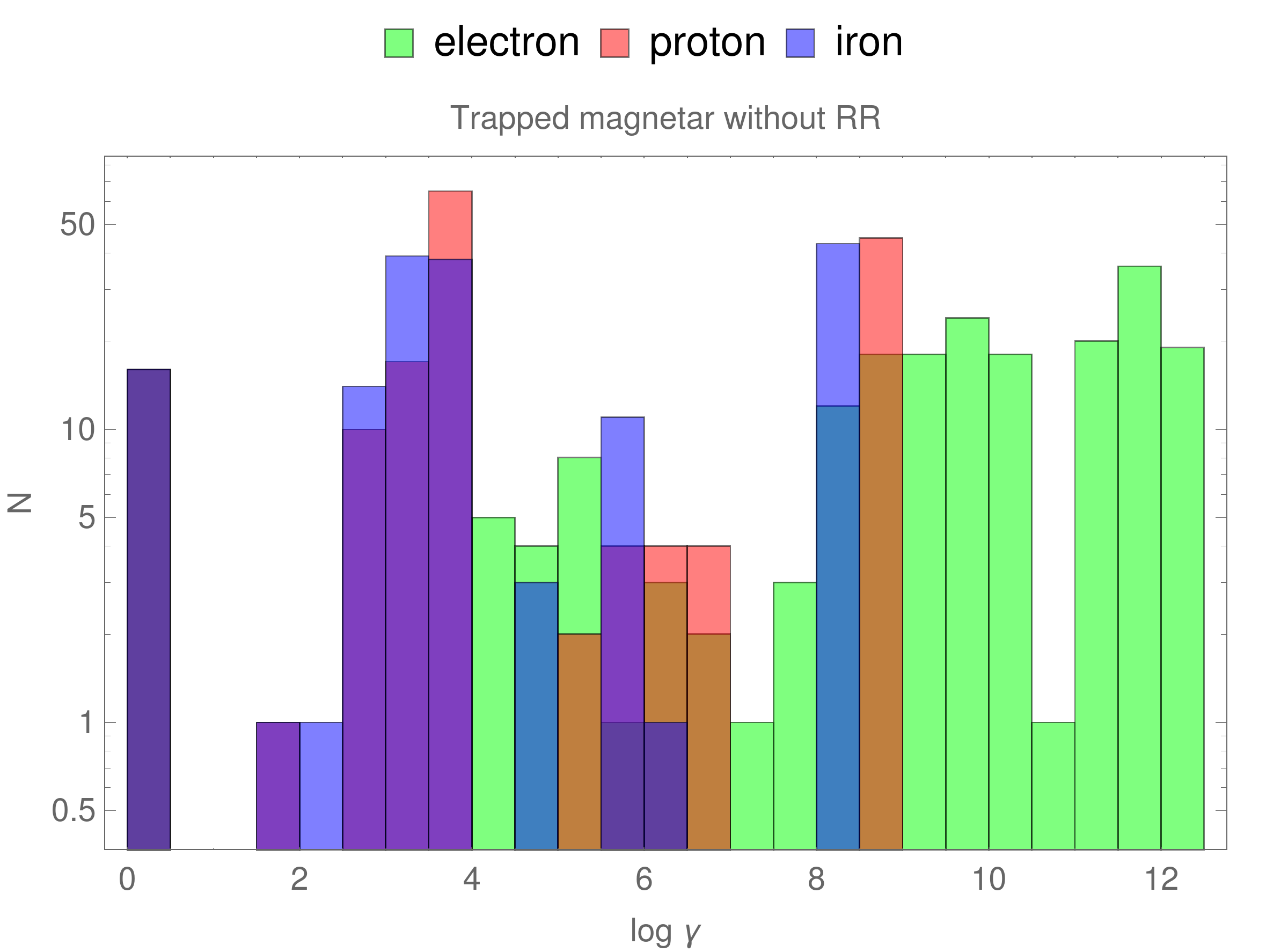}
	\end{tabular}
	\caption{Same as Fig.~\ref{fig:escaped_gamma_final_g1_cfl-2} but for trapped particles.}
	\label{fig:trapped_gamma_final_g1_cfl-2}
\end{figure*}

By default we assume that trapped particles are those not crashing onto the neutron star and not escaping to large distances outside the light cylinder within the simulation time span corresponding to several neutron star periods.
Fig.~\ref{fig:trapped_gamma_final_g1_cfl-2} summarizes the distribution of Lorentz factors for electrons, protons and irons in the LLR approximation and without radiation reaction. Protons and irons are still insensitive to \RR except for magnetars. Electrons are much more sensitive to radiation reaction, decreasing their Lorentz factor by four orders of magnitude for millisecond pulsars, young pulsars and magnetars. Millisecond pulsars produces trapped protons and irons with energies about $\gamma_{\rm f} \sim 10^7$ whereas young pulsars and magnetars one decade more up to $\gamma_{\rm f} \sim 10^8$, no matter if \RR is included or not. Electrons are trapped with similar Lorentz factor although slightly less for millisecond pulsars.

\subsection{Maximum Lorentz factor}

For escaping particles, in the wave zone, the gain in energy is limited by the spherical nature of the electromagnetic field, meaning decreasing in strength with distance like $1/r$. For a null like electromagnetic field \cite{petri_particle_2021} showed that this severely limits the maximum Lorentz factor to values of
\begin{equation}\label{eq:gamma_max}
\gamma_{\rm max} \approx 2 \, (a_{\rm {B_L}}/\pi)^{2/3}
\end{equation}
$a_{\rm {B_L}}$ being the strength parameter as measured at the light cylinder. Radiation reaction remains also negligible in this wave zone. %However, this estimate does not take into account the radial component of the magnetic field that leads to possible resonances \citep{michel_electrodynamics_1999}.
Table~\ref{tab:ParametresRL} summarizes the relevant parameters at the light cylinder for the three kinds of neutrons stars. As a rule of thumb, we found no particle with Lorentz factor exceeding $\gamma_{\rm f} \approx 10^{9.1}$ in the LLR regime. Because the vacuum electromagnetic used in our simulations corresponds to the one producing the strongest parallel electric field (with respect to the magnetic field), no particle should be created and moving with Lorentz factor higher than $10^{9.1}$ within the magnetosphere.
\begin{table*}
	\centering
	\begin{tabular}{lcccc}
		\hline
		Neutron star & $\log a_{\rm {B_L}}$ & \multicolumn{3}{c}{$\log \gamma_{\rm max}$} \\
		\hline
		& & Crashed & Trapped & Escaped \\
		\hline
		millisecond & 9.3 / 6.0 / 5.7 & 7.1 / 7.1 / 6.8 & 6.5 / 7.1 / 6.8 & 9.1 / 6.0 / 5.6 \\
		young       & 7.6 / 4.4 / 4.1 & 6.0 / 7.8 / 7.5 & 7.8 / 7.8 / 7.5 & 8.0 / 4.1 / 3.7 \\
		magnetar    & 7.6 / 4.4 / 4.1 & 6.4 / 8.1 / 7.7 & 8.2 / 8.1 / 7.7 & 7.1 / 3.2 / 2.9 \\
		\hline
	\end{tabular}
	\caption{\label{tab:ParametresRL}Maximum Lorentz factors $\gamma_{\rm max}$ for the three kind of particles: electrons / protons / iron nuclei. The value of the strength parameter at the light cylinder is also given.}
\end{table*}
Eq.~\eqref{eq:gamma_max} is satisfied for non null like electromagnetic waves as those launched by a rotating magnetic dipole. Instead we found a simple linear relation relating the strength parameter $a_{\rm {B_L}}$ to the Lorentz factor such that 
\begin{equation}\label{eq:gamma_max_2}
\gamma_{\rm max} \approx a_{\rm {B_L}} .
\end{equation}
This increase in the acceleration efficiency is imputed to the presence of a still strong radial component of the electric field which was absent in the study of \cite{petri_particle_2021}. 
The simulations performed in this section only followed a small number of particles due to the stringent computation time required to accurately evolve the particle velocity and position. Describing the plasma feedback onto the electromagnetic field would require a much larger number of particles coupled to the evolution of the electromagnetic field via Maxwell equations, leading to a particle-in-cell code. So far, although PIC codes exist and have been adapted to simulate neutron star magnetospheres, none was yet able to handle the parameter space explored in the present work. Therefore let us contrast our results in light of existing kinetic descriptions of the magnetosphere.

\subsection{Comparison to previous works}

%\rev{2. Section 4 is dedicated to applying this method to neutron star magnetospheres. Scientifically this is the part that contains novel results. It would be helpful if the author can put these results in the context of the general neutron star literature, comparing them to previous works on the same topic.}

Several investigations of particle acceleration and radiation reaction around neutron stars have been attempted in the literature. Very rare are however studies employing realistic field strengths for the neutron star due to severe numerical limitations. Let us mention however the pioneer work of \cite{finkbeiner_effects_1989} and \cite{finkbeiner_applicability_1990} who employed a single test particle approach with radiation reaction and found acceleration around the neutron star up to Lorentz factors of about $\gamma_{\rm f} \lesssim 10^9$ for the Crab parameters. In a similar manner, at very large distances, in the wind zone, \cite{michel_electrodynamics_1999} studied particle acceleration without radiation reaction and found asymptotic values of $\gamma_{\rm f} \lesssim 10^9$. The flaw of these studies is that particles evolve in a prescribed external field without possible feedback due to the plasma around the neutron star. A fully kinetic description of the plasma and field started only recently using PIC schemes, earlier simulations having not taken into account radiation reaction. Unfortunately, the flaw of this approach is the use of unrealistically low field strength. Let us however mention some of these works.

\cite{cerutti_particle_2015} studied acceleration for an axisymmetric magnetosphere without radiation reaction. They got a maximum energy for leptons $\gamma_{\rm f} \lesssim 10^3$ related linearly to the magnetisation parameter in the plasma. Thereafter \cite{cerutti_modelling_2016} included the radiation reaction force and got maximum energies one order of magnitude less with $\gamma_{\rm f} \lesssim 10^2$. Dissipation in the striped wind due to magnetic reconnection led \cite{cerutti_dissipation_2020} to the same conclusion. Other PIC simulations performed by \cite{kalapotharakos_three-dimensional_2018} using similar algorithms with radiation reaction found similar results with $\gamma_{\rm f} \lesssim 10^3$ for pairs. Nevertheless, these authors extrapolated to realistic energies by using rescaling techniques for field strengths, time and space scales. How effective and consistent this rescaling operates is not clear as the problem is highly non-linear in a significant radiation reaction regime. General relativity does not significantly change these conclusions as shown by \cite{philippov_ab-initio_2018} who including frame-dragging effects and found $\gamma_{\rm f} \lesssim 500$ by extending their special relativistic results in \cite{philippov_ab_2015}.

When focusing on the near field of a dipole \cite{ferrari_acceleration_1974} found an asymptotic Lorentz factor for electron about $10^8$ and slight larger for proton almost $10^9$ but for faster rotation in a field of an oblique rotating dipole with strength $\numprint{5e6}$~T. When radiation reaction remains irrelevant, their results agree with those of \cite{kulsrud_effect_1972}, demonstrating a linear growth with the field strength parameter. \cite{laue_acceleration_1986} investigates the special case of an orthogonal rotator with radiation reaction and for typical neutron star parameters, they found maximum energy for electrons about $10^9$ and for protons about $10^6$.

Hadron acceleration has been much less discuss in this context but is equally important to understand the origin of ultra-high energy cosmic rays. To this end \cite{guepin_proton_2020} investigated proton and pair acceleration in an aligned neutron star magnetosphere with radiation reaction. They drastically reduced the neutron star radius and the proton over electron mass ratio for computational purposes, therefore they found highest energies for pairs only about $\gamma_{\rm f} \lesssim 700$ and for protons about $\gamma_{\rm f} \lesssim 40$. These state of the art results emphasize the difficulty to tend towards a realistic and self-consistent description of neutron star magnetospheres. The central bottleneck is the particle pusher, requiring to resolve temporally the gyromotion. This drawback is circumvent by employing an approximation called the radiation reaction limit, summarized in section~\ref{sec:EOM}. It is therefore important to assess quantitatively the accuracy and efficiency of this alternative approach as done in the next section.

\section{Comparison with the radiation reaction limit}
\label{sec:VRR}

The results obtained in this section rely on the numerical integration of the LLR equation accounting for realistic parameters introducing a huge gap between the gyro-frequency and the neutron star rotation period. The question arises then on how to improve our algorithm or to speed up the computation by several decades. To this end, in a last section we compare the LLR results to the \RR limit regime to assert the usefulness of the latter.

Integrating the exact LLR equations requires to resolve the gyro-frequency which is very stringent and impossible to use for a large sample of particles as required to perform kinetic simulations such as those done in PIC or Vlasov codes. We therefore checked the accuracy of the much faster radiation reaction limit approximation where the particle velocity is expressed in terms of the local electromagnetic field, eq.~\eqref{eq:VRR}. To this aim, we computed trajectories for electrons and protons in the field of a millisecond pulsar for different magnetic moment inclination angles and different initial particle positions. Because by construction the speed in eq.~\eqref{eq:VRR} is equal to the speed of light~$v_\pm = c$, in the LLR approach particles are kicked with high initial Lorentz factors $\gamma_0=10^3$ and a velocity parallel to $\vec{v}_\pm$ in order to have comparable initial conditions for both sets of runs.

\begin{figure}
	\centering
	\includegraphics[width=\linewidth]{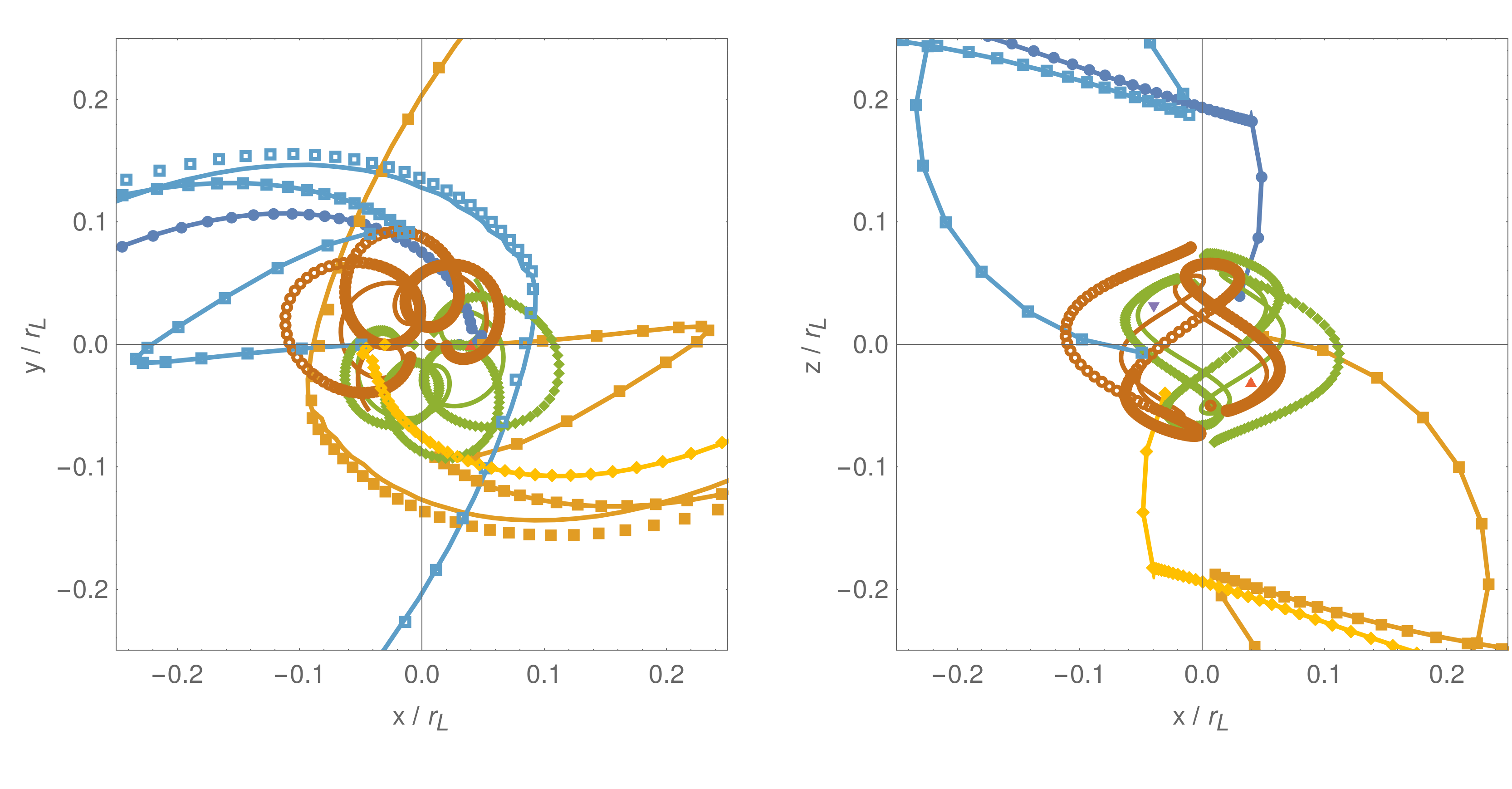}
	\includegraphics[width=\linewidth]{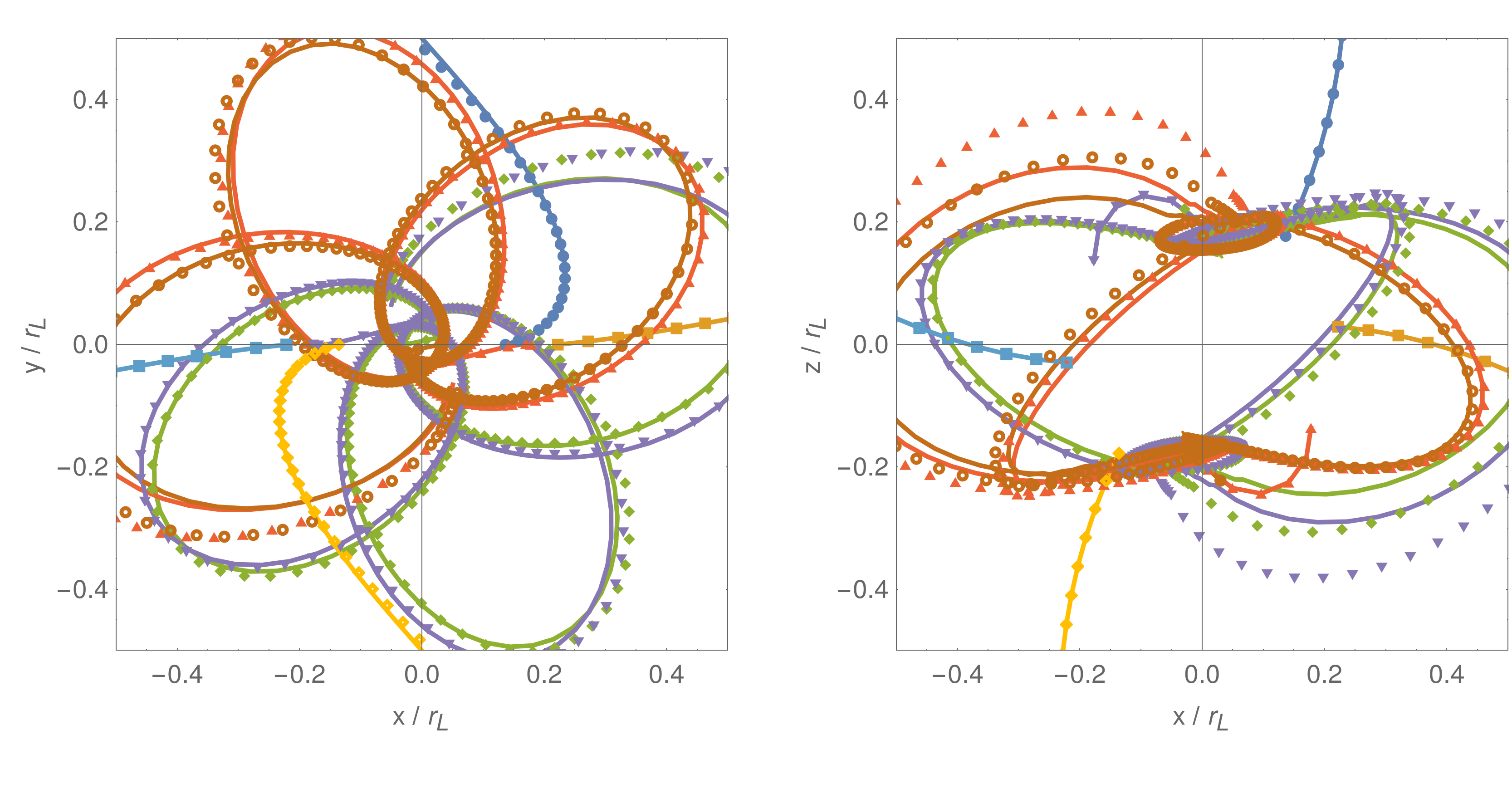} \\
	\includegraphics[width=\linewidth]{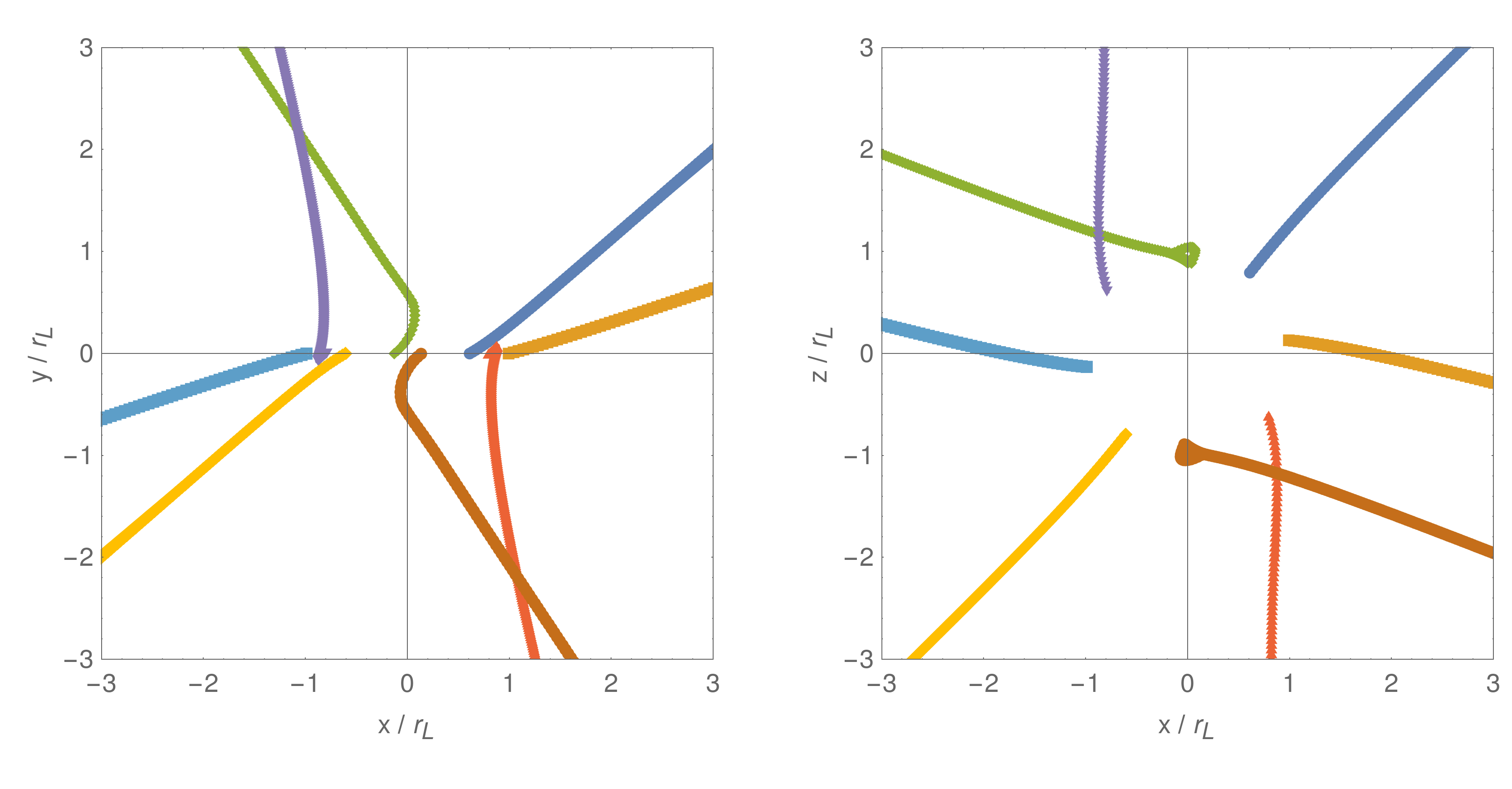} \\
	\caption{A sample of trajectories for electrons obtained with LLR in the field of a millisecond pulsar, in solid lines, and in the radiation reaction limit, marked with dotted symbols. Trajectories are projected along the $xy$ plane on the left column and on the $xz$ plane on the right column. Particles are launched from the stellar surface in the upper row, at a distance $\sqrt{R_* \, \rlight}$ in the middle row and at a distance $r=\rlight$ on the bottom row.}
	\label{fig:electron_comparaison_ar_LLR}
\end{figure}

Fig.~\ref{fig:electron_comparaison_ar_LLR} shows a sample of electron trajectories and demonstrates the reasonable results obtained by this asymptotic regime for a millisecond pulsar. However the precision depends on the particle initial position. For motions starting at the surface, upper row in Fig.~\ref{fig:electron_comparaison_ar_LLR}, some trajectories, in blue, orange and yellow are well reproduced by the \RR limit regime. The accuracy is less good for the brown and green paths although the general trend is conserved. When starting at larger distances from the surface, at $\sqrt{R_* \, \rlight}$ like in the middle row of Fig.~\ref{fig:electron_comparaison_ar_LLR}, we observe better agreement between both regimes. The best results are obtained for particles well away from the surface, starting at $r=\rlight$, lower row in Fig.~\ref{fig:electron_comparaison_ar_LLR}. All trajectories computes in the \RR limit regime overlap with the LLR integration.  

\begin{figure}
	\centering
	\includegraphics[width=\linewidth]{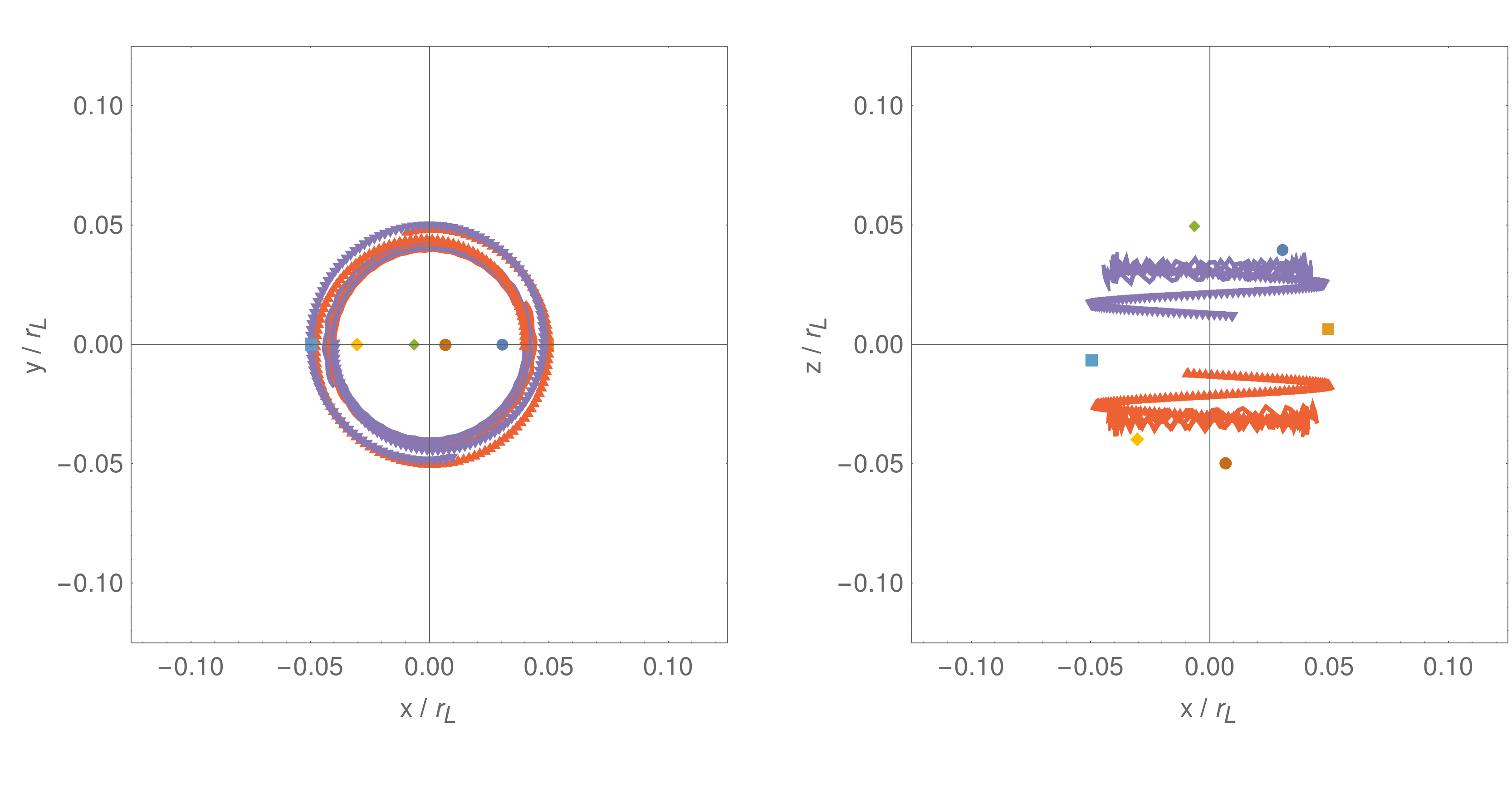}
	\includegraphics[width=\linewidth]{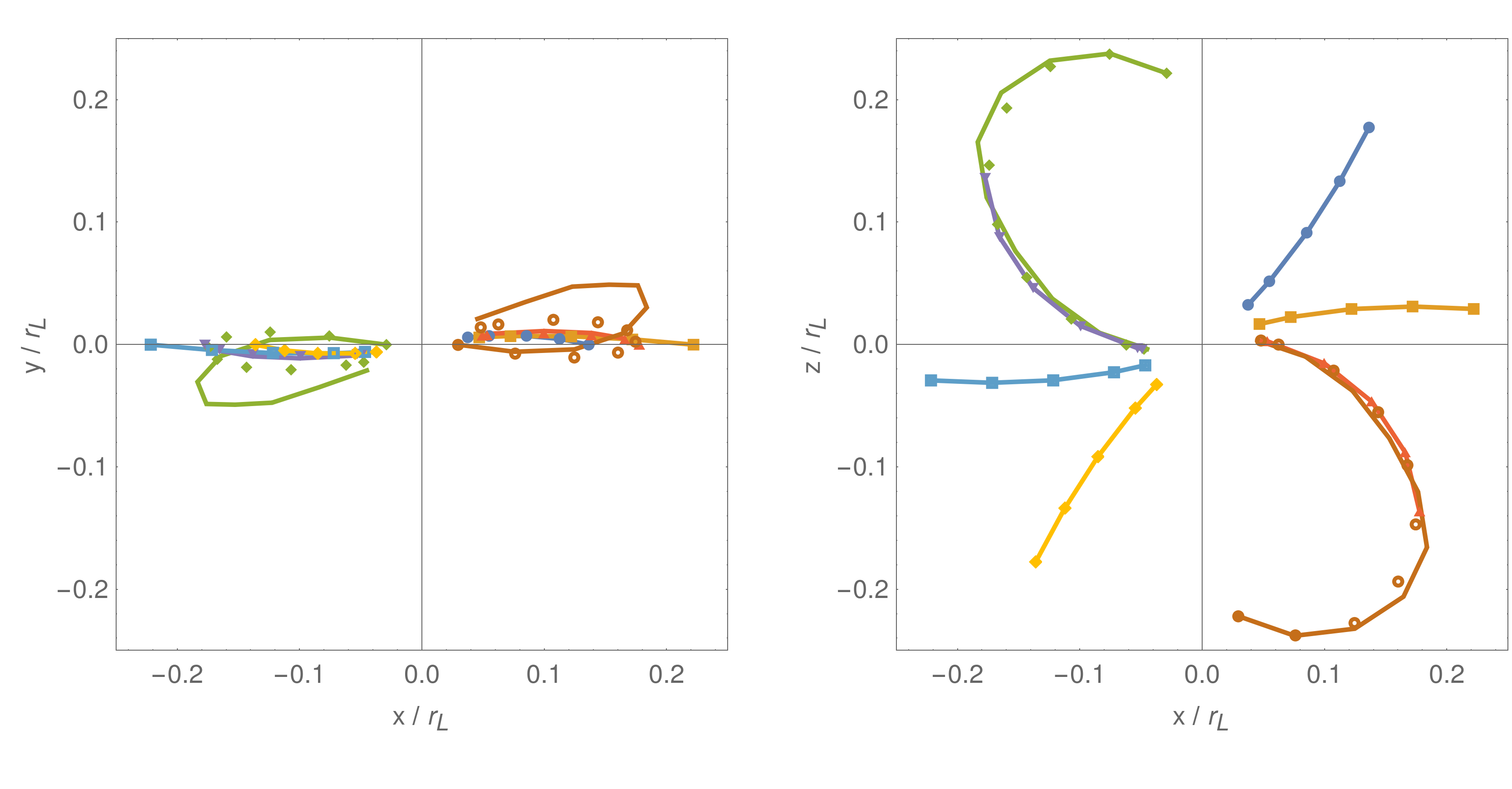}
	\includegraphics[width=\linewidth]{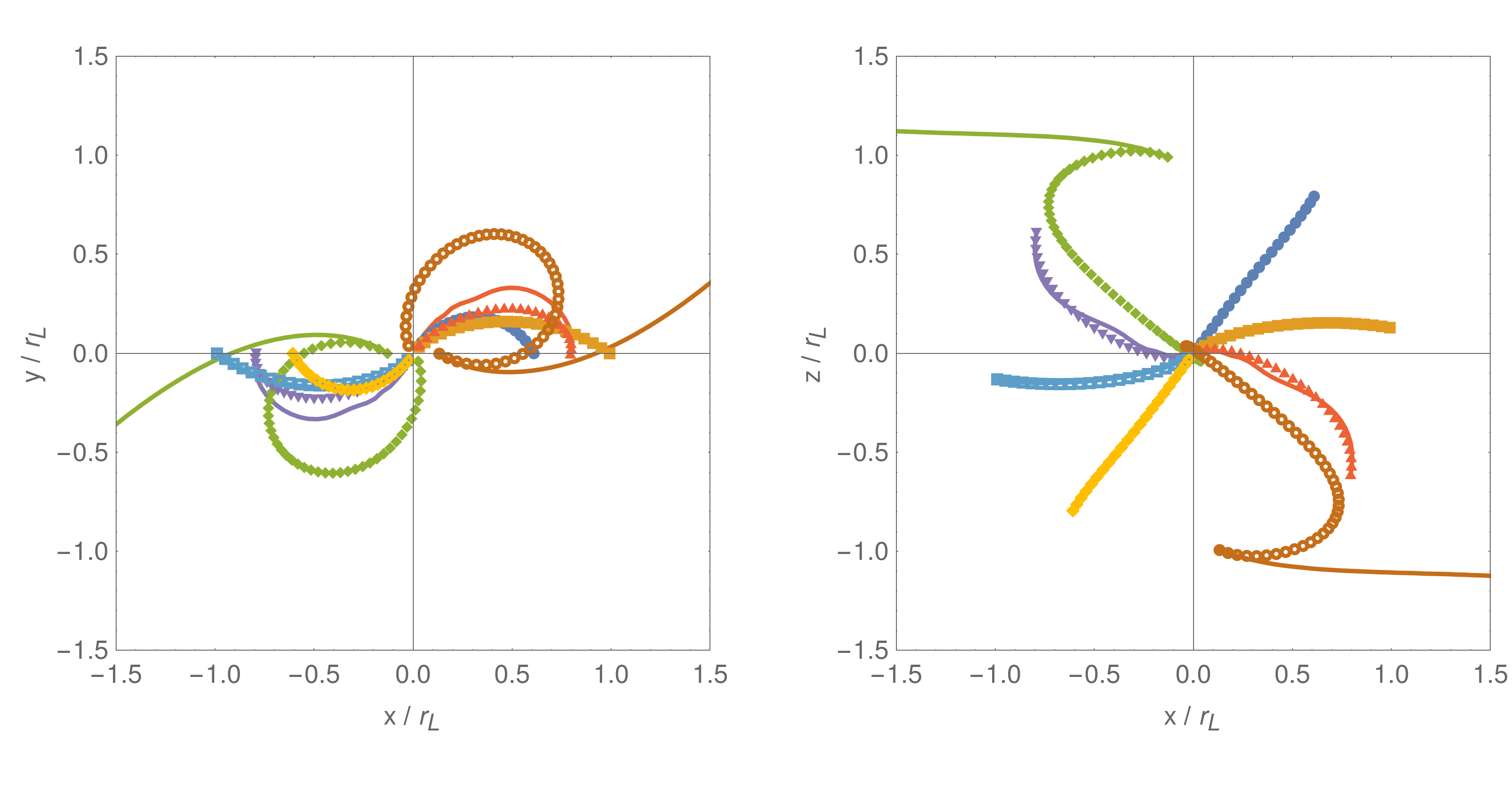}
	\caption{Same as fig.~\ref{fig:electron_comparaison_ar_LLR} but for protons.}
	\label{fig:proton_comparaison_ar_LLR}
\end{figure}

Comparison of trajectories for protons are shown in fig.~\ref{fig:proton_comparaison_ar_LLR}. Here the agreement is satisfactory within the light cylinder, close to the surface, upper row, and at intermediate distances, middle row. For protons starting at the light cylinder radius $r=\rlight$ the results are more contrasted, some trajectories being well reproduced, in blue, yellow and orange colours and some being false like the brown and green colour motions, expected to crash on the surface but escaping in the \RR limit regime.
Irons show trajectories very similar to protons because of the nearly identical mass over charge ratio $q/m$, figures are therefore not shown in this almost identical case.

The \RR limit regime is less accurate than the exact LLR integration scheme but this is partially compensated by the drastic decrease in computational time, lowered by several orders of magnitude in this approximation. Expression \eqref{eq:VRR} could certainly be improved by carefully investigating theses problematic cases but we do not pursue this aim in this work. We demonstrated however that the velocity in eq.\eqref{eq:VRR} offers a valuable compromise between a time consuming full integration of the equation of motion in LLR and a artificial and unrealistic down scaling of the major physical parameters making a neutron star a neutron star.

%\rev{3. Section 5 compares the comprehensive numerical recipe developed in Section 3
%	with a reduced model where the particle velocity is given by the local
%	electromagnetic field. The same comment as point 1 applies. Although the solid
%	trajectories and dotted lines are, in general, close to each other, it would be
%	much more useful to have a quantitative comparison between these two recipes.
%	For example, one can see in Figure 12 \& 13 that sometimes there can be
%	significant deviations between the corresponding curves, and it is not obvious
%	whether it is due to discretization error, initial condition, or a deficiency of
%	the radiation reaction limit method. Furthermore, the computations are done in a
%	Deutsch solution, where the parallel electric field is the maximum for a pulsar
%	magnetosphere. In a plasma-filled magnetosphere the parallel E field is, in
%	general, much weaker. It is therefore more important to quantify how the reduced
%	method deviates from the more reliable solution under such circumstances.}

\begin{figure}
	\centering
	\includegraphics[width=\linewidth]{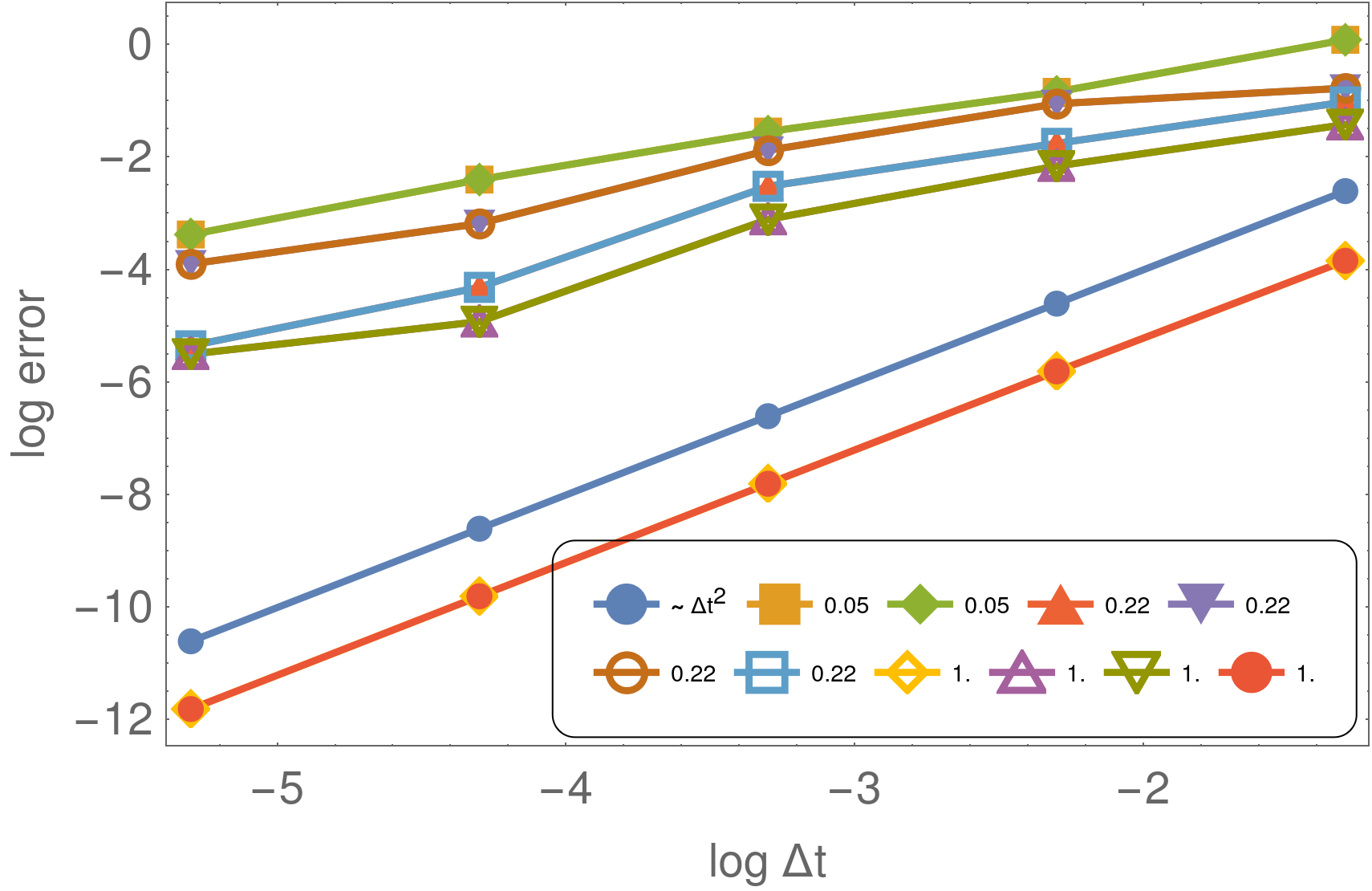}
	\caption{Convergence of the radiation reaction limit regime showing the method of integration to be between first and second order in time depending on the initial position of the particle given by the number $r_0/\rlight$ in the legend. The $\Delta t^2$ decrease in shown in blue filled circles.}
	\label{fig:erreuraristote}
\end{figure}

\begin{figure}
	\centering
	\includegraphics[width=\linewidth]{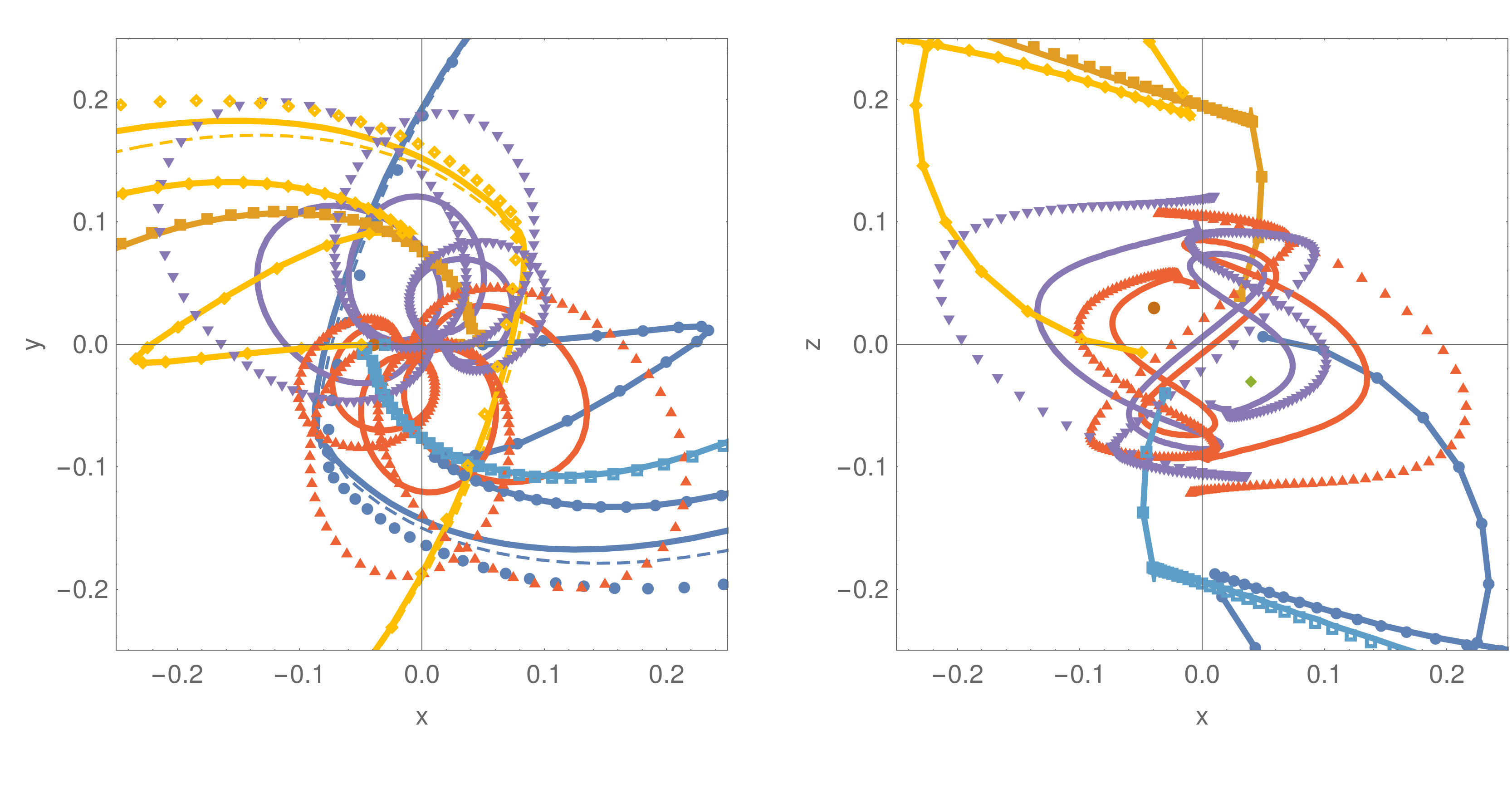}
	\includegraphics[width=\linewidth]{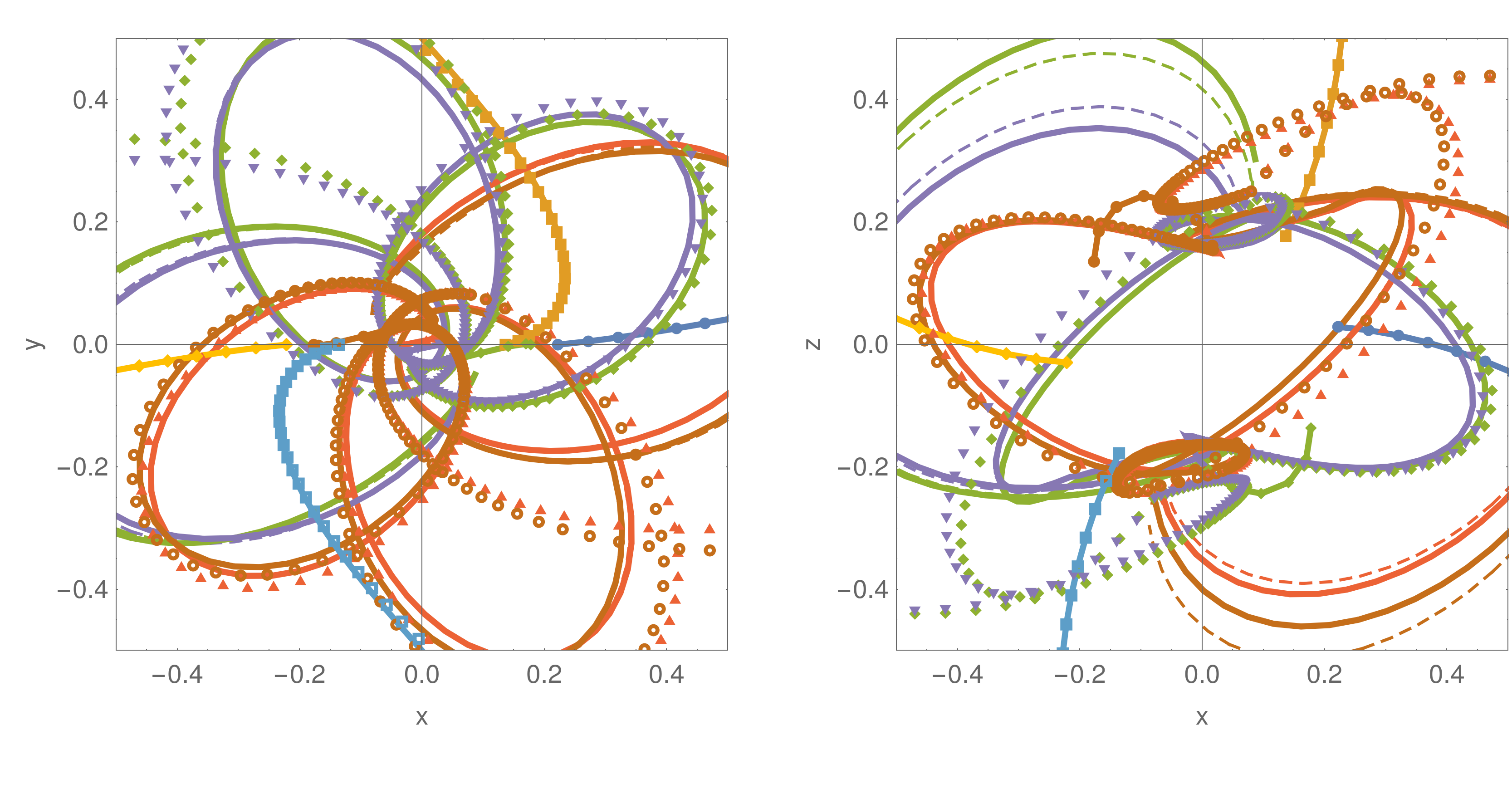}
	\includegraphics[width=\linewidth]{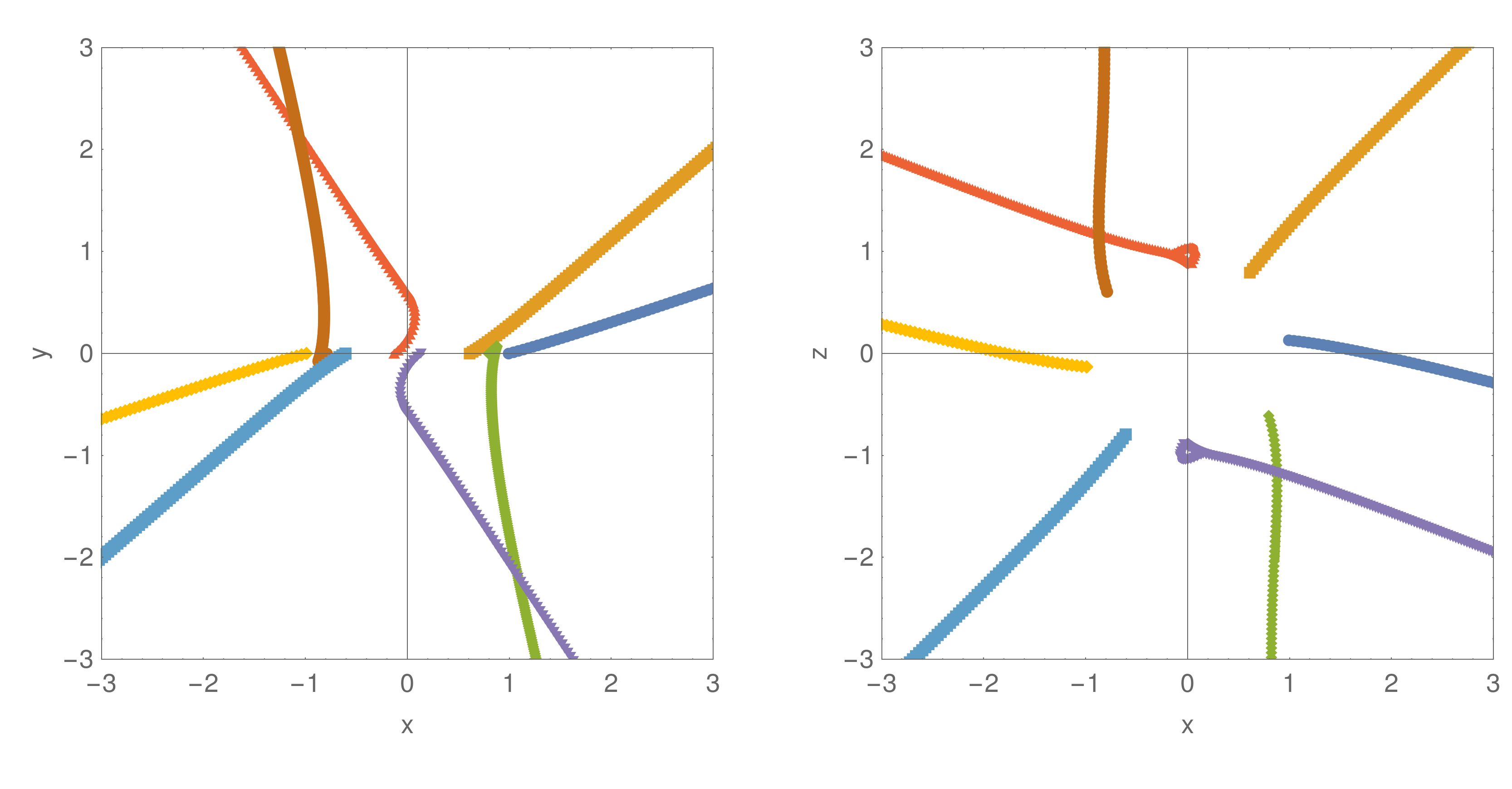}
	\caption{Comparison between the radiation reaction limit, in dot markers, the LLR, in thick solid lines and the approximated LLR motion, in thin dashed lines.}
	\label{fig:electroncomparaisonapproxarllrcfl-23}
\end{figure}

A convergence analysis of the radiation reaction limit integration scheme is shown in Fig.~\ref{fig:erreuraristote} for the relative error. We simulated a sample of twelve particles starting at different locations within the magnetosphere and compared their last position to a reference solution. As no exact analytical solutions are known, we use as a reference numerical solution the one with the smallest time step. The integration scheme oscillates between first and second order in time depending on the initial position of the particle as shown by the number $r_0/\rlight$ in the legend. For reference the $\Delta t^2$ behaviour is shown in blue filled circles. In the previous simulations, we fixed the time step to $\Delta t \approx 10^{-4}$ thus expecting a precision better that 3 digits in all cases. The discrepancy between \LL and \RR regime can therefore not be explained by some discretization effect. We also checked that the initial condition on the velocity does not impact the trajectory in \LL. The explanation must be search in the deficiency of the \RR regime to satisfactorily account for all possible trajectories. Indeed, this regime assumes a radiative friction force opposite to the 3-velocity vector. However, the 3D version of the LLR equation also contains components along $\mathbf{E} \wedge \mathbf{B}$ and along $\mathbf{E}$ and $\mathbf{B}$ when the linear term in velocity is retained. The main discrepancy arises from the neglect of this linear term.
In order to proof this argument, we designed a simplified version of the \LL equation by only retaining in the new analytical solution the part of the \RR force directed along the velocity (which is valid for ultra-relativistic speeds). The expressions for the 4-velocity then become
\begin{subequations}
	\begin{align}
	u_E & = \lambda_{\rm B} \, \frac{u_E^0 \, \cosh(\lambda_E \, \tau) + \tilde{F} \, u_E^0 \, \sinh(\lambda_E \, \tau) / \lambda_E}{\sqrt{(\lambda_E^2 + \lambda_B^2) \, |u_E^0|^2 + (\lambda_{\rm B}^2 -(\lambda_E^2 + \lambda_B^2) \,|u_E^0|^2) \, e^{-2\,\lambda_{\rm B}^2 \, \tau_0\,\tau}}} \\
	u_B & = \lambda_{\rm E} \, \frac{u_B^0 \, \cos(\lambda_B \, \tau) + \tilde{F} \, u_B^0 \, \sin(\lambda_B \, \tau) / \lambda_B}{\sqrt{(\lambda_E^2 + \lambda_B^2) \, |u_B^0|^2 + (\lambda_{\rm E}^2 -(\lambda_E^2 + \lambda_B^2) \,|u_B^0|^2) \, e^{2\,\lambda_{\rm E}^2 \, \tau_0\,\tau}}}	\end{align}
\end{subequations}
replacing equation~\eqref{eq:solution_exacte_totale}.
The results are shown in Fig.~\ref{fig:electroncomparaisonapproxarllrcfl-23} for the exact \LL equation in solid lines, the approximated \LL equation in dashed lines and the radiation reaction regime in dots. We observe some significant differences notably, in the middle right panel. We stress however that \RR regime gives accurate results at low computational time expense for the majority of cases.

\section{Conclusions}
\label{sec:Conclusions}

Strongly magnetized rotating neutron stars are powerful and efficient particle accelerators able to accelerate leptons and hadrons to Lorentz factors as high as $10^9$ for the former and slightly less for the latter. This upper limit remains largely independent on the nature of neutron star: millisecond pulsar, young pulsar or magnetar. We achieved these results by implementing realistic parameters in our particle pusher based on the exact solution of the LLR approximation of the equation of motion.
Through extensive numerical tests, we show that our scheme is second order in proper time.

The simulation results are accurate and robust but at the expense of high computational cost because of the need to resolve the gyro-motion which is many decades smaller than the neutron star spin period. The \RR limit regime offers a good compromise between accuracy and computational cost but the simplistic expression used is unable to reproduce all trajectories satisfactorily. Nevertheless, it could be conceivable to improve this expression by taking into account a finite Lorentz factor and special electromagnetic field configuration when the radiation reaction is negligible due to a weak accelerating electric field. Nevertheless this extension is left for future work.

A straightforward implementation of the above pusher into a PIC code or codes is prevented by the fact that LLR uses the proper time as integration parameter. Its conversion into an inertial observer time is however feasible as shown by \cite{petri_relativistic_2020}. The next logical step would then be to shift from the test particle motion to a fully kinetic plasma simulation where the particle charge and current densities retroact to the electromagnetic field via Maxwell equations.

\begin{acknowledgements}
I am grateful to the referee for helpful comments and suggestions that helped to improved this work.
This work has been supported by the CEFIPRA grant IFC/F5904-B/2018 and ANR-20-CE31-0010. %We acknowledge the High Performance Computing center of the University of Strasbourg for supporting this work by providing scientific support and access to computing resources. 
\end{acknowledgements}

\bibliographystyle{aa}
\bibliography{/home/petri/zotero/Ma_bibliotheque}

\begin{thebibliography}{39}
\expandafter\ifx\csname natexlab\endcsname\relax\def\natexlab#1{#1}\fi

\bibitem[{Abraham(1902)}]{abraham_prinzipien_1902}
Abraham, M. 1902, Annalen der Physik, 315, 105

\bibitem[{Abraham(1904)}]{abraham_zur_1904}
Abraham, M. 1904, Annalen der Physik, 319, 236

\bibitem[{Boghosian(1987)}]{boghosian_covariant_1987}
Boghosian, B.~M. 1987, PhD thesis, publication Title: Ph.D. Thesis ADS Bibcode:
  1987PhDT.......197B

\bibitem[{Boris(1970)}]{boris_relativistic_1970}
Boris, J. 1970, Proceeding of Fourth Conference on Numerical Simulations of
  Plasmas

\bibitem[{Cerutti {et~al.}(2015)Cerutti, Philippov, Parfrey, \&
  Spitkovsky}]{cerutti_particle_2015}
Cerutti, B., Philippov, A., Parfrey, K., \& Spitkovsky, A. 2015, MNRAS, 448,
  606

\bibitem[{Cerutti {et~al.}(2020)Cerutti, Philippov, \&
  Dubus}]{cerutti_dissipation_2020}
Cerutti, B., Philippov, A.~A., \& Dubus, G. 2020, A\&A, 642, A204

\bibitem[{Cerutti {et~al.}(2016)Cerutti, Philippov, \&
  Spitkovsky}]{cerutti_modelling_2016}
Cerutti, B., Philippov, A.~A., \& Spitkovsky, A. 2016, Monthly Notices of the
  Royal Astronomical Society, 457, 2401

\bibitem[{Deutsch(1955)}]{deutsch_electromagnetic_1955}
Deutsch, A.~J. 1955, Annales d'Astrophysique, 18, 1

\bibitem[{Dirac(1938)}]{dirac_classical_1938}
Dirac, P. A.~M. 1938, Proc. R. Soc. Lond. Series A, 167, 148

\bibitem[{Elkina {et~al.}(2014)Elkina, Fedotov, Herzing, \&
  Ruhl}]{elkina_improving_2014}
Elkina, N.~V., Fedotov, A.~M., Herzing, C., \& Ruhl, H. 2014, Phys. Rev. E, 89,
  053315

\bibitem[{Ferrari \& Trussoni(1974)}]{ferrari_acceleration_1974}
Ferrari, A. \& Trussoni, E. 1974, A\&A, 36, 267

\bibitem[{Finkbeiner {et~al.}(1989)Finkbeiner, Herold, Ertl, \&
  Ruder}]{finkbeiner_effects_1989}
Finkbeiner, B., Herold, H., Ertl, T., \& Ruder, H. 1989, A\&A, 225, 479

\bibitem[{Finkbeiner {et~al.}(1990)Finkbeiner, Herold, \&
  Ruder}]{finkbeiner_applicability_1990}
Finkbeiner, B., Herold, H., \& Ruder, H. 1990, A\&A, 238, 462

\bibitem[{Fulton \& Rohrlich(1960)}]{fulton_classical_1960}
Fulton, T. \& Rohrlich, F. 1960, Annals of Physics, 9, 499

\bibitem[{Gordon \& Hafizi(2021)}]{gordon_special_2021}
Gordon, D.~F. \& Hafizi, B. 2021, Comput. Phys. Commun, 258, 107628

\bibitem[{Gordon {et~al.}(2017{\natexlab{a}})Gordon, Hafizi, \&
  Palastro}]{gordon_pushing_2017}
Gordon, D.~F., Hafizi, B., \& Palastro, J. 2017{\natexlab{a}}, AIP Conference
  Proceedings, 1812, 050002, publisher: American Institute of Physics

\bibitem[{Gordon {et~al.}(2017{\natexlab{b}})Gordon, Palastro, \&
  Hafizi}]{gordon_superponderomotive_2017}
Gordon, D.~F., Palastro, J.~P., \& Hafizi, B. 2017{\natexlab{b}}, Phys. Rev. A,
  95, 033403, publisher: American Physical Society

\bibitem[{Guépin {et~al.}(2020)Guépin, Cerutti, \&
  Kotera}]{guepin_proton_2020}
Guépin, C., Cerutti, B., \& Kotera, K. 2020, A\&A, 635, A138, publisher: EDP
  Sciences

\bibitem[{Hadad {et~al.}(2010)Hadad, Labun, Rafelski, Elkina, Klier, \&
  Ruhl}]{hadad_effects_2010}
Hadad, Y., Labun, L., Rafelski, J., {et~al.} 2010, Phys. Rev. D, 82, 096012

\bibitem[{Heintzmann \& Schrüfer(1973)}]{heintzmann_exact_1973}
Heintzmann, H. \& Schrüfer, E. 1973, Physics Letters A, 43, 287

\bibitem[{Herold {et~al.}(1985)Herold, Ertl, \& Ruder}]{herold_generation_1985}
Herold, H., Ertl, T., \& Ruder, H. 1985, Mitteilungen der Astronomischen
  Gesellschaft Hamburg, 63, 174

\bibitem[{Kalapotharakos {et~al.}(2018)Kalapotharakos, Brambilla, Timokhin,
  Harding, \& Kazanas}]{kalapotharakos_three-dimensional_2018}
Kalapotharakos, C., Brambilla, G., Timokhin, A., Harding, A.~K., \& Kazanas, D.
  2018, ApJ, 857, 44

\bibitem[{Kelner {et~al.}(2015)Kelner, Prosekin, \&
  Aharonian}]{kelner_synchro-curvature_2015}
Kelner, S.~R., Prosekin, A.~Y., \& Aharonian, F.~A. 2015, AJ, 149, 33

\bibitem[{Kulsrud(1972)}]{kulsrud_effect_1972}
Kulsrud, R.~M. 1972, The Astrophysical Journal Letters, 174, L25

\bibitem[{Landau \& Lifchitz(1989)}]{landau_physique_1989}
Landau, L. \& Lifchitz, E. 1989, Physique théorique : {Tome} 2, {Théorie} des
  champs (Moscou: Mir)

\bibitem[{Laue \& Thielheim(1986)}]{laue_acceleration_1986}
Laue, H. \& Thielheim, K.~O. 1986, ApJS, 61, 465

\bibitem[{Li {et~al.}(2021)Li, Decyk, Miller, Tableman, Tsung, Vranic, Fonseca,
  \& Mori}]{li_accurately_2021}
Li, F., Decyk, V.~K., Miller, K.~G., {et~al.} 2021, Journal of Computational
  Physics, 438, 110367

\bibitem[{Lorentz(1916)}]{lorentz_theory_1916}
Lorentz, H. A. H.~A. 1916, The theory of electrons and its applications to the
  phenomena of light and radiant heat (Leipzig : B.G. Teubner ; New York : G.E.
  Stechert)

\bibitem[{Mestel {et~al.}(1985)Mestel, Robertson, Wang, \&
  Westfold}]{mestel_axisymmetric_1985}
Mestel, L., Robertson, J.~A., Wang, Y.~M., \& Westfold, K.~C. 1985, Monthly
  Notices of the Royal Astronomical Society, 217, 443, aDS Bibcode:
  1985MNRAS.217..443M

\bibitem[{Michel \& Li(1999)}]{michel_electrodynamics_1999}
Michel, F. \& Li, H. 1999, Physics Reports, 318, 227

\bibitem[{Olver(2010)}]{olver_nist_2010}
Olver, F. W.~J. 2010, {NIST} handbook of mathematical functions (Cambridge ;
  New York: Cambridge University Press : National Institute of Standards and
  Technology (U.S.)), oCLC: ocn502037224

\bibitem[{Philippov \& Spitkovsky(2018)}]{philippov_ab-initio_2018}
Philippov, A.~A. \& Spitkovsky, A. 2018, ApJ, 855, 94

\bibitem[{Philippov {et~al.}(2015)Philippov, Spitkovsky, \&
  Cerutti}]{philippov_ab_2015}
Philippov, A.~A., Spitkovsky, A., \& Cerutti, B. 2015, The Astrophysical
  Journal Letters, 801, L19

\bibitem[{Piazza(2008)}]{piazza_exact_2008}
Piazza, A.~D. 2008, Lett Math Phys, 83, 305

\bibitem[{Pétri(2020)}]{petri_relativistic_2020}
Pétri, J. 2020, J. Plasma Phys., 86, 825860402, publisher: Cambridge
  University Press

\bibitem[{Pétri(2021)}]{petri_particle_2021}
Pétri, J. 2021, Monthly Notices of the Royal Astronomical Society, 503, 2123

\bibitem[{Rohrlich(2007)}]{rohrlich_classical_2007}
Rohrlich, F. 2007, Classical {Charged} {Particles}, 3rd edn. (Singapore ;
  Hackensack, NJ: World Scientific Pub Co Inc)

\bibitem[{Tomczak \& Pétri(2020)}]{tomczak_particle_2020}
Tomczak, I. \& Pétri, J. 2020, J. Plasma Phys., 86, 825860401, publisher:
  Cambridge University Press

\bibitem[{Vranic {et~al.}(2016)Vranic, Martins, Fonseca, \&
  Silva}]{vranic_classical_2016}
Vranic, M., Martins, J.~L., Fonseca, R.~A., \& Silva, L.~O. 2016, Computer
  Physics Communications, 204, 141

\end{thebibliography}

\end{document}